\documentclass[fleqn,usenatbib]{mnras}

\usepackage{newtxtext,newtxmath}

\usepackage[T1]{fontenc}
\usepackage{ae,aecompl}

\usepackage{graphicx}	
\usepackage{amsmath}	
\usepackage{amssymb}	
\usepackage{pdflscape}  
\usepackage{rotating}   
\usepackage{breakurl}

\title[Faint Standards for $ZYJHK$]{Faint Standards for $ZYJHK$\\ from the UKIDSS and VISTA Surveys}

\author[S. K. Leggett et al.]{
S. K. Leggett,$^{1}$\thanks{E-mail sleggett@gemini.edu}
Nicholas J. G. Cross,$^{2}$
and Nigel C. Hambly$^{2}$
\\
$^{1}$Gemini Observatory Northern Operations\\
$^{2}$Institute for Astronomy, University of Edinburgh, Royal Observatory, Blackford Hill, Edinburgh EH9~3HJ, UK\\
}

\date{Accepted XXX. Received YYY; in original form ZZZ}

\pubyear{2015}

\begin{document}
\label{firstpage}
\pagerange{\pageref{firstpage}--\pageref{lastpage}}
\maketitle


\begin{abstract}
The currently defined ``UKIRT Faint Standards'' have $JHK$ magnitudes between 10 and 15, with $K_{\rm median}=11.2$. These stars will be too bright for the next generation of large telescopes.
We have used multi-epoch observations taken as part of the UKIRT Infrared Deep Sky Survey (UKIDSS) 
and the Visible and Infrared Survey Telescope for Astronomy (VISTA) surveys
to identify non-variable stars with $JHK$ magnitudes in the range 16 -- 19. The stars were selected from the UKIDSS Deep  Extragalactic Survey (DXS)  and Ultra Deep Survey (UDS), the WFCAM calibration data (WFCAMCAL08B), 
the VISTA Deep Extragalactic Observations (VIDEO) and  UltraVISTA.
Sources selected from the near-infrared databases were paired with the Pan-STARRS Data Release 2
of optical to near-infrared photometry and the Gaia astrometric Data Release 2.  Colour indices and other measurements  were used to exclude  sources that did not appear to be simple single stars. From an initial selection of 169 sources, we present a final sample of 81 standard stars with $ZYJHK$ magnitudes, or a subset, each with 20 to 600 observations  in each filter. The new standards have
$Ks_{\rm median}=17.5$. The relative photometric uncertainty for the sample is  $< 0.006$ mag and the absolute uncertainty is estimated to be $\lesssim 0.02$ mag. The sources are distributed equatorially and are accessible from both hemispheres. 
\end{abstract}
\begin{keywords}
standards -- methods observational -- techniques photometric
\end{keywords}



\section{Introduction}

Optical and infrared sky surveys have produced data with excellent astrometric and photometric precision. In 2019, we are benefiting from 
data releases by ground- and space-based surveys that were many years in the planning. 

The Gaia mission \citep{Prusti2016} issued Data Release~2 on 2018 April 25. This release contains astrometric results for about 1.7 billion stars brighter than optical magnitude 21; parallaxes and proper motions are given for
about 1.3 billion of these \citep{Brown2018}.
The Sloan Digital Sky Survey \citep[SDSS, ][]{York2000} and the  Panoramic Survey Telescope \& Rapid Response System \citep[Pan-STARRS, ][]{Chambers2016}
have imaged large areas of the Northern sky in blue/green to far-red/near-infrared filters. 
\citet{Tonry2012} give transformations between the SDSS and Pan-STARRS photometric systems. 
After transformation, the rms difference between the SDSS and Pan-STARRS $griz$ photometry is  8~mmag, following a
recalibration of the SDSS photometry with new flat fields and zero points derived from Pan-STARRS 
\citep{Finkbeiner2016}. This is consistent with \citet{Padmanabhan2008} who report a relative calibration accuracy of 0.7 -- 1.3\% for each filter in the SDSS.

Several near-infrared surveys of the sky have been undertaken in the last two decades. 
The 2-micron All Sky Survey \citep[2MASS,][]{Skrutskie2006} was executed between 1997 and 2001, with 1.3~m telescopes in both hemispheres providing complete sky coverage. 
The UKIRT Infrared Deep Sky Survey \citep[UKIDSS,][]{Lawrence2007} 
consisted of several Northern-hemisphere sub-surveys using the WFCAM camera and $ZYJHK$ filters, and was executed between 2007 and 2011 on the 3.8~m UKIRT on Mauna~Kea.
Following UKIDSS, in 2012, the UKIRT Hemisphere Survey \citep[UHS, ][]{Dye2018} began, and aims to provide continuous coverage in the $J$ and $K$ bands over the 
Declination range of zero to $+60$ degrees. A collaboration between the University of Hawaii and 
the United States Naval Observatory is continuing the UKIRT survey and adding the $H$ band \citep{Hodapp2018}.
In the Southern hemisphere, the 4.1~m Visible and Infrared Survey Telescope for Astronomy \citep[VISTA][]{Sutherland2015} started surveying the sky in 2009; the first surveys are complete or nearing completion as of  2019\footnote{\burl{https://www.eso.org/sci/observing/PublicSurveys/sciencePublicSurveys.html}};
VISTA is producing several sub-surveys using the $ZYJHKs$ filters, or a subset. 
\citet{Gonzalez2018}
compare the VISTA and UKIDSS photometric systems using equatorial stars observed with both cameras. The differences are small, and after transformation the rms difference between the measurement sets is 1 -- 3 mmag at $JHK$. 

The near-infrared surveys cover the entire sky and provide  1 -- 10 sources per square arcminute on average. Such stars could  be used to calibrate science data to  $\sim$10\% \citep{Gonzalez2018, Hodgkin2009}.
Photometric standard stars are needed to calibrate data more accurately. 
In the near-infrared these have frequently been provided by UKIRT measurements published by \citet{Hawarden2001} and  \citet{Leggett2006}. 
The UKIRT Faint Standards (FS) 
cover a range in Declination of $-25$ to $+55$~degrees. They
have $JHK$ Vega magnitudes of 10 -- 15 with 
$K_{\rm median}=11.2$ and $\sigma_{\rm median}=0.01$.
The UKIRT FS require integration times of 1 -- 10 seconds when observed on 8~m telescopes such as at the Gemini Observatory. 
On future telescopes with diameters three or more times larger, the integration times will be $< 1$~s;
such short integrations are not only inefficient, they may give rise to calibration problems such as poor linearity corrections \citep[e.g. ][]{Leggett2006}.

In this paper we identify stars in the UKIDSS and VISTA surveys that are fainter than the UKIRT FS by a factor of $\sim$100. We select sources that have a large number of repeat measurements which show them to be non-variable, and which have high precision photometric measurements. We pair the list of candidates with the Pan-STARRS and Gaia databases and use the optical and near-infrared colours to refine the sample and produce a set of well-measured and well-behaved single  stars.

\section{The Surveys}

The Northern hemisphere UKIDSS (with the WFCAM camera) and the Southern hemisphere VISTA use similar-size telescopes, and   similar filters defined according to the Mauna Kea Observatories filter specifications \citep{Tokunaga2002}.
Survey data generated by both telescopes are initially processed by the Cambridge Astronomy Survey Unit (CASU) and then transferred to the Wide-Field Astronomy Unit (WFAU) in Edinburgh for further processing and archiving \citep{Cross2012,Hambly2008}. Table~1 gives a summary of the properties of the surveys used in this work.

\begin{table}
\begin{scriptsize}
	\caption{Overview of the Sky Survey Data}
	\label{Tab1}
	\begin{tabular}{lrccc} 
		\hline
		Survey & Area & Filters & Vega mag &   Ref.\\
       Name & deg$^2$ &        &  $5\sigma$ Limits   &  \\
		\hline
UKIDSS Deep  Extragalactic  & 35.0 & $JHK$ & 21 -- 22 & 1 \\
Survey (DXS) &  &  & & \\
\hline
UKIDSS Ultra & 0.8 & $JHK$ & 23 -- 25 & 1 \\
Deep Survey (UDS)$^a$ &  &  & & \\
\hline
VISTA Deep Extragalactic  & 12.0 & $ZYJHKs$ & 24 -- 25 & 2\\
Observations (VIDEO) & &  & & \\
\hline
VISTA UltraVISTA$^b$ & 1.5 & $YJHKs$ & 22 -- 24 &  3\\
\hline
WFCAM Calibration & 10.4 & $ZYJHK$ & 18 -- 19 & 4\\
WFCAMCAL08  &  &  & & \\
		\hline
	\end{tabular}
	References: (1) \citet{Lawrence2007}, (2) \citet{Jarvis2013},  (3) \citet{McCracken2012}, (4) \citet{Ferreira2015}.\\
$^a$ For the UDS, the source table aperture magnitudes are not aperture corrected and the variability table mean magnitudes are. For a sample of 30 objects, we find an average offset in photometry of $0.190 \pm 0.005$, $0.202 \pm 0.006$, and $0.179 \pm 0.006$, at $J$, $H$ and $K$ respectively, such that the mean magnitudes are brighter.\\
$^b$ For UltraVISTA there are differences in the aperture corrections and zeropoints used for the source table aperture magnitudes and the variability table mean magnitudes. For a sample of 19 objects we find {\tt meanMag} $-$ {\tt aperMag3} $= 0.020 \pm 0.005$, $0.052 \pm 0.006$, $-0.025 \pm 0.010$, and $-0.004 \pm 0.012$ at $Y$,  $J$, $H$ and $Ks$ respectively.
\end{scriptsize}
\end{table}

Each of the surveys has a source table (e.g. \verb+dxsSource+) which contains band-merged detections from deep images. 
In addition, for the surveys used here, there is a variability table (e.g. \verb+dxsVariability+) which contains the light-curve statistics for each primary source that is detected in at least one epoch. Details of the multi-epoch table structure and processing can be found in \cite{Cross2009}.
The number of good observations for each filter and each source is given as {\tt nGoodObs} in the variability table; these are measurements which are not flagged as blended, saturated, etc. using the post-processing error bit flag ({\tt ppErrBits}), i.e. a good observation has {\tt ppErrBits=0}\footnote{\burl{http://wsa.roe.ac.uk/ppErrBits.html}}. The source and variability tables share the same primary key {\bf sourceID}.

Our goal is to identify stars that are not variable, that have well-measured  $JHK$ magnitudes, and are fainter than $K = 16$. Therefore we only used surveys with variability tables for all of the $J$, $H$, and $K$ filters. 
This means we include the UKIDSS Deep Extragalactic Survey (DXS), Ultra Deep Survey (UDS) and WFCAMCAL calibration data, but do not include the UKIDSS Galactic Clusters, Galactic Plane or Large Area Surveys (GCS, GPS, LAS). Similarly we use the VISTA Deep Extragalactic Observations Survey (VIDEO) and UltraVISTA surveys, but do not include the VISTA Kilo-Degree Infrared Galaxy, Magellanic Clouds, Variables in the Via Lactea,  or Hemisphere Surveys (VIKING, VMC, VVV, VHS). The DXS, UDS, WFCAMCAL and VIDEO surveys are complete.
The UltraVISTA survey is being continued\footnote{\burl{https://www.eso.org/sci/observing/PublicSurveys/sciencePublicSurveys.html\#vistacycle2}} with a final data release planned for 2021\footnote{\burl{https://www.eso.org/sci/observing/PublicSurveys/docs/UltraVISTA2_SMP_03022017.pdf}}.

The variability flag for each source ({\tt variableClass}) is determined from repeat measurements and is stored in the variability  table \citep{Cross2009}. The value is determined by the significance of the weighted average of the intrinsic noise over the expected noise, as given by a noise model for each pointing and each band. The weighting of the intrinsic noise is based on the number of good observations: $w_f=\frac{N_{obs,f}-N_{min}}{N_{obs,max}-N_{min}}$, where $N_{min}=5$ is the minimum number of observations necessary to measure variability, $N_{obs,f}$ is the number of good observations in that band, and $N_{obs,max}$ is the maximum number of good observations for the star in any band.  The default classification is non-variable ({\tt variableClass} $= 0$), and  if a star has fewer than 5 good observations in all bands it will be classified as non-variable; none of the objects selected here fall into this category. If a star has fewer than 5 observations in all bands but one, that band will determine the classification. If there are hundreds of observations in one band and tens in the others, the band with hundreds will have a strong weighting compared to the others. The star is classified as variable ({\tt variableClass} $= 1$) if the ratio of the weighted intrinsic noise to the expected noise is $> 3$.

We adopt the mean photometric magnitudes and uncertainties given in the  variability tables as the reference calibration data in this work, and not the source table aperture magnitudes, for the following reasons.
In some cases, the source tables do not go as deep as the stacked variability images; this is the case for the WFCAMCAL08 database  where it was important to avoid source blending for calibration purposes.  For the
Ultra Deep Survey the source table photometry {\tt AperMag} 
is not aperture-corrected, while the variability photometry  {\tt MeanMag} is (Table 1, Section 3). For the most recently processed survey, UltraVISTA, there are small differences between the variability and source table photometry due to different
aperture corrections 
and Vega-to-AB zeropoint
corrections (Table 1, Section 3).
These issues will be corrected in a forthcoming UltraVISTA release. 
The variability photometry is processed in the same way for all the surveys used here, and so provides a self-consistent sample.

\section{The SQL Selections}

We used the WFCAM Science Archive\footnote{\burl{http://wsa.roe.ac.uk}} and the VISTA 
Science Archive\footnote{\burl{http://surveys.roe.ac.uk/vsa}} to query the DXS, UDS, UltraVISTA, VIDEO and WFCAMCAL databases. The most recent data releases available at the time of writing were used: Data Release~11 of the DXS and UDS was used, Data Release~4 for UltraVISTA,  Data Release~5 for VIDEO, and WFCAMCAL08B.
For all surveys we selected for the photometric source to be the same as the source in the variability catalogue and for non-variable sources, for example {\tt SELECT FROM dxsSource AS s,	dxsVariability AS v WHERE s.sourceID=v.sourceID AND v.variableClass=0}. 

We selected for point sources using the surveys' morphological classifications. For the shallower surveys  -- DXS and WFCAMCAL -- the classification scheme uses 
a statistic which describes how point-like each object is with respect to an empirically derived, idealized radial profile representing the point source function (PSF) for the frame \citep{Irwin2004, Hambly2008}. The deeper surveys -- UDS, UltraVISTA and VIDEO --  
use the TERAPIX SWARP image resampling tool \citep{Bertin2002} and the CLASS\_STAR statistic\footnote{\burl{https://sextractor.readthedocs.io/en/latest/ClassStar.html}}
generated by the SEXTRACTOR software  \citep{Bertin1996}, together with magnitude cuts in each band 
\citep{Liske2003, Warren2007}. We found that we could select a good-sized sample of objects (more than ten) from each of the  DXS, UDS, and WFCAMCAL surveys by selecting for stars only, using
{\tt s.mergedClass=-1}. For  VIDEO we relaxed the selection to stars and probable stars, with {\tt  s.mergedClass in (-1, -2)}, in order to get a useful sample.
For UltraVISTA we found that the mergedClass statistic in the source table classifies all sources with multi-filter photometry as galaxies. A visual inspection of the sources in the stacked images showed that the PSFs in the $Y$ and $J$ images are extended. We expect that future releases of the ongoing UltraVISTA survey will include improved morphological classifications.   
For this work, for UltraVISTA, we selected for possible stars by using the class statistic determined from the $H$ and $K$ images, {\tt s.hclassStat>0.7 AND s.ksclassStat>0.7}, and we use additional profile and colour selections to separate galaxies from stars (Section 6). 


We selected for precise photometry by restricting the uncertainty in the mean magnitude to $\leq 0.006$ mag.
The photometric uncertainties are given  in the variability catalogue by 
the rms value of the multiple measurements  ({\tt MagRms}) and the median absolute deviation of the magnitude  
 ({\tt MagMAD}): we  selected for sources where these values scaled as expected with {\tt nGoodObs}, for example {\tt (v.jMagMAD/SQRT(v.jnGoodObs - 1))<=0.004 AND  
 (v.jMagRms/SQRT(v.jnGoodObs - 1))<=0.006}. 

For the DXS, VIDEO and WFCAMCAL surveys we selected for consistency between the source and variability table magnitudes by limiting the difference
to $2.5~\sigma$ where $\sigma$ is determined from the variability rms and the source aperture magnitude error, for example, {\tt
(s.jAperMag3 - v.jMeanMag) < 2.5 * (SQRT(
s.jAperMag3err *  s.jAperMag3err + 
(v.jMagRms/SQRT(v.jnGoodObs - 1)) 
* (v.jMagRms/SQRT(v.jnGoodObs - 1))))}. 
For the UDS and UltraVISTA surveys, where there are systematic differences between the source and variability table magnitudes (Section 2 and Table 1), we selected for objects with a small range around the average offset; for example for the UDS we used {\tt(s.jAperMag3 - v.jMeanMag < 0.20) AND 
(s.jAperMag3 - v.jMeanMag > 0.18)}, and for UltraVISTA we used
{\tt ((s.jAperMag3 - v.jMeanMag) <= -0.04) AND ((s.jAperMag3 - v.jMeanMag)  >= -0.08)}.

A limit on target declination was also implemented. In Sections 6 and 7 we use  Pan-STARRS optical data, together with the near-infrared data, to further refine the star/galaxy separation and to remove sources which may be multiple, as evidenced by unusual colours. For this reason we restricted our searches to declination $> -30^{\circ}$.

The constraints on brightness and number of observations in each filter varied with each survey. 
Sources with $K > 16$ or $K > 16.5$  were selected from the DXS, UltraVISTA and
WFCAMCAL, while sources with $K > 17.5$ were selected from the deeper UDS and VIDEO surveys (Table 1). For the data to be of calibration quality, we adopt a minimum number of measurements of 
20 for each object in each filter.
Less than twenty $H$-band observations were obtained for some objects in the DXS, and for that survey we restricted the search to a minimum of 20 observations in $J$ and $K$ and $5$ in $H$, in order to use the colour information. The minimum number of observations for the other surveys ranged from 20 to 100 for each filter. 
While the $JHK$ filters are our priority in this work, we also record the $Z$ and $Y$ magnitudes where available.

The SQL queries used for each survey in their complete form are given in the Appendix. The searches produced 169 sources: 34 from the DXS, 51 from the UDS, 25 from UltraVISTA, 47 from the VIDEO survey, and 12 from WFCAMCAL.

\section{The WFCAM-UKIDSS and VISTA Photometric Systems}

Both WFCAM and VISTA use filters specified by the Mauna Kea Observatories system  \citep{Tokunaga2002}. There are small differences in the filters as delivered, and differences in the site and telescope optics, which lead to small differences between the native UKIDSS and VISTA photometric systems. \citet{Gonzalez2018} use a large sample of reddening-free stars with photometric errors $< 0.1$~mag,  observed with both cameras, to derive the following color transformations between VISTA and WFCAM:
$$ Z_V - Z_W = -(0.037\pm 0.008)\times (J - K)_W + (0.040\pm 0.005)$$
$$ Y_V - Y_W = -(0.010\pm 0.003)\times (J - K)_W - (0.048\pm 0.002)$$
$$ J_V - J_W = -(0.028\pm 0.002)\times (J - K)_W - (0.004\pm 0.001)$$
$$ H_V - H_W = -(0.037\pm 0.001)\times (J - K)_W + (0.025\pm 0.001)$$
$$ Ks_V - K_W = (0.017\pm 0.003)\times (J - K)_W - (0.022\pm 0.002)$$
VISTA survey observations are continuing, and the photometry presented here can be enhanced by future UltraVISTA data releases, hence
we convert the UKIDSS photometry to the VISTA system and use that as our reference system in this work.

The  UKIDSS UDS field (Data Release 11) and the VISTA VIDEO  XMM-Newton field (Data Release 5) overlap by 0.1~deg in Right Ascension and 0.9~deg in Declination. A search for non-variable point sources in the common region with   {\tt AperMag3err} 
$\leq 0.006$ 
produced a sample of 43 stars with $H$ and $K$ magnitudes between 13 and 16 (no $J$ measurements were available). Allowing for small color transformations (see above), we find a mean difference between the UDS and VIDEO survey magnitudes of 
0.010~mag at $H$ and $K$. These results suggest that the color-transformed standard star photometry presented in this paper is robust at the 1\% level.

\section{Matching to Pan-STARRS and Gaia}

The coordinates of the sources produced by our SQL searches of the WFCAM and VISTA science archives, described above,  were matched to 
Data Release 2 of the Pan-STARRS data archive\footnote {\burl{https//catalogs.mast.stsci.edu/}} and
Data Release 2 of the Gaia data 
archive\footnote{\burl{https//gea.esac.esa.int/archive/}}, which were the most recent releases at the time of writing.
The Pan-STARRS $grizy$ magnitudes ranged from 18 to 23. 

For about one-third of the sample, one or two of the Pan-STARRS filters have an aperture and PSF magnitude that differ by more than $2.5\sigma$. About 60\% of these we identify as possible galaxies below. The majority of the remainder have PSF and aperture magnitude differences that are not large in absolute terms, and it is possible that the uncertainties are slightly underestimated in this Pan-STARRS release.

Sources with Pan-STARRS AB magnitudes $r \lesssim 21$ and  $i \lesssim 20.5$ were detected by Gaia,  consistent with the $G < 21$ limit of Gaia Data Release 2 \citep{Brown2018}. Of those, about 85\%  have proper motion measurements (all $\lesssim~30$~mas~yr$^{-1}$). 
About one-third of the sources with proper motion measurements have a trigonometric parallax measurement that is positive and has an uncertainty smaller than the parallax measurement. Fifteen   of the 169 objects found in our searches have a Gaia parallax measurement that is significant. 
The next release of Gaia data, expected in the third quarter of 2020, will include additional and improved astrometry\footnote{\burl{https://www.cosmos.esa.int/web/gaia/release}}.

\section{Star/Galaxy Separation}

Our goal here is to provide as clean a sample as possible of truly point source calibrators, so that, even if imaged at very high resolution, the aperture corresponding to the photometry is unambiguous. We therefore further prune the sample by excluding possible galaxies.


\begin{figure}
\vskip -0.15in
	\includegraphics[angle=-90,width=0.53\textwidth]{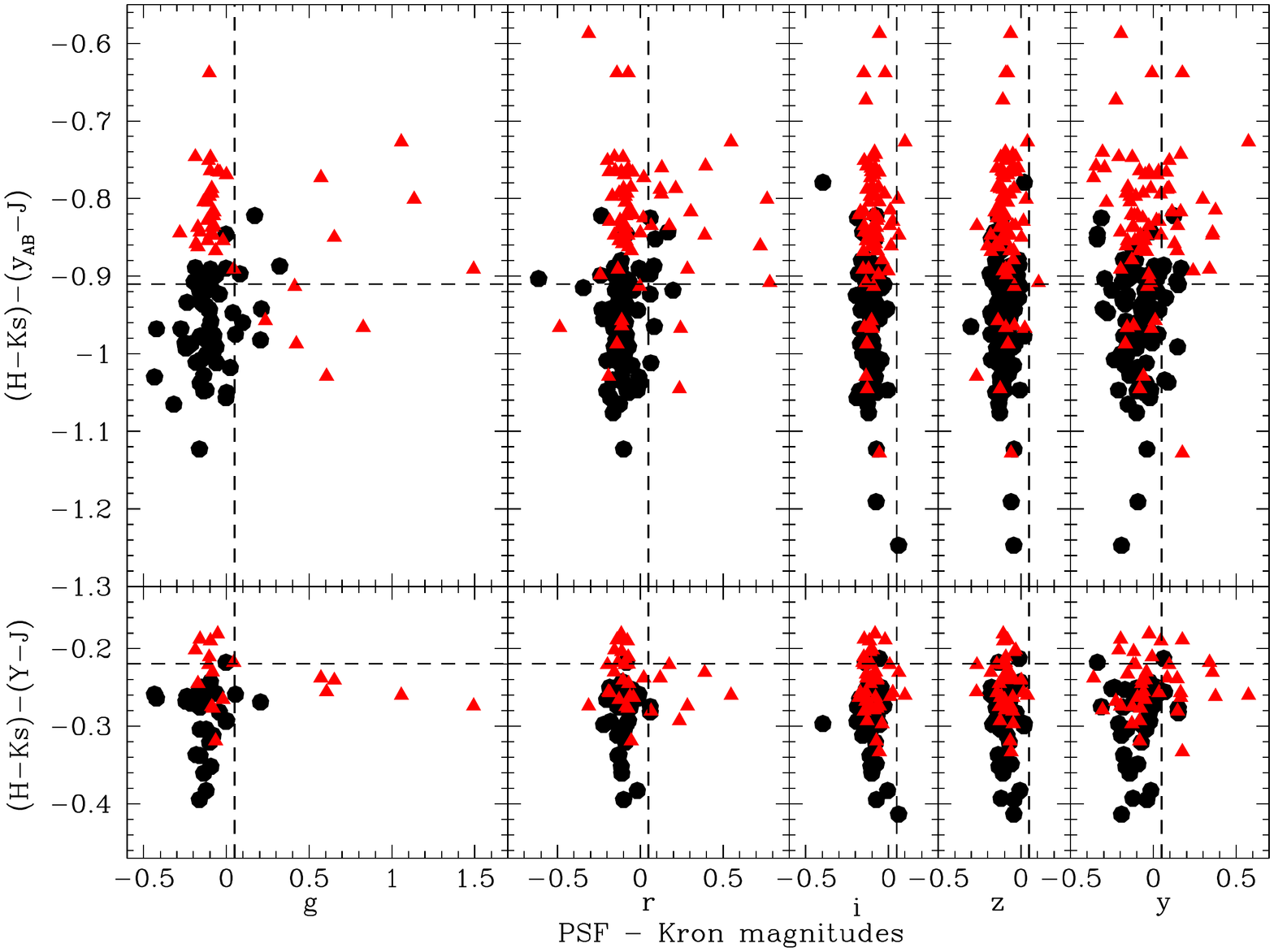}
\vskip -0.1in
   \caption{Red triangles are identified as possible galaxies either by the PSF $-$ Kron magnitude or the colour index. The limits on these values are indicated by the dashed lines, see Section 6.}
    \label{fig1}
\end{figure}

\citet{Davies2018} determined that stars and galaxies can be separated in VISTA data by using a near-infrared colour index. \citet{Davies2018} select for galaxies by applying the colour cut $(H - Ks) - (Y - J) > -0.26$, and verified the selection by visual inspection of objects brighter than $Y = 21.2$. Not all of our sample has a $Y$-band measurement and we determined  the $(H - Ks) - (y_{AB} - J)$ color, replacing the VISTA $Y_{Vega}$ with  the Pan-STARRS $y_{AB}$ magnitude. Calibrating that index against the VISTA index,  we adopt as a stellar indicator
$(H - Ks) - (Y - J) < -0.22$ or 
$(H - Ks) - (y_{AB} - J) < -0.91$.

We also explored the use of the Pan-STARRS PSF and Kron magnitudes, searching for the brighter Kron magnitudes that would be expected if the source was extended \citep[e.g.][their Figure 19]{Chambers2016}. \citet{Chambers2016} find that stars can be selected by  PSF $-$ Kron $<$ 0.05 magnitude, for magnitudes as faint as $\sim$ 21. 

Figure 1 combines these two indicators, plotting $(H - Ks) - (Y - J)$ and  $(H - Ks) - (y_{AB} - J)$ against PSF $-$ Kron magnitudes for each of the Pan-STARRS filters. The star/galaxy cuts are shown in the Figure; we excluded as possible galaxies sources which are either too red in one or both of the color indices, or too bright in any of the  PSF $-$ Kron colours. The sources that remain are shown in black in Figure 1, and lie to the lower left, or have error bars that would place them in the lower left, of each panel (error bars are omitted from the Figure for clarity). These cuts identified 77 of the 169 sources as possible galaxies, and we omit them from the standard star sample. The Appendix Table B1 lists the possible galaxies together with Pan-STARRS and VISTA-system photometry, and Gaia data where available.

\section{Exclusion of Multiple Systems and Other Contaminants}

We further refine the likely-star sample of 92 objects by excluding sources with atypical colours. These may be multiple systems, or the colours may be compromised in some way. To produce a conservative sample, we reject outliers from colour sequences prescribed by the majority of the sample. Figure 2 shows bluer colors --- $g_{AB} - r_{AB}$, $r_{AB} - i_{AB}$, $i_{AB} - z_{AB}$  --- and Figure 3  redder colours --- $z_{AB} - y_{AB}$, $y_{AB} - J_{Vega}$, $J_{Vega} - H_{Vega}$. We identify 7 sources as outliers; these are shown as open circles in Figures 2 and 3 and listed in Appendix Table C1.

 \begin{figure}
\vskip -0.5in
	\includegraphics[angle=0,width=0.5\textwidth]{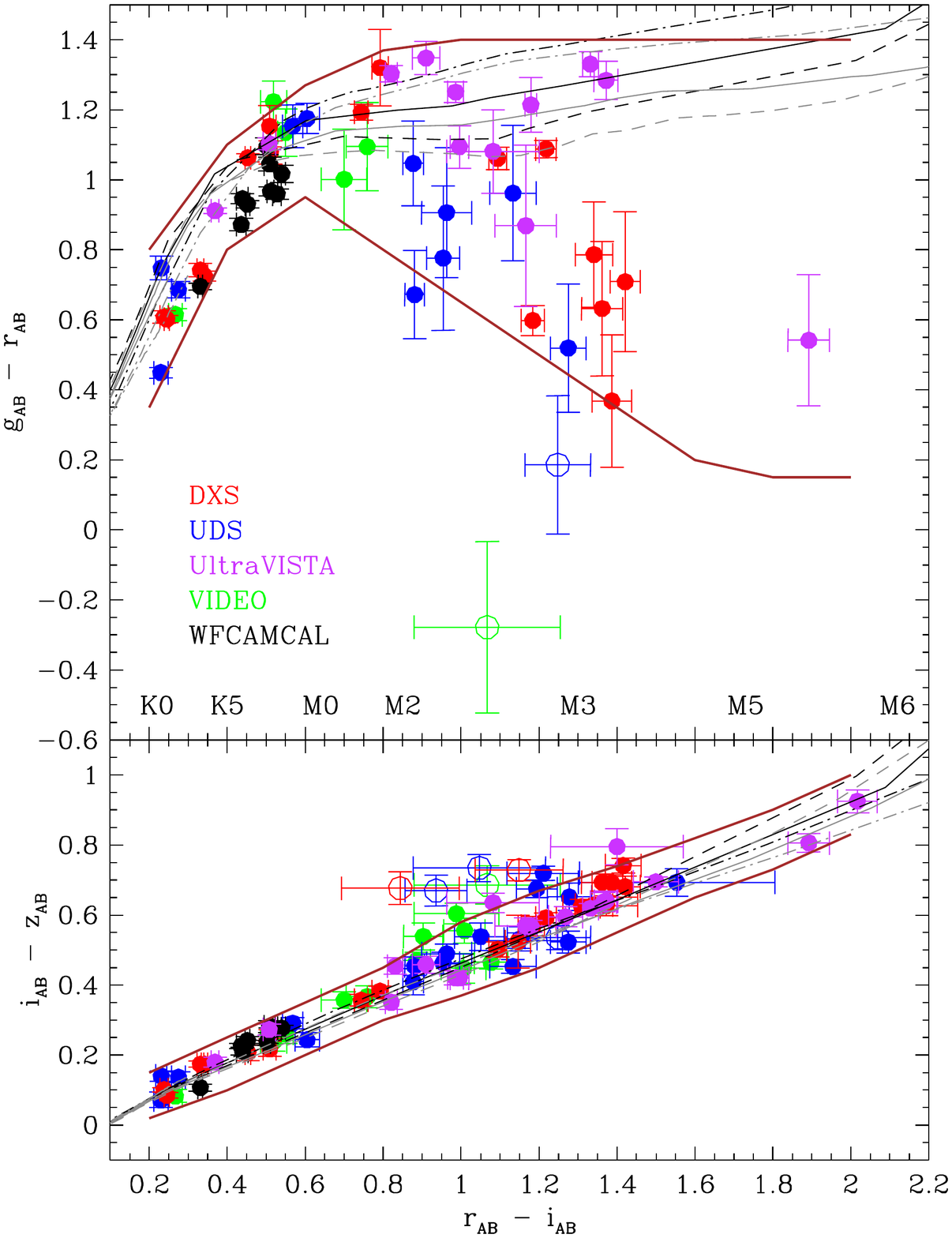}
\vskip -0.3in
\caption{Colours of the likely stellar sources. No reddening correction has been applied. Open circles are objects with one or more colours that deviate from the loci identified by the sample.  The thinner lines are model isochrones from the   
BT-SETTL set (black, \citet{Allard2012,Baraffe2015}) and the PARSEC set (grey, \citet{Bressan2012, Girardi2000}).
For these sequences, age in Gyr and [m/H] are: 5, 0.0 (solid line); 1.0, $+0.3$ (dashed line); 10, $-0.5$ (dash-dot line). The thick brown lines indicate the regions adopted here as defining normal star colours. Two sources are excluded in the top panel, being either too blue in $g_{AB} - r_{AB}$ or in $r_{AB} - i_{AB}$. One of these is also an outlier in the lower panel, along with four other sources that appear too blue in $r_{AB} - i_{AB}$
or too red in $i_{AB} - z_{AB}$.  Spectral types along the $x$ axis are from the relationship between  $r_{AB} - i_{AB}$ and type given by \citet{Covey2007}.
}
    \label{fig2}
\end{figure}

\begin{figure}
\vskip -0.5in
	\includegraphics[angle=0,width=0.5\textwidth]{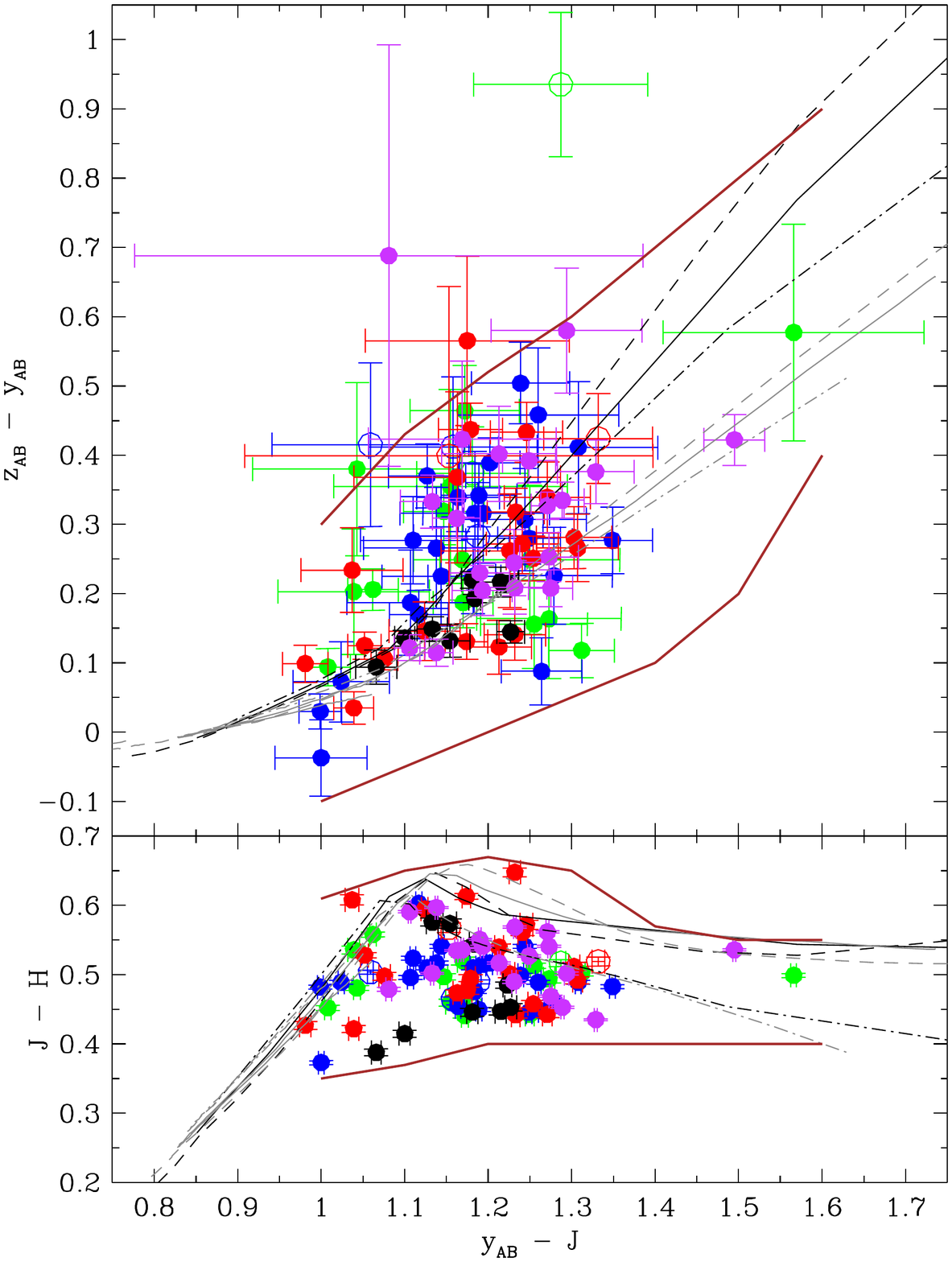}
\vskip -0.3in
   \caption{Colours of the likely stellar sources with symbols and lines as in Figure 2. No reddening correction has been applied. In addition to the 6 outliers identified in Figure 2, one additional outlier is identified in the upper panel, as too red in $z_{AB} - y_{AB}$ or too blue in $y_{AB} - J$.}
    \label{fig3}
\end{figure}

Figures 2 and 3 include color sequences  produced by stellar model atmospheres. We used the  BT-SETTL models \citep{Allard2012,Baraffe2015} and the PARSEC models  \citep{Bressan2012, Girardi2000}, for a range of metallicity as described in the Figure caption.
For both model sets the $g - r$ colour deviates significantly from observation for M-type stars; this may be due to some issue common to the model atmospheres or it may be due to errors in the $g$ bandpass adopted by both modelling groups, which could introduce a colour term.  The colors and absolute magnitudes (where available), combined with the models, imply that the sample consists of dwarf stars with masses between 0.1 and 1.0~M$_{\odot}$.

Lastly, for the DXS, VIDEO and WFCAMCAL surveys, where there is no offset between the source table aperture magnitudes and the
variability table mean magnitudes (Sections 2, 3), 
we excluded sources where these values deviated significantly.
The earlier selections excluded discrepant sources based on the variability rms, now we exclude sources based on the variability 
MAD values.  We excluded sources where the aperture and mean magnitudes differed by $> 0.03$~mag {\it and} the difference was significant by $> 2.5~\sigma$. Four additional objects
were rejected by this criterion and these are listed in Appendix Table C2. All of these rejected sources are at the faint end of the associated survey. 

After exclusion of these 11 objects, the final sample consists of 81 stars ranging in spectral type from K2 to M6.

\section{Final  Sample}

Table 2 compiles the Pan-STARRS, UKIDSS/VISTA and Gaia data for the final sample of 81 stars. Spectral types are also given; these are estimated from  the $r_{AB} - i_{AB}$ color \citep{Covey2007} where available. Where this color was not available the type was estimated by interpolating the $i_{AB} - z_{AB}$ or $y_{AB} - J_{Vega}$ colors, using stars with all three colors to define the relationships.

\newgeometry{width=20cm,height=13cm}
\begin{landscape}
\begin{table*}
\begin{tiny}
	\caption{Data for Final Sample}
			\setlength\tabcolsep{1pt}
	\begin{tabular}{rrcrrrrrrrrrrrrrrrr}
		\hline
 \multicolumn{2}{c}{IR Survey} & Spectral &    \multicolumn{6}{c}{Gaia} &    \multicolumn{5}{c}{Pan-STARRS meanPSF AB mags} & 
   \multicolumn{5}{c}{VISTA Vega mags}\\
   Name & RA$^{\circ}$     & Type		&	RA$^{\circ}$ 	&	Decl.$^{\circ}$	& Parallax  & $\mu$ RA		      &	$\mu$ Decl.	 	& G			&	$g$ &	$r$ 	&	$i$	& $z$ 	& $y$ 	& $Z$		& $Y$	&	$J$	&	$H$	& $Ks$	\\
     & Decl.$^{\circ}$	&		&	$\pm$ mas		&	$\pm$ mas   	& $\pm$  mas & $\pm$ mas yr$^{-1}$ &	$\pm$ mas yr$^{-1}$ 	& 		& $\pm$ 	& $\pm$ 	& $\pm$ 	& $\pm$  	& $\pm$ 	&  $\pm$	&	 $\pm$		&	 $\pm$				&	 $\pm$		&	 $\pm$			\\
\hline
UDS	&	34.0342392	&	M3.5	&		&		&		&		&		&		&		&	21.7683	&	20.5566	&	19.8377	&	19.5578	&		&		&	18.3094	&	17.8626	&	17.5573	\\
	&	-5.1728814	&		&		&		&		&		&		&		&		&	0.0918	&	0.0133	&	0.0158	&	0.0340	&		&		&	0.0020	&	0.0018	&	0.0032	\\
UDS	&	34.0585392	&	M3	&		&		&		&		&		&		&		&	21.6694	&	20.5205	&	19.9937	&	19.6780	&		&		&	18.4943	&	18.0184	&	17.7527	\\
	&	-5.3530358	&		&		&		&		&		&		&		&		&	0.1112	&	0.0192	&	0.0112	&	0.0886	&		&		&	0.0020	&	0.0020	&	0.0032	\\
UDS	&	34.0906904	&	M2.5	&	34.090724762816	&	-5.187250050454	&		&		&		&	20.8948	&	22.0173	&	21.1110	&	20.1476	&	19.6591	&	19.5706	&		&		&	18.3070	&	17.8586	&	17.5862	\\
	&	-5.1871909	&		&	1.72235	&	2.62332	&		&		&		&		&	0.1755	&	0.0610	&	0.0198	&	0.0199	&	0.0440	&		&		&	0.0020	&	0.0020	&	0.0031	\\
UDS	&	34.1166206	&	M0	&	34.116703956194	&	-5.433148535694	&		&	31.4547	&	-14.0542	&	20.5484	&	21.6757	&	20.5011	&	19.8962	&	19.6520	&	19.4654	&		&		&	18.3581	&	17.8624	&	17.6619	\\
	&	-5.4330768	&		&	1.13832	&	1.14117	&		&	1.9471	&	1.7836	&		&	0.0312	&	0.0299	&	0.0088	&	0.0175	&	0.0736	&		&		&	0.0020	&	0.0018	&	0.0032	\\
UDS	&	34.1459342	&	M3	&		&		&		&		&		&		&		&	21.9087	&	20.6313	&	19.9788	&	19.5901	&		&		&	18.3878	&	17.8711	&	17.5971	\\
	&	-5.4192513	&		&		&		&		&		&		&		&		&	0.0484	&	0.0303	&	0.0248	&	0.0264	&		&		&	0.0021	&	0.0018	&	0.0033	\\
UDS	&	34.1559576	&	M3	&		&		&		&		&		&		&	22.4436	&	21.4817	&	20.3488	&	19.8964	&	19.6713	&		&		&	18.4886	&	17.9790	&	17.7385	\\
	&	-5.1864870	&		&		&		&		&		&		&		&	0.1855	&	0.0578	&	0.0138	&	0.0136	&	0.0520	&		&		&	0.0021	&	0.0020	&	0.0033	\\
UDS	&	34.1991747	&	M2	&	34.199178767942	&	-4.947697715516	&		&		&		&	20.8574	&	22.0767	&	21.0300	&	20.1533	&	19.7429	&	19.4767	&		&		&	18.3392	&	17.8223	&	17.5819	\\
	&	-4.9476391	&		&	2.11700	&	2.55827	&		&		&		&		&	0.1193	&	0.0229	&	0.0130	&	0.0348	&	0.0801	&		&		&	0.0021	&	0.0019	&	0.0033	\\
UDS	&	34.2232606	&	M2.5	&		&		&		&		&		&		&	22.0736	&	21.2984	&	20.3444	&	19.8807	&	19.6561	&		&		&	18.5122	&	17.9708	&	17.7333	\\
	&	-4.7491396	&		&		&		&		&		&		&		&	0.2023	&	0.0397	&	0.0146	&	0.0271	&	0.0661	&		&		&	0.0022	&	0.0021	&	0.0034	\\
UDS	&	34.2308454	&	M3	&		&		&		&		&		&		&		&	21.7580	&	20.5639	&	19.8892	&	19.5468	&		&		&	18.3580	&	17.9076	&	17.6218	\\
	&	-5.3911233	&		&		&		&		&		&		&		&		&	0.0466	&	0.0234	&	0.0114	&	0.0442	&		&		&	0.0021	&	0.0020	&	0.0032	\\
UDS	&	34.3103343	&	M4.5	&		&		&		&		&		&		&		&		&	20.9380	&	20.1853	&	19.9076	&		&		&	18.5593	&	18.0758	&	17.7539	\\
	&	-5.4172864	&		&		&		&		&		&		&		&		&		&	0.0301	&	0.0446	&	0.0168	&		&		&	0.0022	&	0.0018	&	0.0034	\\
UDS	&	34.3240632	&	M5	&		&		&		&		&		&		&		&		&	20.9455	&	20.1946	&	19.7373	&		&		&	18.4768	&	17.9876	&	17.6696	\\
	&	-5.4334966	&		&		&		&		&		&		&		&		&		&	0.0271	&	0.0264	&	0.0933	&		&		&	0.0023	&	0.0020	&	0.0034	\\
UDS	&	34.3475622	&	M3	&		&		&		&		&		&		&	22.1287	&	21.6100	&	20.3351	&	19.8115	&	19.5857	&		&		&	18.3072	&	17.8462	&	17.5767	\\
	&	-5.4844981	&		&		&		&		&		&		&		&	0.1780	&	0.0413	&	0.0208	&	0.0229	&	0.0656	&		&		&	0.0020	&	0.0018	&	0.0032	\\
UDS	&	34.3750969	&	M4	&		&		&		&		&		&		&		&	22.2721	&	20.7193	&	20.0259	&	19.7191	&		&		&	18.4748	&	17.9338	&	17.6445	\\
	&	-5.4848715	&		&		&		&		&		&		&		&		&	0.2496	&	0.0288	&	0.0261	&	0.0687	&		&		&	0.0022	&	0.0019	&	0.0034	\\
UDS	&	34.4389254	&	M3	&		&		&		&		&		&		&		&	21.5469	&	20.3938	&	19.8619	&	19.4921	&		&		&	18.3653	&	17.8549	&	17.6085	\\
	&	-5.4868910	&		&		&		&		&		&		&		&		&	0.0350	&	0.0244	&	0.0254	&	0.0394	&		&		&	0.0021	&	0.0020	&	0.0033	\\
UDS	&	34.5319836	&	M2	&		&		&		&		&		&		&	21.7700	&	21.0978	&	20.2171	&	19.7631	&	19.4859	&		&		&	18.3762	&	17.8516	&	17.6315	\\
	&	-5.1015107	&		&		&		&		&		&		&		&	0.1245	&	0.0211	&	0.0142	&	0.0218	&	0.0591	&		&		&	0.0020	&	0.0018	&	0.0033	\\
UDS	&	34.6700469	&	M5.5	&		&		&		&		&		&		&		&		&	20.8852	&	20.0903	&	19.5863	&		&		&	18.3469	&	17.8485	&	17.5109	\\
	&	-5.0787756	&		&		&		&		&		&		&		&		&		&	0.0378	&	0.0342	&	0.0477	&		&		&	0.0024	&	0.0020	&	0.0036	\\
UDS	&	34.6776197	&	M2.5	&		&		&		&		&		&		&		&	21.5419	&	20.4913	&	19.9533	&	19.6368	&		&		&	18.4426	&	17.9281	&	17.6586	\\
	&	-4.9671258	&		&		&		&		&		&		&		&		&	0.0643	&	0.0223	&	0.0325	&	0.0645	&		&		&	0.0022	&	0.0019	&	0.0035	\\
UDS	&	34.7079081	&	K3	&	34.707936489500	&	-4.903956579988	&		&	6.0938	&	-0.6828	&	19.8705	&	20.3459	&	19.6602	&	19.3849	&	19.2484	&	19.1746	&		&		&	18.1506	&	17.6618	&	17.5555	\\
	&	-4.9039390	&		&	1.49577	&	1.02274	&		&	2.2974	&	1.7844	&		&	0.0188	&	0.0144	&	0.0105	&	0.0102	&	0.0573	&		&		&	0.0017	&	0.0017	&	0.0029	\\
UDS	&	34.7425103	&	K2	&	34.742551524244	&	-5.478149266917	&		&	-1.6880	&	-2.0504	&	19.7838	&	20.4217	&	19.6735	&	19.4427	&	19.3034	&	19.2729	&		&		&	18.2738	&	17.7916	&	17.6957	\\
	&	-5.4781043	&		&	0.84654	&	0.62984	&		&	1.3900	&	1.0100	&		&	0.0316	&	0.0109	&	0.0086	&	0.0109	&	0.0226	&		&		&	0.0019	&	0.0018	&	0.0029	\\
UDS	&	34.7745424	&	M5	&		&		&		&		&		&		&		&		&	20.9764	&	20.1684	&	19.7567	&		&		&	18.4493	&	17.9604	&	17.6400	\\
	&	-4.8889668	&		&		&		&		&		&		&		&		&		&	0.0321	&	0.0280	&	0.0913	&		&		&	0.0022	&	0.0020	&	0.0034	\\
UDS	&	34.8172698	&	M4	&		&		&		&		&		&		&		&		&	20.8266	&	20.1456	&	19.8074	&		&		&	18.6426	&	18.1886	&	17.9029	\\
	&	-4.7942724	&		&		&		&		&		&		&		&		&		&	0.0220	&	0.0338	&	0.0365	&		&		&	0.0022	&	0.0019	&	0.0033	\\
UDS	&	34.8182276	&	K7	&	34.818219802904	&	-5.106312522351	&		&		&		&	20.6559	&	21.8021	&	20.6494	&	20.0812	&	19.7897	&	19.6196	&		&		&	18.5033	&	17.9004	&	17.7308	\\
	&	-5.1062739	&		&	1.77157	&	1.21047	&		&		&		&		&	0.0558	&	0.0243	&	0.0099	&	0.0119	&	0.0328	&		&		&	0.0021	&	0.0019	&	0.0034	\\
UDS	&	34.8710949	&	K2	&	34.871103834372	&	-5.115517311973	&		&	6.0251	&	-7.9278	&	19.6855	&	20.0623	&	19.6130	&	19.3828	&	19.3110	&	19.3479	&		&		&	18.3481	&	17.9747	&	17.8797	\\
	&	-5.1154861	&		&	0.67365	&	0.56550	&		&	0.9709	&	0.8974	&		&	0.0075	&	0.0131	&	0.0133	&	0.0175	&	0.0523	&		&		&	0.0016	&	0.0019	&	0.0029	\\
VIDEO	&	35.0542019	&	M0	&		&		&		&		&		&		&	21.9615	&	20.9614	&	20.2612	&	19.9043	&	19.7169	&		&	19.0018	&	18.5465	&	18.0267	&	17.8310	\\
	&	-4.4813805	&		&		&		&		&		&		&		&	0.1321	&	0.0562	&	0.0170	&	0.0159	&	0.0324	&		&	0.0021	&	0.0028	&	0.0047	&	0.0043	\\
VIDEO	&	35.1152774	&	M4.5	&		&		&		&		&		&		&		&		&	20.8536	&	20.1628	&	19.6987	&		&	19.0704	&	18.5271	&	18.0857	&	17.8255	\\
	&	-4.9003215	&		&		&		&		&		&		&		&		&		&	0.0245	&	0.0260	&	0.0604	&		&	0.0020	&	0.0021	&	0.0032	&	0.0049	\\
VIDEO	&	35.3795003	&	M2.5	&		&		&		&		&		&		&		&	21.5975	&	20.5879	&	20.0334	&	19.6534	&		&	19.1025	&	18.6101	&	18.1286	&	17.9108	\\
	&	-4.6057616	&		&		&		&		&		&		&		&		&	0.0449	&	0.0275	&	0.0215	&	0.1227	&		&	0.0023	&	0.0023	&	0.0028	&	0.0049	\\
VIDEO	&	35.6598904	&	K7	&	35.659947796971	&	-4.635110377207	&	2.9760	&	-1.5893	&	-8.0207	&	20.4536	&	21.5459	&	20.4112	&	19.8614	&	19.5991	&	19.3927	&		&	18.7513	&	18.3307	&	17.7733	&	17.6104	\\
	&	-4.6351333	&		&	0.97484	&	1.29221	&	1.2256	&	2.1361	&	3.3777	&		&	0.0663	&	0.0149	&	0.0130	&	0.0262	&	0.0153	&		&	0.0016	&	0.0020	&	0.0030	&	0.0032	\\
VIDEO	&	35.6610332	&	M2.5	&		&		&		&		&		&		&		&	21.5461	&	20.5580	&	19.9542	&	19.7055	&		&	19.0260	&	18.5364	&	18.0755	&	17.8501	\\
	&	-4.5453215	&		&		&		&		&		&		&		&		&	0.1076	&	0.0137	&	0.0262	&	0.0518	&		&	0.0019	&	0.0021	&	0.0025	&	0.0039	\\
VIDEO	&	35.6959814	&	K7	&		&		&		&		&		&		&	22.1388	&	20.9156	&	20.3969	&	20.1589	&	19.9562	&		&	19.3279	&	18.9173	&	18.3808	&	18.1880	\\
	&	-4.5264450	&		&		&		&		&		&		&		&	0.0521	&	0.0287	&	0.0167	&	0.0209	&	0.0883	&		&	0.0026	&	0.0036	&	0.0043	&	0.0049	\\
VIDEO	&	35.7281547	&	M3.5	&		&		&		&		&		&		&		&		&	21.0309	&	20.3774	&	20.0223	&		&	19.3492	&	18.8672	&	18.4067	&	18.1372	\\
	&	-5.2037306	&		&		&		&		&		&		&		&		&		&	0.0243	&	0.0327	&	0.1362	&		&	0.0027	&	0.0029	&	0.0041	&	0.0053	\\
VIDEO	&	35.8660837	&	M3	&		&		&		&		&		&		&		&	21.5821	&	20.5765	&	20.1285	&	19.8094	&		&	19.1724	&	18.6624	&	18.1648	&	17.9286	\\
	&	-4.1075563	&		&		&		&		&		&		&		&		&	0.0493	&	0.0305	&	0.0296	&	0.0382	&		&	0.0024	&	0.0024	&	0.0042	&	0.0050	\\
VIDEO	&	35.9623324	&	M2	&		&		&		&		&		&		&		&	21.1034	&	20.2167	&	19.7468	&	19.6287	&		&	18.8035	&	18.3172	&	17.8155	&	17.5798	\\
	&	-4.2237615	&		&		&		&		&		&		&		&		&	0.0327	&	0.0131	&	0.0196	&	0.0342	&		&	0.0017	&	0.0021	&	0.0034	&	0.0045	\\
VIDEO	&	35.9681759	&	M2.5	&		&		&		&		&		&		&		&	21.4215	&	20.3440	&	19.8795	&	19.6303	&		&	18.8680	&	18.3799	&	17.9367	&	17.7016	\\
	&	-4.2573895	&		&		&		&		&		&		&		&		&	0.0414	&	0.0137	&	0.0135	&	0.1066	&		&	0.0021	&	0.0028	&	0.0028	&	0.0042	\\
VIDEO	&	35.9729949	&	M1	&	35.972994037179	&	-5.204738379604	&		&	5.0723	&	2.6179	&	20.7741	&	21.9219	&	20.8275	&	20.0676	&	19.7011	&	19.5449	&		&	18.7733	&	18.2901	&	17.7774	&	17.5529	\\
	&	-5.2047350	&		&	2.08523	&	1.37865	&		&	2.9317	&	3.4628	&		&	0.1150	&	0.0521	&	0.0119	&	0.0286	&	0.0575	&		&	0.0025	&	0.0022	&	0.0028	&	0.0039	\\
VIDEO	&	36.0194131	&	M6	&		&		&		&		&		&		&		&		&	21.4811	&	20.5903	&	20.0130	&		&	19.1804	&	18.4466	&	17.9480	&	17.6283	\\
	&	-4.8728386	&		&		&		&		&		&		&		&		&		&	0.1512	&	0.0271	&	0.1545	&		&	0.0022	&	0.0024	&	0.0021	&	0.0044	\\
VIDEO	&	36.7075066	&	K2	&	36.707528955400	&	-4.530284911233	&		&	-0.4147	&	-9.8394	&	20.0023	&	20.4814	&	19.8664	&	19.5993	&	19.5159	&	19.4218	&	18.9232	&	18.7517	&	18.4144	&	17.9624	&	17.8677	\\
	&	-4.5302703	&		&	0.79715	&	0.72506	&		&	1.7904	&	1.5098	&		&	0.0114	&	0.0133	&	0.0118	&	0.0144	&	0.0234	&	0.0017	&	0.0016	&	0.0022	&	0.0034	&	0.0040	\\
CAL	&	87.7352518	&	K3	&	87.735239861645	&	15.881631420747	&		&	-0.8779	&	-2.5404	&	18.2273	&	18.9028	&	18.2084	&	17.8773	&	17.7695	&	17.6758	&	17.2730	&	17.0544	&	16.6104	&	16.2224	&	16.0999	\\
	&	15.8816295	&		&	0.15854	&	0.14113	&		&	0.3188	&	0.2820	&		&	0.0078	&	0.0054	&	0.0026	&	0.0087	&	0.0229	&	0.0070	&	0.0035	&	0.0031	&	0.0041	&	0.0060	\\
CAL	&	88.0439399	&	K5	&	88.043961153500	&	16.212597884799	&	0.3641	&	2.3316	&	-5.6403	&	18.7329	&	19.6508	&	18.7046	&	18.2656	&	18.0549	&	17.9226	&	17.4837	&	17.2840	&	16.7693	&	16.1954	&	16.0186	\\
	&	16.2125856	&		&	0.59256	&	0.52929	&	0.3362	&	0.8564	&	0.7802	&		&	0.0083	&	0.0118	&	0.0087	&	0.0058	&	0.0226	&	0.0095	&	0.0044	&	0.0048	&	0.0036	&	0.0051	\\
UVISTA	&	149.7053794	&	M3	&	149.705377978507	&	2.486235956106	&		&		&		&	20.7270	&	21.9821	&	21.1130	&	19.9467	&	19.3778	&	19.1248	&		&	18.3682	&	17.8524	&	17.3109	&	17.0759	\\
	&	2.4862294	&		&	6.20199	&	11.01370	&		&		&		&		&	0.2167	&	0.0765	&	0.0180	&	0.0154	&	0.0320	&		&	0.0013	&	0.0009	&	0.0010	&	0.0010	\\
UVISTA	&	149.7087571	&	M2.5	&	149.708761363556	&	2.78887908216031	&		&		&		&	20.6709	&	22.0938	&	20.8438	&	19.8575	&	19.4370	&	19.1278	&		&	18.4555	&	17.9656	&	17.4313	&	17.2116	\\
	&	2.7888722	&		&	3.41061	&	5.09192	&		&		&		&		&	0.0243	&	0.0166	&	0.0112	&	0.0171	&	0.0227	&		&	0.0011	&	0.0008	&	0.0012	&	0.0018	\\
		\hline
	\end{tabular}
\\
\end{tiny}
\end{table*}
\end{landscape}
\restoregeometry

\newgeometry{width=20cm,height=13cm}
\begin{landscape}
\begin{table*}
\begin{tiny}
	\contcaption{Data for Final Sample}
			\setlength\tabcolsep{1pt}
	\begin{tabular}{rrcrrrrrrrrrrrrrrrr}
		\hline
 \multicolumn{2}{c}{IR Survey} & Spectral &    \multicolumn{6}{c}{Gaia} &    \multicolumn{5}{c}{Pan-STARRS meanPSF AB mags} & 
   \multicolumn{5}{c}{VISTA Vega mags}\\
   Name & RA$^{\circ}$     & Type		&	RA$^{\circ}$ 	&	Decl.$^{\circ}$	& Parallax  & $\mu$ RA		      &	$\mu$ Decl.	 	& G			&	$g$ &	$r$ 	&	$i$	& $z$ 	& $y$ 	& $Z$		& $Y$	&	$J$	&	$H$	& $Ks$			\\
     & Decl.$^{\circ}$	&		&	$\pm$ mas		&	$\pm$ mas   	& $\pm$  mas & $\pm$ mas yr$^{-1}$ &	$\pm$ mas yr$^{-1}$ 	& 		& $\pm$ 	& $\pm$ 	& $\pm$ 	& $\pm$  	& $\pm$ 	&  $\pm$	&	 $\pm$		&	 $\pm$				&	 $\pm$		&	 $\pm$			\\
\hline
UVISTA	&	149.7234241	&	M3.5	&		&		&		&		&		&		&		&	22.7104	&	21.3098	&	20.5154	&	19.9353	&		&	19.2576	&	18.6410	&	18.1390	&	17.8376	\\
	&	2.7511805	&		&		&		&		&		&		&		&		&	0.1649	&	0.0422	&	0.0297	&	0.0848	&		&	0.0012	&	0.0012	&	0.0019	&	0.0021	\\
UVISTA	&	149.7264829	&	M3.5	&	149.726469124665	&	2.590250565341	&		&		&		&	20.8861	&	22.7546	&	21.4708	&	20.0987	&	19.4587	&	19.0669	&		&	18.3791	&	17.8180	&	17.2913	&	17.0327	\\
	&	2.5902461	&		&	6.90065	&	7.34634	&		&		&		&		&	0.0478	&	0.0270	&	0.0148	&	0.0238	&	0.0208	&		&	0.0013	&	0.0007	&	0.0011	&	0.0012	\\
UVISTA	&	149.7307359	&	M3.5	&		&		&		&		&		&		&		&	21.8113	&	20.4456	&	19.8085	&	19.4070	&		&	18.7503	&	18.1943	&	17.6766	&	17.4237	\\
	&	2.7399894	&		&		&		&		&		&		&		&		&	0.0307	&	0.0130	&	0.0246	&	0.0639	&		&	0.0010	&	0.0008	&	0.0013	&	0.0019	\\
UVISTA	&	149.7313560	&	M2	&	149.731351713864	&	2.692886882358	&	2.2849	&	-5.1377	&	-2.4384	&	19.9948	&	21.4538	&	20.1508	&	19.3297	&	18.9789	&	18.7493	&		&	18.0241	&	17.5590	&	17.0082	&	16.7933	\\
	&	2.6928870	&		&	0.63176	&	0.62314	&	0.8261	&	1.2600	&	1.1541	&		&	0.0214	&	0.0114	&	0.0106	&	0.0185	&	0.0312	&		&	0.0008	&	0.0006	&	0.0009	&	0.0009	\\
UVISTA	&	149.731965	&	M4	&		&		&		&		&		&		&		&		&	21.7754	&	21.0125	&	20.3254	&		&	19.8434	&	19.2441	&	18.7651	&	18.4652	\\
	&	2.7327131	&		&		&		&		&		&		&		&		&		&	0.0568	&	0.0570	&	0.2994	&		&	0.0022	&	0.0020	&	0.0025	&	0.0033	\\
UVISTA	&	149.7332239	&	M5.5	&		&		&		&		&		&		&	22.8180	&	22.2761	&	20.3844	&	19.5777	&	19.2018	&		&	18.4389	&	17.8730	&	17.4382	&	17.1655	\\
	&	2.7947267	&		&		&		&		&		&		&		&	0.1809	&	0.0487	&	0.0215	&	0.0164	&	0.0430	&		&	0.0009	&	0.0008	&	0.0013	&	0.0016	\\
UVISTA	&	149.7355822	&	M2.5	&		&		&		&		&		&		&	22.3015	&	21.2084	&	20.2115	&	19.7921	&	19.5830	&		&	18.8246	&	18.3514	&	17.7820	&	17.5625	\\
	&	2.3205009	&		&		&		&		&		&		&		&	0.0571	&	0.0209	&	0.0122	&	0.0141	&	0.0365	&		&	0.0010	&	0.0007	&	0.0012	&	0.0014	\\
UVISTA	&	149.7360512	&	K4	&	149.736045993165	&	2.487259844103	&	-0.2321	&	-7.3325	&	-5.4547	&	19.5169	&	20.3606	&	19.4492	&	19.0804	&	18.8997	&	18.7779	&		&	18.0802	&	17.6720	&	17.0809	&	16.9594	\\
	&	2.4872598	&		&	0.64011	&	0.37359	&	0.9176	&	0.8382	&	0.8325	&		&	0.0066	&	0.0042	&	0.0089	&	0.0113	&	0.0148	&		&	0.0008	&	0.0006	&	0.0009	&	0.0010	\\
UVISTA	&	149.7476065	&	M2.5	&		&		&		&		&		&		&	22.9265	&	21.8455	&	20.7642	&	20.1292	&	19.8838	&		&	19.1804	&	18.6534	&	18.1630	&	17.9131	\\
	&	2.7272746	&		&		&		&		&		&		&		&	0.0222	&	0.1171	&	0.0209	&	0.0178	&	0.0443	&		&	0.0016	&	0.0014	&	0.0017	&	0.0027	\\
UVISTA	&	149.7564883	&	M3.5	&	149.756487939153	&	2.727999832329	&		&		&		&	20.5141	&	22.3374	&	21.0074	&	19.6753	&	19.0564	&	18.7287	&		&	17.9949	&	17.4583	&	16.8964	&	16.6443	\\
	&	2.7279976	&		&	2.92877	&	5.22498	&		&		&		&		&	0.0312	&	0.0179	&	0.0104	&	0.0098	&	0.0172	&		&	0.0011	&	0.0008	&	0.0010	&	0.0012	\\
UVISTA	&	149.7597522	&	M6	&		&		&		&		&		&		&		&	22.6823	&	20.6647	&	19.7397	&	19.3184	&		&	18.4764	&	17.8232	&	17.2872	&	16.9664	\\
	&	2.7827541	&		&		&		&		&		&		&		&		&	0.0444	&	0.0250	&	0.0208	&	0.0301	&		&	0.0012	&	0.0009	&	0.0014	&	0.0015	\\
UVISTA	&	149.7742789	&	K7	&	149.774279147971	&	2.668706544853	&	-0.3755	&	-6.9845	&	-12.8963	&	19.8543	&	20.9477	&	19.8446	&	19.3370	&	19.0638	&	18.9487	&		&	18.2324	&	17.8107	&	17.2140	&	17.0529	\\
	&	2.6687115	&		&	0.63748	&	0.58222	&	0.7819	&	1.0446	&	0.9791	&		&	0.0150	&	0.0085	&	0.0144	&	0.0084	&	0.0180	&		&	0.0011	&	0.0008	&	0.0010	&	0.0010	\\
UVISTA	&	149.7772394	&	M3	&		&		&		&		&		&		&		&	21.7842	&	20.5162	&	19.9210	&	19.4984	&		&	18.8640	&	18.3290	&	17.7934	&	17.5463	\\
	&	2.7783479	&		&		&		&		&		&		&		&		&	0.0347	&	0.0261	&	0.0117	&	0.1116	&		&	0.0011	&	0.0011	&	0.0014	&	0.0018	\\
UVISTA	&	149.7923194	&	M4	&	149.792307399543	&	2.583274196954	&		&		&		&	20.7672	&		&	21.4016	&	19.9016	&	19.2066	&	18.8721	&		&	18.1225	&	17.5830	&	17.1304	&	16.8712	\\
	&	2.5832843	&		&	2.90648	&	4.29789	&		&		&		&		&		&	0.0565	&	0.0098	&	0.0096	&	0.0238	&		&	0.0012	&	0.0006	&	0.0011	&	0.0010	\\
UVISTA	&	149.8156044	&	M2	&	149.815608819914	&	2.769757311106	&		&		&		&	20.5042	&	21.9587	&	20.6105	&	19.7011	&	19.2435	&	18.9104	&		&	18.2690	&	17.7772	&	17.2745	&	17.0546	\\
	&	2.7697500	&		&	2.38224	&	4.30129	&		&		&		&		&	0.0351	&	0.0316	&	0.0148	&	0.0193	&	0.0332	&		&	0.0009	&	0.0010	&	0.0010	&	0.0013	\\
UVISTA	&	149.8172140	&	M2	&	149.817198369447	&	2.771754406963	&		&		&		&	20.3904	&	22.0959	&	20.4851	&	19.6542	&	19.1997	&	18.9953	&		&	18.2823	&	17.8016	&	17.2609	&	17.0372	\\
	&	2.7717426	&		&	2.10211	&	3.85719	&		&		&		&		&	0.0587	&	0.0162	&	0.0122	&	0.0203	&	0.0301	&		&	0.0009	&	0.0008	&	0.0013	&	0.0014	\\
UVISTA	&	149.8251519	&	M3	&	149.825158826105	&	2.760822268756	&		&		&		&	20.7489	&	22.2857	&	21.0721	&	19.8930	&	19.3248	&	19.1166	&		&	18.3345	&	17.8417	&	17.3732	&	17.1457	\\
	&	2.7608222	&		&	4.45694	&	7.53329	&		&		&		&		&	0.0777	&	0.0076	&	0.0134	&	0.0105	&	0.0236	&		&	0.0010	&	0.0007	&	0.0012	&	0.0014	\\
CAL	&	276.7225557	&	K7	&	276.722547796086	&	4.058718034922	&	0.7440	&	1.2330	&	-1.1142	&	18.8017	&	19.9119	&	18.8163	&	18.3138	&	18.0759	&	17.8826	&	17.4972	&	17.2159	&	16.6992	&	16.1551	&	15.9994	\\
	&	4.0587376	&		&	0.25266	&	0.27647	&	0.2437	&	0.4722	&	0.5167	&		&	0.0217	&	0.0091	&	0.0033	&	0.0028	&	0.0053	&	0.0081	&	0.0042	&	0.0036	&	0.0034	&	0.0052	\\
CAL	&	276.8280740	&	K5	&	276.828044639757	&	3.977392480520	&	0.4415	&	-1.0556	&	-4.3463	&	18.4684	&	19.3584	&	18.4862	&	18.0496	&	17.8239	&	17.6885	&	17.2663	&	17.0325	&	16.5879	&	16.1728	&	16.0038	\\
	&	3.9774041	&		&	0.15888	&	0.16831	&	0.1767	&	0.3142	&	0.3214	&		&	0.0131	&	0.0119	&	0.0033	&	0.0070	&	0.0089	&	0.0074	&	0.0035	&	0.0036	&	0.0033	&	0.0051	\\
CAL	&	277.2399424	&	K7	&	277.239920836661	&	4.095191453128	&	0.4534	&	-1.9862	&	-5.2619	&	18.9085	&	19.9814	&	18.9638	&	18.4238	&	18.1466	&	17.9265	&	17.5724	&	17.2598	&	16.7456	&	16.3002	&	16.0376	\\
	&	4.0952285	&		&	0.25057	&	0.29023	&	0.3339	&	0.4730	&	0.4341	&		&	0.0228	&	0.0087	&	0.0042	&	0.0067	&	0.0097	&	0.0085	&	0.0041	&	0.0037	&	0.0037	&	0.0052	\\
CAL	&	277.2699877	&	K7	&	277.269972705599	&	4.522125403396	&	0.3592	&	-4.1032	&	-2.9879	&	18.9311	&	19.9941	&	18.9491	&	18.4395	&	18.1591	&	18.0144	&	17.6120	&	17.3401	&	16.7872	&	16.3344	&	16.1180	\\
	&	4.5221425	&		&	0.26166	&	0.26634	&	0.3066	&	0.5879	&	0.6956	&		&	0.0158	&	0.0124	&	0.0083	&	0.0151	&	0.0058	&	0.0081	&	0.0040	&	0.0037	&	0.0037	&	0.0056	\\
DXS	&	333.0711763	&	M1	&	333.071203143918	&	0.830304033314	&	0.1228	&	21.2977	&	-2.2154	&	19.6927	&	20.9885	&	19.7961	&	19.0524	&	18.6951	&	18.5724	&		&		&	17.3587	&	16.8191	&	16.5811	\\
	&	0.8303505	&		&	0.61772	&	0.78507	&	0.5681	&	1.0132	&	1.1778	&		&	0.0180	&	0.0141	&	0.0108	&	0.0117	&	0.0366	&		&		&	0.0027	&	0.0029	&	0.0046	\\
DXS	&	333.0715557	&	M3	&	333.071561677249	&	0.916173667634	&		&		&		&	20.9872	&	21.9931	&	21.3950	&	20.2114	&	19.6434	&	19.3805	&		&		&	18.1551	&	17.6539	&	17.3931	\\
	&	0.9162355	&		&	27.75785	&	57.45897	&		&		&		&		&	0.0362	&	0.0237	&	0.0173	&	0.0216	&	0.0790	&		&		&	0.0035	&	0.0036	&	0.0054	\\
DXS	&	333.0718253	&	K7	&	333.071813320305	&	1.188906051043	&	1.2559	&	-0.0237	&	-4.0006	&	20.2384	&	21.3538	&	20.2010	&	19.6916	&	19.4740	&	19.2395	&		&		&	18.2026	&	17.5949	&	17.4465	\\
	&	1.1889522	&		&	1.68843	&	1.04016	&	1.9038	&	1.6055	&	1.9193	&		&	0.0572	&	0.0135	&	0.0086	&	0.0194	&	0.0583	&		&		&	0.0042	&	0.0055	&	0.0059	\\
DXS	&	333.2839533	&	M3	&	333.283923388425	&	-0.410674164956	&		&		&		&	20.7836	&		&	21.3054	&	19.9960	&	19.3718	&	19.0911	&		&		&	17.7879	&	17.2757	&	17.0097	\\
	&	-0.4105822	&		&	1.74142	&	1.61698	&		&		&		&		&		&	0.0441	&	0.0187	&	0.0091	&	0.0436	&		&		&	0.0030	&	0.0050	&	0.0048	\\
DXS	&	333.2928434	&	M3	&	333.292847124344	&	0.734269316808	&	3.8847	&	-11.7293	&	-21.1246	&	20.3301	&	21.8557	&	20.7682	&	19.5504	&	18.9582	&	18.6398	&		&		&	17.4070	&	16.9648	&	16.6916	\\
	&	0.7343005	&		&	0.79955	&	0.90143	&	0.9405	&	1.4193	&	1.5556	&		&	0.0122	&	0.0223	&	0.0140	&	0.0034	&	0.0338	&		&		&	0.0028	&	0.0033	&	0.0048	\\
DXS	&	333.3028790	&	M4	&	333.302887172419	&	0.754442344733	&		&		&		&	20.6888	&	21.9774	&	21.2677	&	19.8465	&	19.1663	&	18.8274	&		&		&	17.5557	&	17.1132	&	16.8090	\\
	&	0.7544734	&		&	7.12835	&	11.75424	&		&		&		&		&	0.1965	&	0.0374	&	0.0103	&	0.0277	&	0.0427	&		&		&	0.0035	&	0.0037	&	0.0054	\\
DXS	&	333.3594622	&	M3	&		&		&		&		&		&		&		&	21.5006	&	20.3548	&	19.8311	&	19.4634	&		&		&	18.3002	&	17.8259	&	17.5602	\\
	&	0.7512320	&		&		&		&		&		&		&		&		&	0.0285	&	0.0143	&	0.0738	&	0.0992	&		&		&	0.0034	&	0.0034	&	0.0057	\\
DXS	&	333.3824122	&	M4		&	& & & & & & & 21.7048	&	20.2887	&	19.5484	&		18.9835	&		&		&	17.8077	&	17.3309	&	16.9979	\\
	&	0.7424491	&			&	& & & & & & & 0.0436	&	0.0132	&	0.0174	&		0.1214	&		&		&	0.0028	&	0.0037	&	0.0048	\\
DXS	&	333.4515020	&	M3	&	333.451575578969	&	0.727800712380	&	-0.7282	&	20.6346	&	3.6240	&	20.6231	&	21.9420	&	21.1563	&	19.8150	&	19.1856	&	18.9350	&		&		&	17.6807	&	17.2226	&	16.9545	\\
	&	0.7278254	&		&	0.86841	&	1.12594	&	1.3384	&	2.9172	&	2.2283	&		&	0.1427	&	0.0470	&	0.0100	&	0.0124	&	0.0300	&		&		&	0.0044	&	0.0032	&	0.0049	\\
DXS	&	333.4987510	&	M4	&		&		&		&		&		&		&	21.9939	&	21.6259	&	20.2389	&	19.5468	&	19.1100	&		&		&	17.9313	&	17.4356	&	17.1440	\\
	&	0.7349581	&		&		&		&		&		&		&		&	0.1835	&	0.0487	&	0.0139	&	0.0187	&	0.0334	&		&		&	0.0034	&	0.0043	&	0.0052	\\
DXS	&	333.7094552	&	K5	&	333.709446334135	&	0.524569397526	&	0.3710	&	-5.8741	&	-3.8524	&	19.2601	&	20.2572	&	19.1939	&	18.7410	&	18.5351	&	18.3885	&		&		&	17.2650	&	16.6699	&	16.5297	\\
	&	0.5245924	&		&	0.38446	&	0.40323	&	0.4482	&	0.7133	&	0.7032	&		&	0.0115	&	0.0043	&	0.0057	&	0.0215	&	0.0366	&		&		&	0.0024	&	0.0032	&	0.0064	\\
DXS	&	333.9366041	&	M1	&	333.936644901728	&	-0.947205672205	&	1.9967	&	0.3402	&	-7.7910	&	19.7115	&	21.1755	&	19.8556	&	19.0643	&	18.6820	&	18.5408	&		&		&	17.3086	&	16.6614	&	16.4942	\\
	&	-0.9471891	&		&	0.64218	&	0.51022	&	0.7047	&	1.0896	&	0.9006	&		&	0.1083	&	0.0193	&	0.0091	&	0.0108	&	0.0351	&		&		&	0.0024	&	0.0058	&	0.0050	\\
DXS	&	334.6493784	&	M4	&		&		&		&		&		&		&		&	21.5236	&	20.1513	&	19.5239	&	19.0910	&		&		&	17.8448	&	17.2731	&	16.9818	\\
	&	1.3819891	&		&		&		&		&		&		&		&		&	0.0793	&	0.0153	&	0.0244	&	0.0358	&		&		&	0.0039	&	0.0054	&	0.0051	\\
DXS	&	334.8182190	&	M4	&	334.818255423593	&	0.392802658682	&		&		&		&	20.7669	&	21.9217	&	21.2902	&	19.9285	&	19.2348	&	18.9691	&		&		&	17.6622	&	17.1697	&	16.8747	\\
	&	0.3928450	&		&	2.87510	&	1.83692	&		&		&		&		&	0.1852	&	0.0516	&	0.0094	&	0.0230	&	0.0427	&		&		&	0.0034	&	0.0044	&	0.0054	\\
DXS	&	334.9251383	&	K3	&	334.925138353308	&	0.727415208209	&		&	-0.8993	&	-5.5026	&	19.1814	&	19.8406	&	19.0993	&	18.7683	&	18.5935	&	18.4693	&		&		&	17.4172	&	16.8889	&	16.7754	\\
	&	0.7274321	&		&	0.51157	&	0.39138	&		&	0.6877	&	0.6893	&		&	0.0174	&	0.0075	&	0.0031	&	0.0080	&	0.0182	&		&		&	0.0030	&	0.0016	&	0.0047	\\
DXS	&	335.0303119	&	K2	&	335.030321608241	&	0.445380424500	&		&	5.6377	&	-5.3554	&	18.9125	&	19.4517	&	18.8504	&	18.6046	&	18.5203	&	18.4213	&		&		&	17.4399	&	17.0128	&	16.9227	\\
	&	0.4454373	&		&	0.76013	&	0.40567	&		&	0.7263	&	0.6658	&		&	0.0076	&	0.0056	&	0.0032	&	0.0067	&	0.0260	&		&		&	0.0040	&	0.0035	&	0.0049	\\
DXS	&	335.4746997	&	M2.5	&	335.474729551077	&	0.509174939517	&		&	12.0091	&	-1.3009	&	20.1233	&	21.5299	&	20.4693	&	19.3749	&	18.8733	&	18.6013	&		&		&	17.3604	&	16.7993	&	16.5495	\\
	&	0.5091841	&		&	0.91754	&	0.67359	&		&	1.2168	&	1.4291	&		&	0.0234	&	0.0223	&	0.0070	&	0.0201	&	0.0375	&		&		&	0.0031	&	0.0034	&	0.0049	\\
DXS	&	335.5234567	&	K3	&	335.523439344855	&	0.287394859518	&		&	4.9240	&	-7.9009	&	18.9031	&	19.5952	&	18.8704	&	18.5275	&	18.3543	&	18.2482	&		&		&	17.1721	&	16.6742	&	16.5423	\\
	&	0.2874459	&		&	0.50264	&	0.27839	&		&	0.5269	&	0.5531	&		&	0.0122	&	0.0083	&	0.0023	&	0.0114	&	0.0113	&		&		&	0.0036	&	0.0036	&	0.0052	\\
DXS	&	335.5823067	&	K7	&	335.582301746661	&	-0.144465229042	&		&	4.2254	&	-7.0727	&	19.4810	&	20.5273	&	19.4493	&	18.9384	&	18.6777	&	18.5472	&		&		&	17.3728	&	16.7596	&	16.6027	\\
	&	-0.1444080	&		&	0.43214	&	0.50392	&		&	0.9150	&	0.7789	&		&	0.0318	&	0.0079	&	0.0062	&	0.0110	&	0.0226	&		&		&	0.0031	&	0.0034	&	0.0042	\\
DXS	&	335.6150576	&	K2	&	335.615033009253	&	0.509092433091	&		&	-1.6997	&	-8.0080	&	18.7713	&	19.3124	&	18.7038	&	18.4660	&	18.3652	&	18.3305	&		&		&	17.2914	&	16.8690	&	16.7655	\\
	&	0.5091194	&		&	0.39047	&	0.28177	&		&	0.5121	&	0.5160	&		&	0.0146	&	0.0096	&	0.0035	&	0.0051	&	0.0227	&		&		&	0.0027	&	0.0036	&	0.0054	\\
		\hline
	\end{tabular}
\end{tiny}
\end{table*}
\end{landscape}
\restoregeometry

\section{Calibration Data}

Tables 3 and 4 list the photometric calibration data for $Z$ and $Y$, and Tables 5 and 6 
list the photometric calibration data for $JHK(s)$,
selected from the WFCAM (UKIDSS) and VISTA archives respectively. The transformations between the WFCAM and VISTA systems are given in Section 4. The uncertainties in the magnitudes which have been converted to the VISTA system includes the uncertainty in the transformation.

\begin{figure}
\vskip -1.4in
\hskip -0.1in
	\includegraphics[angle=-90,width=0.53\textwidth]{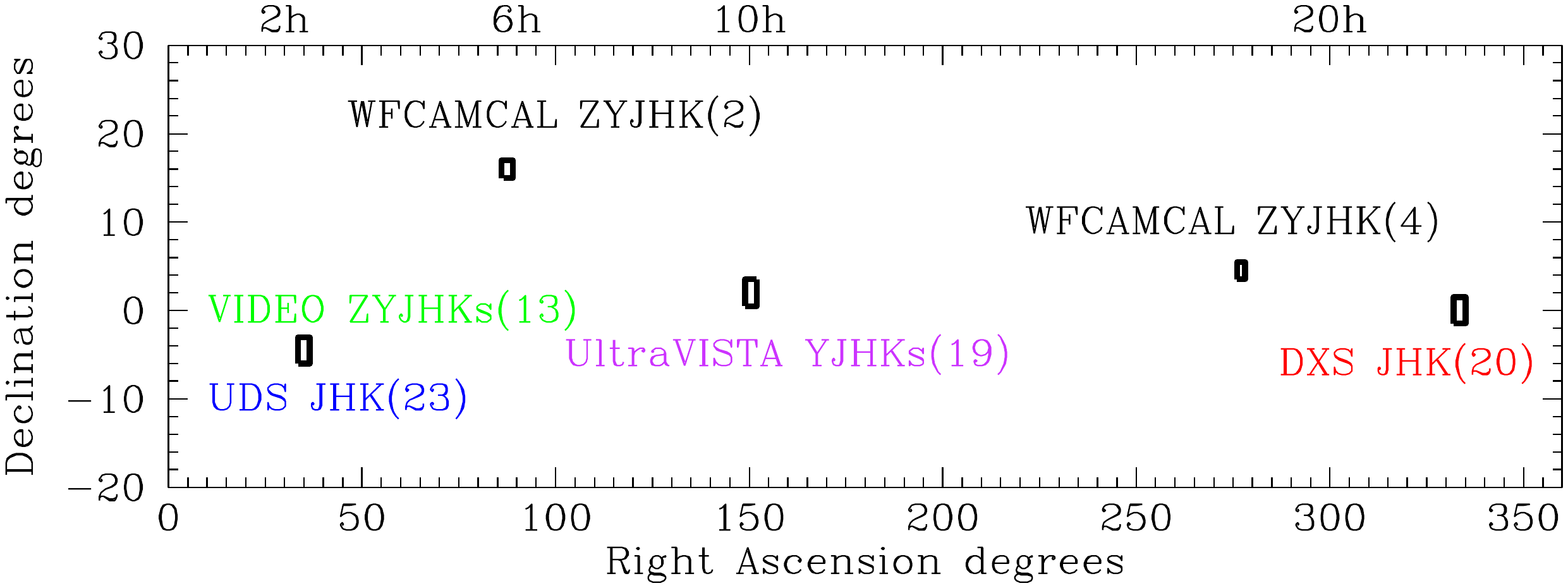}
\vskip -0.1in
    \caption{Sky chart showing the location and filter coverage of the final selection of standards. The number of stars at each location is given in parentheses.}
    \label{fig4}
\end{figure}

\begin{figure}
\vskip -0.5in
\hskip -0.1in
	\includegraphics[angle=0,width=0.5\textwidth]{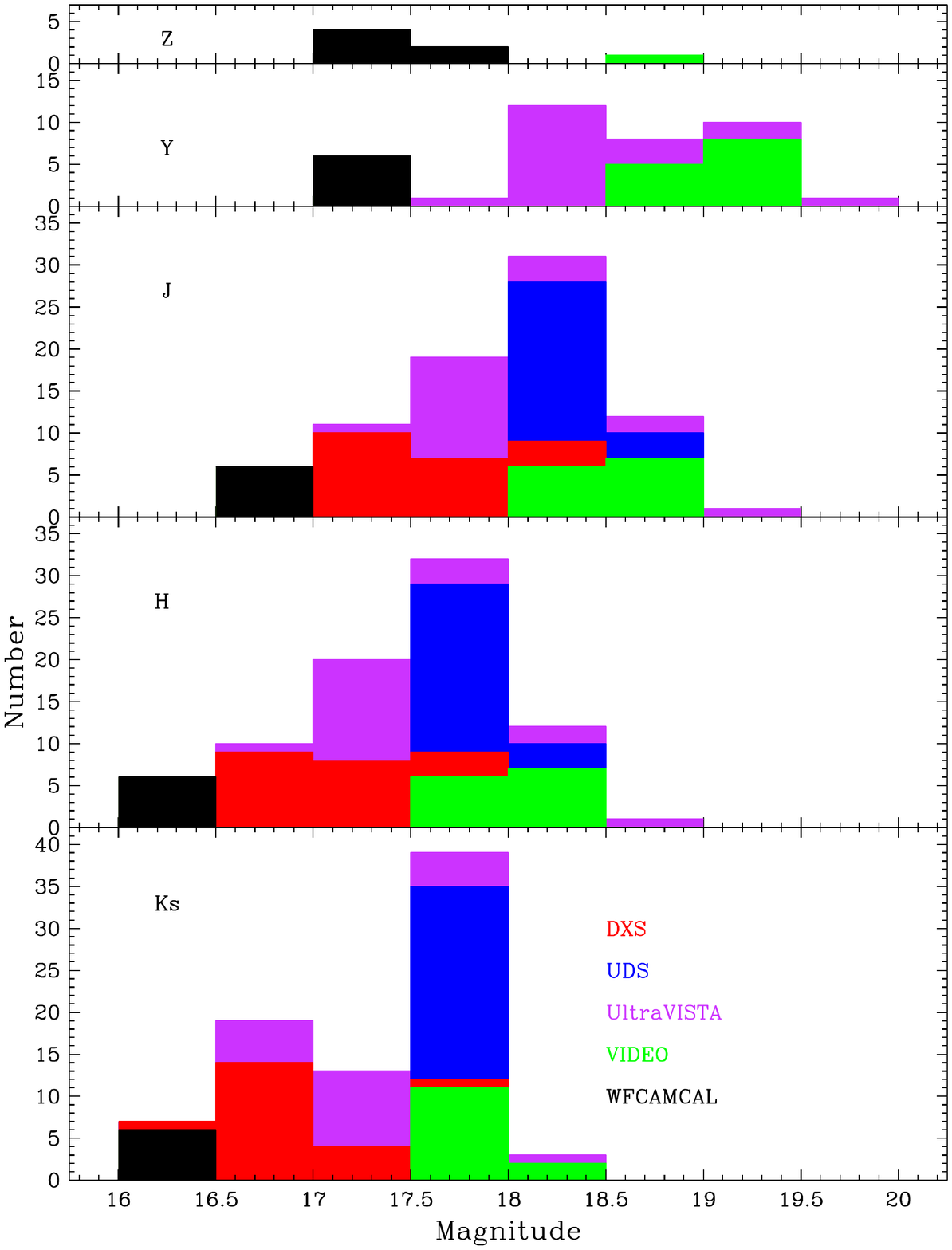}
\vskip -0.4in
    \caption{Histogram of magnitude distribution in each filter for each survey, for the final selection of standard stars.}
    \label{fig5}
\end{figure}

We only include stars with twenty or more good observations in the variability tables, for each filter, for use as calibrators.  
Figure 4 illustrates the location on the sky of the new photometric standards, and Figure 5 shows the magnitude distribution for each filter.

The mean magnitude from the variability table, with an uncertainty given by  $1.484 \times MagMAD/\sqrt{nGoodObs - 1}$,
is the fundamental photometric reference for calibration purposes.
The photometry presented here has a relative uncertainty of $< 0.006$ mag.
\cite{Gonzalez2018} and \citet{Hodgkin2009} compare the WFCAM and UKIDSS photometric systems, compare each system to the 2MASS system, and compare the WFCAM system to the FS system which is defined by a different camera on UKIRT.  The authors also examine the VISTA and WFCAM colours of samples of A0 stars, which by definition are zero. These various comparisons suggest that the VISTA and WFCAM photometric accuracy is $\lesssim 0.02$~mag, which also applies to the sample of stars presented here.

\section{Conclusions}
    
The traditional near-infrared photometric standard stars published by \citet{Hawarden2001} and \citet{Leggett2006} are too bright for efficient observing by 
current 8- to 10-m class telescopes, and future extremely large 30- to 40-m class telescopes. In order to provide a set of fainter standards for community use,
we queried the DXS, UDS, WFCAMCAL, VIDEO and UltraVISTA surveys using the  WFCAM and VISTA Science Archives. 
Non-variable, probable stars were identified
with $JHK(s)$ magnitudes between 16 and 19.
The initial sample of 169 sources  was further refined by excluding 77 objects that may be galaxies, as indicated by red/near-infrared colours or  PAN-STARRS Kron magnitudes. Eleven of the remaining sources were excluded due to  atypical colours,
or discrepant aperture and mean magnitudes from the source and variability tables, respectively. In this way we have produced a sample of 81 non-variable objects with precise photometry,
that are likely to be single K and M stars.
The new standard stars are distributed equatorially 
and are accessible from both hemispheres. 
Table~\ref{tab:final} collates the  calibration data on the VISTA $ZYJHKs$ photometric system. The table also gives a unique running identification number for convenience --- we refer to this as the Very Faint Standard, or VFS, identification number. We also give standard IAU names based on sexagesimal coordinate strings. Finder charts are presented in Appendix~D.

\begin{table*}
\begin{scriptsize}
	\caption{Standard Stars from WFCAM: $ZY$}
	\begin{tabular}{rrrrrrrrrrrrrr}
		\hline
  Survey & RA$^{\circ}$     & 
  n $Z$ & $Z$ mean & $Z$ Aper & $Z$ rms & $Z$ MAD & 
  n $Y$ & $Y$ mean & $Y$ Aper & $Y$ rms & $Y$ MAD & 
  $Z_{\rm VISTA}$ & $Y_{\rm VISTA}$ \\
      & Decl.$^{\circ}$	&
      & $\pm$ mag	& $\pm$  mag	&  mag&  mag &
    & $\pm$  mag	& $\pm$  mag	&  mag&  mag &

     $\pm$  mag	& $\pm$    mag   \\  
\hline
CAL	&	87.7352518	&	106	&	17.2521	&	17.2224	&	0.0732	&	0.0177	&	121	&	17.0335	&	17.0391	&	0.0339	&	0.0183	&	17.2730	&	17.0544	\\
	&	15.8816295	&		&	0.0026	&	0.0149	&		&		&		&	0.0025	&	0.0138	&		&		&	0.0070	&	0.0035	\\
CAL	&	88.0439399	&	86	&	17.4721	&	17.4935	&	0.1718	&	0.0327	&	103	&	17.2724	&	17.2371	&	0.0411	&	0.0218	&	17.4837	&	17.2840	\\
	&	16.2125856	&		&	0.0053	&	0.0178	&		&		&		&	0.0032	&	0.0154	&		&		&	0.0095	&	0.0044	\\
CAL	&	276.7225557	&	133	&	17.4837	&	17.4682	&	0.0588	&	0.0218	&	131	&	17.2023	&	17.1913	&	0.0544	&	0.0227	&	17.4972	&	17.2159	\\
	&	4.0587376	&		&	0.0028	&	0.0173	&		&		&		&	0.0030	&	0.0151	&		&		&	0.0081	&	0.0042	\\
CAL	&	276.8280740	&	140	&	17.2483	&	17.2810	&	0.0845	&	0.0203	&	138	&	17.0144	&	17.0237	&	0.0425	&	0.0181	&	17.2663	&	17.0325	\\
	&	3.9774041	&		&	0.0026	&	0.0155	&		&		&		&	0.0023	&	0.0137	&		&		&	0.0074	&	0.0035	\\
CAL	&	277.2399424	&	135	&	17.5592	&	17.5412	&	0.0638	&	0.0294	&	137	&	17.2465	&	17.2558	&	0.0497	&	0.0221	&	17.5724	&	17.2598	\\
	&	4.0952285	&		&	0.0038	&	0.0182	&		&		&		&	0.0028	&	0.0160	&		&		&	0.0085	&	0.0041	\\
CAL	&	277.2699877	&	135	&	17.5972	&	17.5926	&	0.0667	&	0.0257	&	133	&	17.3253	&	17.2961	&	0.0560	&	0.0218	&	17.6120	&	17.3401	\\
	&	4.5221425	&		&	0.0033	&	0.0183	&		&		&		&	0.0028	&	0.0157	&		&		&	0.0081	&	0.0040	\\
  		\hline
	\end{tabular}
\end{scriptsize}
\end{table*}

\begin{table*}
\begin{scriptsize}
	\caption{Standard Stars from VISTA: $ZY$}
	\begin{tabular}{rrrrrrrrrrrr}
		\hline
  Survey & RA$^{\circ}$     & 
  n $Z$ & $Z$ mean & $Z$ Aper & $Z$ rms & $Z$ MAD & 
  n $Y$ & $Y$ mean & $Y$ Aper & $Y$ rms & $Y$ MAD  \\
      & Decl.$^{\circ}$	&
      & $\pm$ mag	& $\pm$  mag	&  mag&   mag&
    & $\pm$  mag	& $\pm$  mag	&  mag&  mag 
   \\  
\hline	
VIDEO	&	35.0542019	&		&		&		&		&		&	78	&	19.0018	&	18.9692	&	0.0327	&	0.0126	\\
	&	-4.4813805	&		&		&		&		&		&		&	0.0021	&	0.0098	&		&		\\
VIDEO	&	35.1152774	&		&		&		&		&		&	78	&	19.0704	&	19.0412	&	0.0219	&	0.0118	\\
	&	-4.9003215	&		&		&		&		&		&		&	0.0020	&	0.0088	&		&		\\
VIDEO	&	35.3795003	&		&		&		&		&		&	78	&	19.1025	&	19.0890	&	0.0294	&	0.0133	\\
	&	-4.6057616	&		&		&		&		&		&		&	0.0023	&	0.0089	&		&		\\
VIDEO	&	35.6598904	&		&		&		&		&		&	77	&	18.7513	&	18.7303	&	0.0292	&	0.0094	\\
	&	-4.6351333	&		&		&		&		&		&		&	0.0016	&	0.0052	&		&		\\
VIDEO	&	35.6610332	&		&		&		&		&		&	78	&	19.0260	&	18.9936	&	0.0234	&	0.0111	\\
	&	-4.5453215	&		&		&		&		&		&		&	0.0019	&	0.0075	&		&		\\
VIDEO	&	35.6959814	&		&		&		&		&		&	78	&	19.3279	&	19.2979	&	0.0228	&	0.0154	\\
	&	-4.5264450	&		&		&		&		&		&		&	0.0026	&	0.0105	&		&		\\
VIDEO	&	35.7281547	&		&		&		&		&		&	78	&	19.3492	&	19.3460	&	0.0236	&	0.0160	\\
	&	-5.2037306	&		&		&		&		&		&		&	0.0027	&	0.0112	&		&		\\
VIDEO	&	35.8660837	&		&		&		&		&		&	78	&	19.1724	&	19.1566	&	0.0343	&	0.0143	\\
	&	-4.1075563	&		&		&		&		&		&		&	0.0024	&	0.0106	&		&		\\
VIDEO	&	35.9623324	&		&		&		&		&		&	78	&	18.8035	&	18.7944	&	0.0206	&	0.0099	\\
	&	-4.2237615	&		&		&		&		&		&		&	0.0017	&	0.0088	&		&		\\
VIDEO	&	35.9681759	&		&		&		&		&		&	77	&	18.8680	&	18.8437	&	0.0210	&	0.0122	\\
	&	-4.2573895	&		&		&		&		&		&		&	0.0021	&	0.0288	&		&		\\
VIDEO	&	35.9729949	&		&		&		&		&		&	78	&	18.7733	&	18.8435	&	0.0211	&	0.0147	\\
	&	-5.2047350	&		&		&		&		&		&		&	0.0025	&	0.0290	&		&		\\
VIDEO	&	36.0194131	&		&		&		&		&		&	78	&	19.1804	&	19.1611	&	0.0243	&	0.0128	\\
	&	-4.8728386	&		&		&		&		&		&		&	0.0022	&	0.0144	&		&		\\
VIDEO	&	36.7075066	&	74	&	18.9232	&	18.8959	&	0.0460	&	0.0099	&	75	&	18.7517	&	18.7244	&	0.0157	&	0.0092	\\
	&	-4.5302703	&		&	0.0017	&	0.0072	&		&		&		&	0.0016	&	0.0074	&		&		\\
UVISTA	&		149.7053794	&	&	&	&	&	&	90	&	18.3682	&	18.3562	&	0.0122	&	0.0084	\\
	&	2.4862294	&	&	&	&	&	&		&	0.0013	&	0.0013	&		&		\\
UVISTA	&		149.7087571	&	&	&	&	&	&	75	&	18.4555	&	18.4382	&	0.0085	&	0.0063	\\
	&	2.7888722	&	&	&	&	&	&		&	0.0011	&	0.0020	&		&		\\
UVISTA	&		149.7234241	&	&	&	&	&	&	75	&	19.2576	&	19.2364	&	0.0104	&	0.0071	\\
	&	2.7511805	&	&	&	&	&	&		&	0.0012	&	0.0029	&		&		\\
UVISTA		&	149.7264829	&	&	&	&	&	&	140	&	18.3791	&	18.3591	&	0.0133	&	0.0101	\\
	&	2.5902461	&	&	&	&	&	&		&	0.0013	&	0.0014	&		&		\\
UVISTA	&		149.7307359	&	&	&	&	&	&	75	&	18.7503	&	18.7310	&	0.0091	&	0.0058	\\
	&	2.7399894	&	&	&	&	&	&		&	0.0010	&	0.0023	&		&		\\
UVISTA	&		149.7313560	&	&	&	&	&	&	143	&	18.0241	&	18.0070	&	0.0094	&	0.0066	\\
	&	2.6928870	&	&	&	&	&	&		&	0.0008	&	0.0011	&		&		\\
UVISTA	&		149.7319650	&	&	&	&	&	&	75	&	19.8434	&	19.8107	&	0.0181	&	0.0126	\\
	&	2.7327131	&	&	&	&	&	&		&	0.0022	&	0.0034	&		&		\\
UVISTA	&		149.7332239	&	&	&	&	&	&	71	&	18.4389	&	18.4214	&	0.0086	&	0.0053	\\
	&	2.7947267	&	&	&	&	&	&		&	0.0009	&	0.0020	&		&		\\
UVISTA	&		149.7355822	&	&	&	&	&	&	140	&	18.8246	&	18.8101	&	0.0121	&	0.0077	\\
	&	2.3205009	&	&	&	&	&	&		&	0.0010	&	0.0017	&		&		\\
UVISTA		&	149.7360512	&	&	&	&	&	&	147	&	18.0802	&	18.0609	&	0.0109	&	0.0069	\\
	&	2.4872598	&	&	&	&	&	&		&	0.0008	&	0.0012	&		&		\\
UVISTA	&		149.7476065	&	&	&	&	&	&	75	&	19.1804	&	19.1586	&	0.0140	&	0.0093	\\
	&	2.7272746	&	&	&	&	&	&		&	0.0016	&	0.0025	&		&		\\
UVISTA		&	149.7564883	&	&	&	&	&	&	75	&	17.9949	&	17.9755	&	0.0098	&	0.0063	\\
	&	2.7279976	&	&	&	&	&	&		&	0.0011	&	0.0014	&		&		\\
UVISTA		&	149.7597522	&	&	&	&	&	&	75	&	18.4764	&	18.4621	&	0.0097	&	0.0069	\\
	&	2.7827541	&	&	&	&	&	&		&	0.0012	&	0.0021	&		&		\\
UVISTA	&		149.7742789	&	&	&	&	&	&	79	&	18.2324	&	18.2136	&	0.0106	&	0.0064	\\
	&	2.6687115	&	&	&	&	&	&		&	0.0011	&	0.0013	&		&		\\
UVISTA	&		149.7772394	&	&	&	&	&	&	75	&	18.8640	&	18.8447	&	0.0097	&	0.0065	\\
	&	2.7783479	&	&	&	&	&	&		&	0.0011	&	0.0025	&		&		\\
UVISTA		&	149.7923194	&	&	&	&	&	&	140	&	18.1225	&	18.0959	&	0.0123	&	0.0095	\\
	&	2.5832843	&	&	&	&	&	&		&	0.0012	&	0.0012	&		&		\\
UVISTA		&	149.8156044	&	&	&	&	&	&	75	&	18.2690	&	18.2479	&	0.0093	&	0.0052	\\
	&	2.7697500	&	&	&	&	&	&		&	0.0009	&	0.0018	&		&		\\
UVISTA		&	149.8172140	&	&	&	&	&	&	75	&	18.2823	&	18.2632	&	0.0075	&	0.0054	\\
	&	2.7717426	&	&	&	&	&	&		&	0.0009	&	0.0018	&		&		\\
UVISTA	&		149.8251519	&	&	&	&	&	&	75	&	18.3345	&	18.3074	&	0.0102	&	0.0060	\\
	&	2.7608222	&	&	&	&	&	&		&	0.0010	&	0.0019	&		&		\\
  		\hline
	\end{tabular}
\\
Note: UltraVISTA $Y$-band source
table aperture magnitudes are systematically brighter than the variability table mean magnitudes by  $\approx 0.02$ mag, due to differences in the aperture correction and zeropoints. All mean magnitudes given in this paper are self-consistent.
\end{scriptsize}
\end{table*}

\newgeometry{width=20cm,height=13cm}
\begin{landscape}
\begin{table*}
\begin{tiny}
	\caption{Standard Stars from WFCAM: $JHK$}
			\setlength\tabcolsep{3pt}
	\begin{tabular}{rrrrrrrrrrrrrrrrrrrr}
		\hline
  Survey & RA$^{\circ}$     & 
  n $J$ & $J$ mean & $J$ Aper & $J$ rms & $J$ MAD & 
  n $H$ & $H$ mean & $H$ Aper & $H$ rms & $H$ MAD & 
    n $K$ & $K$ mean & $K$ Aper & $K$ rms & $K$ MAD & 
  $J_{\rm VISTA}$ & $H_{\rm VISTA}$  & $Ks_{\rm VISTA}$ \\
      & Decl.$^{\circ}$	&
      & $\pm$ mag	& $\pm$  mag	&  mag&  mag &
    & $\pm$  mag	& $\pm$  mag	&  mag&  mag &
   & $\pm$  mag	& $\pm$  mag	&  mag&  mag &
     $\pm$  mag	& $\pm$    mag   & $\pm$    mag  \\  
\hline
UDS	&	34.0342392	&	349	&	18.3349	&	18.5313	&	0.0177	&	0.0108	&	225	&	17.8661	&	18.0749	&	0.0224	&	0.0130	&	513	&	17.5663	&	17.7429&	0.0267	&	0.0141	&	18.3094	&	17.8626	&	17.5573	\\
	&	-5.1728814	&		&	0.0009	&	0.0025	&		&		&		&	0.0013	&	0.0022	&		&		&		&	0.0009	&	0.0019	&		&		&	0.0020	&	0.0018	&	0.0032	\\
UDS	&	34.0585392	&	350	&	18.5195	&	18.7181	&	0.0180	&	0.0108	&	220	&	18.0215	&	18.2202	&	0.0238	&	0.0158	&	522	&	17.7618	&	17.9339	&	0.0286	&	0.0173	&	18.4943	&	18.0184	&	17.7527	\\
	&	-5.3530358	&		&	0.0009	&	0.0028	&		&		&		&	0.0016	&	0.0024	&		&		&		&	0.0011	&	0.0021	&		&		&	0.0020	&	0.0020	&	0.0032	\\
UDS	&	34.0906904	&	351	&	18.3316	&	18.5312	&	0.0201	&	0.0117	&	225	&	17.8608	&	18.0636	&	0.0253	&	0.0154	&	516	&	17.5957	&	17.7785	&	0.0273	&	0.0147	&	18.3070	&	17.8586	&	17.5862	\\
	&	-5.1871909	&		&	0.0009	&	0.0026	&		&		&		&	0.0015	&	0.0023	&		&		&		&	0.0010	&	0.0020	&		&		&	0.0020	&	0.0020	&	0.0031	\\
UDS	&	34.1166206	&	350	&	18.3820	&	18.5803	&	0.0187	&	0.0119	&	221	&	17.8637	&	18.0607	&	0.0226	&	0.0137	&	522	&	17.6719	&	17.8423	&	0.0295	&	0.0191	&	18.3581	&	17.8624	&	17.6619	\\
	&	-5.4330768	&		&	0.0009	&	0.0027	&		&		&		&	0.0014	&	0.0023	&		&		&		&	0.0012	&	0.0020	&		&		&	0.0020	&	0.0018	&	0.0032	\\
UDS	&	34.1459342	&	352	&	18.4145	&	18.6036&	0.0196	&	0.0125	&	221	&	17.8761	&	18.0775	&	0.0231	&	0.0124	&	524	&	17.6053	&	17.7783	&	0.0264	&	0.0173	&	18.3878	&	17.8711	&	17.5971	\\
	&	-5.4192513	&		&	0.0010	&	0.0028	&		&		&		&	0.0012	&	0.0023	&		&		&		&	0.0011	&	0.0020	&		&		&	0.0021	&	0.0018	&	0.0033	\\
UDS	&	34.1559576	&	352	&	18.5141	&	18.7003	&	0.0228	&	0.0129	&	224	&	17.9824	&	18.1923	&	0.0271	&	0.0152	&	516	&	17.7474	&	17.9236 & 0.0304 &	0.0199 & 18.4886 &	17.9790	&	17.7385	\\
	&	-5.1864870	&		&	0.0010	&	0.0029	&		&		&		&	0.0015	&	0.0025	&		&		&		&	0.0013	&	0.0022	&		&		&	0.0021	&	0.0020	&	0.0033	\\
UDS	&	34.1991747	&	353	&	18.3649	&	18.5502	&	0.0199	&	0.0138	&	221	&	17.8259	&	18.0177	&	0.0248	&	0.0143	&	524	&	17.5907	&	17.7792	&	0.0254	&	0.0170	&	18.3392	&	17.8223	&	17.5819	\\
	&	-4.9476391	&		&	0.0011	&	0.0028	&		&		&		&	0.0014	&	0.0022	&		&		&		&	0.0011	&	0.0020	&		&		&	0.0021	&	0.0019	&	0.0033	\\
UDS	&	34.2232606	&	386	&	18.5385	&	18.7258	&	0.0227	&	0.0139	&	257	&	17.9753	&	18.1769	&	0.0291	&	0.0184	& 604	&	17.7418	&	17.9272	&	0.0344	&	0.0215	&	18.5122	&	17.9708	&	17.7333	\\
	&	-4.7491396	&		&	0.0010	&	0.0028	&		&		&		&	0.0017	&	0.0023	&		&		&		&	0.0013	&	0.0022	&		&		&	0.0022	&	0.0021	&	0.0034	\\
UDS	&	34.2308454	&	351	&	18.3830	&	18.5733	&	0.0206	&	0.0130	&	222	&	17.9104	&	18.1110	&	0.0315	&	0.0162	&	535	&	17.6311	&	17.8114	&	0.0264	&	0.0186	&	18.3580	&	17.9076	&	17.6218	\\
	&	-5.3911233	&		&	0.0010	&	0.0023	&		&		&		&	0.0016	&	0.0021	&		&		&		&	0.0012	&	0.0019	&		&		&	0.0021	&	0.0020	&	0.0032	\\
UDS	&	34.3103343	&	348	&	18.5864	&	18.7776	&	0.0259	&	0.0143	&	227	&	18.0813	&	18.2905	&	0.0252	&	0.0132	&	533	&	17.7619	&	17.9373	&	0.0285	&	0.0169	&	18.5593	&	18.0758	&	17.7539	\\
	&	-5.4172864	&		&	0.0011	&	0.0030	&		&		&		&	0.0013	&	0.0026	&		&		&		&	0.0011	&	0.0021	&		&		&	0.0022	&	0.0018	&	0.0034	\\
UDS	&	34.3240632	&	345	&	18.5039	&	18.6955	&	0.0211	&	0.0149	&	227	&	17.9932	&	18.2027	&	0.0273	&	0.0154	&	530	&	17.6775	&	17.8517	&	0.0298	&	0.0174	&	18.4768	&	17.9876	&	17.6696	\\
	&	-5.4334966	&		&	0.0012	&	0.0029	&		&		&		&	0.0015	&	0.0024	&		&		&		&	0.0011	&	0.0020	&		&		&	0.0023	&	0.0020	&	0.0034	\\
UDS	&	34.3475622	&	345	&	18.3321	&	18.5255	&	0.0178	&	0.0110	&	227	&	17.8488	&	18.0557	&	0.0205	&	0.0128	&	531	&	17.5860	&	17.7628	&	0.0243	&	0.0150	&	18.3072	&	17.8462	&	17.5767	\\
	&	-5.4844981	&		&	0.0009	&	0.0026	&		&		&		&	0.0013	&	0.0022	&		&		&		&	0.0010	&	0.0019	&		&		&	0.0020	&	0.0018	&	0.0032	\\
UDS	&	34.3750969	&	345	&	18.5027	&	18.6924	&	0.0183	&	0.0109	&	227	&	17.9403	&	18.1475	&	0.0221	&	0.0140	&	531	&	17.6520	&	17.8285	&	0.0240	&	0.0149	&	18.4748	&	17.9338	&	17.6445	\\
	&	-5.4848715	&		&	0.0009	&	0.0028	&		&		&		&	0.0014	&	0.0023	&		&		&		&	0.0010	&	0.0020	&		&		&	0.0022	&	0.0019	&	0.0034	\\
UDS	&	34.4389254	&	339	&	18.3909	&	18.5846&	0.0219	&	0.0131	&	222	&	17.8585	&	18.0665	&	0.0250	&	0.0155	&	517	&	17.6173	&	17.8000	&	0.0284	&	0.0192	&	18.3653	&	17.8549	&	17.6085	\\
	&	-5.4868910	&		&	0.0011	&	0.0027	&		&		&		&	0.0015	&	0.0022	&		&		&		&	0.0013	&	0.0020	&		&		&	0.0021	&	0.0020	&	0.0033	\\
UDS	&	34.5319836	&	439	&	18.4015	&	18.5871	&	0.0195	&	0.0119	&	270	&	17.8547	&	18.0507	&	0.0263	&	0.0148	&	627	&	17.6405	&	17.8118	&	0.0321	&	0.0213	&	18.3762	&	17.8516	&	17.6315	\\
	&	-5.1015107	&		&	0.0008	&	0.0022	&		&		&		&	0.0013	&	0.0019	&		&		&		&	0.0013	&	0.0018	&		&		&	0.0020	&	0.0018	&	0.0033	\\
UDS	&	34.6700469	&	374	&	18.3749	&	18.5585	&	0.0244	&	0.0165	&	247	&	17.8552	&	18.0493	&	0.0668	&	0.0160	&	576	&	17.5183	&	17.7034	&	0.0378	&	0.0265	&	18.3469	&	17.8485	&	17.5109	\\
	&	-5.0787756	&		&	0.0013	&	0.0024	&		&		&		&	0.0015	&	0.0020	&		&		&		&	0.0016	&	0.0019	&		&		&	0.0024	&	0.0020	&	0.0036	\\
UDS	&	34.6776197	&	404	&	18.4691	&	18.6542	&	0.0231	&	0.0140	&	244	&	17.9327	&	18.1342	&	0.0330	&	0.0149	&	594	&	17.6670	&	17.8461	&	0.0447	&	0.0257	&	18.4426	&	17.9281	&	17.6586	\\
	&	-4.9671258	&		&	0.0010	&	0.0023	&		&		&		&	0.0014	&	0.0020	&		&		&		&	0.0016	&	0.0019	&		&		&	0.0022	&	0.0019	&	0.0035	\\
UDS	&	34.7079081	&	347	&	18.1715	&	18.3579&	0.0152	&	0.0095	&	229	&	17.6592	&	17.8669	&	0.0208	&	0.0123	&	533	&	17.5672	&	17.7433	&	0.0286	&	0.0166	&	18.1506	&	17.6618	&	17.5555	\\
	&	-4.9039390	&		&	0.0008	&	0.0022	&		&		&		&	0.0012	&	0.0018	&		&		&		&	0.0011	&	0.0018	&		&		&	0.0017	&	0.0017	&	0.0029	\\
UDS	&	34.7425103	&	345	&	18.2942	&	18.4874	&	0.0200	&	0.0129	&	229	&	17.7883	&	17.9945	&	0.0249	&	0.0142	&	529	&	17.7077	&	17.8798	&	0.0293	&	0.0193	&	18.2738	&	17.7916	&	17.6957	\\
	&	-5.4781043	&		&	0.0010	&	0.0028	&		&		&		&	0.0014	&	0.0023	&		&		&		&	0.0012	&	0.0022	&		&		&	0.0019	&	0.0018	&	0.0029	\\
UDS	&	34.7745424	&	371	&	18.4765	&	18.6584	&	0.0215	&	0.0125	&	260	&	17.9661	&	18.1642	&	0.0264	&	0.0162	&	595	&	17.6479	&	17.8183	&	0.0336	&	0.0210	&	18.4493	&	17.9604	&	17.6400	\\
	&	-4.8889668	&		&	0.0010	&	0.0024	&		&		&		&	0.0015	&	0.0021	&		&		&		&	0.0013	&	0.0019	&		&		&	0.0022	&	0.0020	&	0.0034	\\
UDS	&	34.8172698	&	354	&	18.6678	&	18.8510	&	0.0221	&	0.0148	&	220	&	18.1916	&	18.3840	&	0.0243	&	0.0143	&	521	&	17.9121	&	18.0983	&	0.0347	&	0.0194	&	18.6426	&	18.1886	&	17.9029	\\
	&	-4.7942724	&		&	0.0012	&	0.0029	&		&		&		&	0.0014	&	0.0027	&		&		&		&	0.0013	&	0.0024	&		&		&	0.0022	&	0.0019	&	0.0033	\\
UDS	&	34.8182276	&	404	&	18.5294	&	18.7202	&	0.0220	&	0.0122	&	249	&	17.9046	&	18.1129	&	0.0277	&	0.0150	&	619	&	17.7394	&	17.9188	&	0.0338	&	0.0221	&	18.5033	&	17.9004	&	17.7308	\\
	&	-5.1062739	&		&	0.0009	&	0.0025	&		&		&		&	0.0014	&	0.0020	&		&		&		&	0.0013	&	0.0020	&		&		&	0.0021	&	0.0019	&	0.0034	\\
UDS	&	34.8710949	&	366	&	18.3653	&	18.5585	&	0.0178	&	0.0112	&	228	&	17.9672	&	18.1734	&	0.0259	&	0.0157	&	536	&	17.8937	&	18.0786	&	0.0387	&	0.0240	&	18.3481	&	17.9747	&	17.8797	\\
	&	-5.1154861	&		&	0.0009	&	0.0025	&		&		&		&	0.0015	&	0.0023	&		&		&		&	0.0015	&	0.0023	&		&		&	0.0016	&	0.0019	&	0.0029	\\
CAL	&	87.7352518	&	136	&	16.6289	&	16.6552	&	0.0591	&	0.0219	&	132	&	16.2164	&	16.1917	&	0.0522	&	0.0301	&	136	&	16.1131	&	16.1481	&	0.0639	&	0.0424	&	16.6104	&	16.2224	&	16.0999	\\
	&	15.8816295	&		&	0.0028	&	0.0160	&		&		&		&	0.0039	&	0.0139	&		&		&		&	0.0054	&	0.0229	&		&		&	0.0031	&	0.0041	&	0.0060	\\
CAL	&	88.0439399	&	130	&	16.7948	&	16.7686	&	0.0543	&	0.0338	&	125	&	16.1988	&	16.1922	&	0.0559	&	0.0251	&	133	&	16.0276	&	16.0285	&	0.0613	&	0.0313	&	16.7693	&	16.1954	&	16.0186	\\
	&	16.2125856	&		&	0.0044	&	0.0170	&		&		&		&	0.0033	&	0.0137	&		&		&		&	0.0040	&	0.0214	&		&		&	0.0048	&	0.0036	&	0.0051	\\
CAL	&	276.7225557	&	156	&	16.7231	&	16.7228	&	0.0678	&	0.0261	&	157	&	16.1566	&	16.1678	&	0.0612	&	0.0271	&	167	&	16.0093	&	16.0034	&	0.0665	&	0.0370	&	16.6992	&	16.1551	&	15.9994	\\
	&	4.0587376	&		&	0.0031	&	0.0174	&		&		&		&	0.0032	&	0.0151	&		&		&		&	0.0043	&	0.0252	&		&		&	0.0036	&	0.0034	&	0.0052	\\
CAL	&	276.8280740	&	158	&	16.6085	&	16.6102	&	0.0485	&	0.0270	&	157	&	16.1697	&	16.1597	&	0.0612	&	0.0255	&	169	&	16.0157	&	16.0139	&	0.0750	&	0.0381	&	16.5879	&	16.1728	&	16.0038	\\
	&	3.9774041	&		&	0.0032	&	0.0162	&		&		&		&	0.0030	&	0.0149	&		&		&		&	0.0044	&	0.0254	&		&		&	0.0036	&	0.0033	&	0.0051	\\
CAL	&	277.2399424	&	156	&	16.7699	&	16.8010	&	0.0727	&	0.0270	&	154	&	16.3020	&	16.2767	&	0.0543	&	0.0294	&	168	&	16.0474	&	16.0938	&	0.0716	&	0.0369	&	16.7456	&	16.3002	&	16.0376	\\
	&	4.0952285	&		&	0.0032	&	0.0187	&		&		&		&	0.0035	&	0.0169	&		&		&		&	0.0042	&	0.0263	&		&		&	0.0037	&	0.0037	&	0.0052	\\
CAL	&	277.2699877	&	157	&	16.8103	&	16.8261	&	0.0741	&	0.0277	&	156	&	16.3346	&	16.3425	&	0.0621	&	0.0291	&	167	&	16.1284	&	16.1660	&	0.0715	&	0.0422	&	16.7872	&	16.3344	&	16.1180	\\
	&	4.5221425	&		&	0.0033	&	0.0181	&		&		&		&	0.0035	&	0.0171	&		&		&		&	0.0049	&	0.0262	&		&		&	0.0037	&	0.0037	&	0.0056	\\
  		\hline
	\end{tabular}
\\
Note: UDS aperture magnitudes are not aperture corrected and are systematically fainter than the mean magnitudes by $\approx 0.2$ mag.\\
\end{tiny}
\end{table*}
\end{landscape}
\restoregeometry

\newgeometry{width=20cm,height=13cm}
\begin{landscape}
\begin{table*}
\begin{tiny}
	\contcaption{Standard Stars from WFCAM: $JHK$}
			\setlength\tabcolsep{2pt}
	\begin{tabular}{rrrrrrrrrrrrrrrrrrrr}
		\hline
  Survey & RA$^{\circ}$     & 
  n $J$ & $J$ mean & $J$ Aper & $J$ rms & $J$ MAD & 
  n $H$ & $H$ mean & $H$ Aper & $H$ rms & $H$ MAD & 
    n $K$ & $K$ mean & $K$ Aper & $K$ rms & $K$ MAD & 
  $J_{\rm VISTA}$ & $H_{\rm VISTA}$  & $Ks_{\rm VISTA}$ \\
      & Decl.$^{\circ}$	&
      & $\pm$ mag	& $\pm$  mag	&  mag&  mag &
    & $\pm$  mag	& $\pm$  mag	&  mag&  mag &
   & $\pm$  mag	& $\pm$  mag	&  mag&  mag &
     $\pm$  mag	& $\pm$    mag   & $\pm$    mag  \\  
\hline
DXS	&	333.0711763	&	26	&	17.3850	&	17.3946	&	0.0128	&	0.0067	&		&		&		&		&		&	22	&	16.5895	&	16.5916	&	0.0227	&	0.0105	&	17.3587	&	16.8191	&	16.5811	\\
	&	0.8303505	&		&	0.0020	&	0.0071	&		&		&		&		&		&		&		&		&	0.0034	&	0.0068	&		&		&	0.0027	&	0.0029	&	0.0046	\\
DXS	&	333.0715557	&	26	&	18.1809	&	18.1990	&	0.0202	&	0.0099	&		&		&		&		&		&	22	&	17.4019	&	17.4092	&	0.0224	&	0.0138	&	18.1551	&	17.6539	&	17.3931	\\
	&	0.9162355	&		&	0.0029	&	0.0106	&		&		&		&		&		&		&		&		&	0.0045	&	0.0115	&		&		&	0.0035	&	0.0036	&	0.0054	\\
DXS	&	333.0718253	&	26	&	18.2282	&	18.2278	&	0.0212	&	0.0125	&		&		&		&		&		&	22	&	17.4554	&	17.4472	&	0.0250	&	0.0155	&	18.2026	&	17.5949	&	17.4465	\\
	&	1.1889522	&		&	0.0037	&	0.0109	&		&		&		&		&		&		&		&		&	0.0050	&	0.0129	&		&		&	0.0042	&	0.0055	&	0.0059	\\
DXS	&	333.2839533	&	24	&	17.8141	&	17.8167	&	0.0133	&	0.0075	&		&		&		&		&		&	27	&	17.0182	&	17.0135	&	0.0258	&	0.0126	&	17.7879	&	17.2757	&	17.0097	\\
	&	-0.4105822	&		&	0.0023	&	0.0089	&		&		&		&		&		&		&		&		&	0.0037	&	0.0085	&		&		&	0.0030	&	0.0050	&	0.0048	\\
DXS	&	333.2928434	&	24	&	17.4314	&	17.4294	&	0.0159	&	0.0069	&	29	&	16.9668	&	16.9427	&	0.0145	&	0.0111	&	24	&	16.7012	&	16.7006	&	0.0245	&	0.0124	&	17.4070	&	16.9648	&	16.6916	\\
	&	0.7343005	&		&	0.0021	&	0.0072	&		&		&		&	0.0031	&	0.0059	&		&		&		&	0.0038	&	0.0074	&		&		&	0.0028	&	0.0033	&	0.0048	\\
DXS	&	333.3028790	&	24	&	17.5811	&	17.5772	&	0.0156	&	0.0095	&	29	&	17.1164	&	17.1209	&	0.0175	&	0.0124	&	24	&	16.8181	&	16.8164	&	0.0194	&	0.0145	&	17.5557	&	17.1132	&	16.8090	\\
	&	0.7544734	&		&	0.0029	&	0.0077	&		&		&		&	0.0035	&	0.0062	&		&		&		&	0.0045	&	0.0077	&		&		&	0.0035	&	0.0037	&	0.0054	\\
DXS	&	333.3594622	&	24	&	18.3253	&	18.3266	&	0.0170	&	0.0095	&	29	&	17.8289	&	17.8372	&	0.0277	&	0.0114	&	24	&	17.5693	&	17.5466	&	0.0278	&	0.0156	&	18.3002	&	17.8259	&	17.5602	\\
	&	0.7512320	&		&	0.0029	&	0.0114	&		&		&		&	0.0032	&	0.0095	&		&		&		&	0.0048	&	0.0123	&		&		&	0.0034	&	0.0034	&	0.0057	\\
DXS	& 333.3824122 & 	24	& 17.8349 & 17.8492 & 0.0114 &	0.0064 & 29 & 17.3366 & 17.3456 & 0.0197 &	0.0122 & 24 & 17.0058 & 17.0084 & 0.0220 &	0.0115 & 17.8077 & 17.3309 &  16.9979 \\
    & 0.7424491 & & 0.0020 & 0.0089 &   &    &  & 0.0034 & 0.0071 & &   &   & 0.0036 & 0.0087 & &  &   0.0028  & 0.0037 &  0.0048 \\
DXS	&	333.4515020	&	24	&	17.7055	&	17.6906	&	0.0147	&	0.0131	&	29	&	17.2250	&	17.2250	&	0.0161	&	0.0104	&	24	&	16.9639	&	16.9546	&	0.0193	&	0.0126	&	17.6807	&	17.2226	&	16.9545	\\
	&	0.7278254	&		&	0.0040	&	0.0082	&		&		&		&	0.0029	&	0.0070	&		&		&		&	0.0039	&	0.0086	&		&		&	0.0044	&	0.0032	&	0.0049	\\
DXS	&	333.4987510	&	24	&	17.9579	&	17.9683	&	0.0163	&	0.0092	&	29	&	17.4404	&	17.4481	&	0.0223	&	0.0146	&	24	&	17.1523	&	17.1722	&	0.0199	&	0.0134	&	17.9313	&	17.4356	&	17.1440	\\
	&	0.7349581	&		&	0.0029	&	0.0094	&		&		&		&	0.0041	&	0.0075	&		&		&		&	0.0042	&	0.0096	&		&		&	0.0034	&	0.0043	&	0.0052	\\
DXS	&	333.7094552	&	25	&	17.2900	&	17.2783	&	0.0111	&	0.0052	&	27	&	16.6727	&	16.6579	&	0.0205	&	0.0103	&	22	&	16.5390	&	16.5553	&	0.0227	&	0.0175	&	17.2650	&	16.6699	&	16.5297	\\
	&	0.5245924	&		&	0.0016	&	0.0066	&		&		&		&	0.0030	&	0.0050	&		&		&		&	0.0057	&	0.0066	&		&		&	0.0024	&	0.0032	&	0.0064	\\
DXS	&	333.9366041	&	24	&	17.3359	&	17.3307	&	0.0087	&	0.0045	&		&		&		&		&		&	24	&	16.5021	&	16.5124	&	0.0240	&	0.0126	&	17.3086	&	16.6614	&	16.4942	\\
	&	-0.9471891	&		&	0.0014	&	0.0070	&		&		&		&		&		&		&		&		&	0.0039	&	0.0063	&		&		&	0.0024	&	0.0058	&	0.0050	\\
DXS	&	334.6493784	&	33	&	17.8736	&	17.8719	&	0.0140	&	0.0128	&	20	&	17.2809	&	17.2940	&	0.0240	&	0.0154	&	44	&	16.9888	&	16.9910	&	0.0264	&	0.0169	&	17.8448	&	17.2731	&	16.9818	\\
	&	1.3819891	&		&	0.0033	&	0.0087	&		&		&		&	0.0053	&	0.0077	&		&		&		&	0.0038	&	0.0086	&		&		&	0.0039	&	0.0054	&	0.0051	\\
DXS	&	334.8182190	&	32	&	17.6887	&	17.6897	&	0.0187	&	0.0107	&		&		&		&		&		&	34	&	16.8830	&	16.8723	&	0.0267	&	0.0168	&	17.6622	&	17.1697	&	16.8747	\\
	&	0.3928450	&		&	0.0029	&	0.0087	&		&		&		&		&		&		&		&		&	0.0043	&	0.0083	&		&		&	0.0034	&	0.0044	&	0.0054	\\
DXS	&	334.9251383	&	31	&	17.4395	&	17.4246	&	0.0122	&	0.0093	&	21	&	16.8880	&	16.8930	&	0.0102	&	0.0033	&	34	&	16.7863	&	16.7858	&	0.0329	&	0.0148	&	17.4172	&	16.8889	&	16.7754	\\
	&	0.7274321	&		&	0.0025	&	0.0074	&		&		&		&	0.0011	&	0.0061	&		&		&		&	0.0038	&	0.0078	&		&		&	0.0030	&	0.0016	&	0.0047	\\
DXS	&	335.0303119	&	29	&	17.4586	&	17.4514	&	0.0215	&	0.0135	&	22	&	17.0071	&	17.0150	&	0.0140	&	0.0103	&	30	&	16.9359	&	16.9255	&	0.0234	&	0.0150	&	17.4399	&	17.0128	&	16.9227	\\
	&	0.4454373	&		&	0.0038	&	0.0076	&		&		&		&	0.0033	&	0.0067	&		&		&		&	0.0041	&	0.0086	&		&		&	0.0040	&	0.0035	&	0.0049	\\
DXS	&	335.4746997	&	45	&	17.3876	&	17.3823	&	0.0200	&	0.0111	&	33	&	16.8051	&	16.8104	&	0.0159	&	0.0119	&	46	&	16.5574	&	16.5396	&	0.0301	&	0.0170	&	17.3604	&	16.7993	&	16.5495	\\
	&	0.5091841	&		&	0.0025	&	0.0072	&		&		&		&	0.0031	&	0.0054	&		&		&		&	0.0038	&	0.0067	&		&		&	0.0031	&	0.0034	&	0.0049	\\
DXS	&	335.5234567	&	31	&	17.1940	&	17.1661	&	0.0169	&	0.0117	&	26	&	16.6729	&	16.6621	&	0.0166	&	0.0116	&	32	&	16.5534	&	16.5204	&	0.0242	&	0.0165	&	17.1721	&	16.6742	&	16.5423	\\
	&	0.2874459	&		&	0.0032	&	0.0066	&		&		&		&	0.0034	&	0.0051	&		&		&		&	0.0044	&	0.0068	&		&		&	0.0036	&	0.0036	&	0.0052	\\
DXS	&	335.5823067	&	28	&	17.3988	&	17.4093	&	0.0155	&	0.0086	&		&		&		&		&		&	33	&	16.6114	&	16.6016	&	0.0201	&	0.0106	&	17.3728	&	16.7596	&	16.6027	\\
	&	-0.1444080	&		&	0.0024	&	0.0074	&		&		&		&		&		&		&		&		&	0.0028	&	0.0070	&		&		&	0.0031	&	0.0034	&	0.0042	\\
DXS	&	335.6150576	&	31	&	17.3103	&	17.3109	&	0.0124	&	0.0083	&	26	&	16.8637	&	16.8537	&	0.0144	&	0.0114	&	32	&	16.7785	&	16.7719	&	0.0269	&	0.0180	&	17.2914	&	16.8690	&	16.7655	\\
	&	0.5091194	&		&	0.0023	&	0.0070	&		&		&		&	0.0034	&	0.0056	&		&		&		&	0.0048	&	0.0078	&		&		&	0.0027	&	0.0036	&	0.0054	\\
  		\hline
	\end{tabular}
\end{tiny}
\end{table*}
\end{landscape}
\restoregeometry

\newgeometry{width=20cm,height=13cm}
\begin{landscape}
\begin{table*}
\begin{tiny}
	\caption{Standard Stars from VISTA: $JHKs$}
			\setlength\tabcolsep{2pt}
	\begin{tabular}{rrrrrrrrrrrrrrrrr}
		\hline
  Survey & RA$^{\circ}$     & 
  n $J$ & $J$ mean & $J$ Aper & $J$ rms & $J$ MAD & 
  n $H$ & $H$ mean & $H$ Aper & $H$ rms & $H$ MAD & 
  n $Ks$ & $Ks$ mean & $Ks$ Aper & $Ks$ rms & $Ks$ MAD  \\
      & Decl.$^{\circ}$	&
      & $\pm$ mag	& $\pm$  mag	&  mag&   mag &
       & $\pm$ mag	& $\pm$  mag	&  mag&   mag &
    & $\pm$  mag	& $\pm$  mag	&  mag&  mag 
   \\  
\hline	
VIDEO	&	35.0542019	&	39	&	18.5465	&	18.5352	&	0.0178	&	0.0117	&	40	&	18.0267	&	18.0154	&	0.0331	&	0.0198	&	48	&	17.8310	&	17.8200	&	0.0394	&	0.0199	\\
	&	-4.4813805	&		&	0.0028	&	0.0093	&		&		&		&	0.0047	&	0.0087	&		&		&		&	0.0043	&	0.0097	&		&		\\
VIDEO	&	35.1152774	&	39	&	18.5271	&	18.5256	&	0.0174	&	0.0085	&	40	&	18.0857	&	18.0768	&	0.0243	&	0.0134	&	48	&	17.8255	&	17.8019	&	0.0366	&	0.0225	\\
	&	-4.9003215	&		&	0.0021	&	0.0081	&		&		&		&	0.0032	&	0.0080	&		&		&		&	0.0049	&	0.0088	&		&		\\
VIDEO	&	35.3795003	&	39	&	18.6101	&	18.5996	&	0.0138	&	0.0096	&	40	&	18.1286	&	18.1186	&	0.0174	&	0.0120	&	48	&	17.9108	&	17.8928	&	0.0369	&	0.0225	\\
	&	-4.6057616	&		&	0.0023	&	0.0081	&		&		&		&	0.0028	&	0.0077	&		&		&		&	0.0049	&	0.0085	&		&		\\
VIDEO	&	35.6598904	&	39	&	18.3307	&	18.3223	&	0.0343	&	0.0083	&	40	&	17.7733	&	17.7600	&	0.0237	&	0.0126	&	48	&	17.6104	&	17.5946	&	0.0293	&	0.0146	\\
	&	-4.635133	&		&	0.0020	&	0.0050	&		&		&		&	0.0030	&	0.0047	&		&		&		&	0.0032	&	0.0054	&		&		\\
VIDEO	&	35.6610332	&	39	&	18.5364	&	18.5202	&	0.0155	&	0.0087	&	40	&	18.0755	&	18.0640	&	0.0200	&	0.0104	&	48	&	17.8501	&	17.8392	&	0.0403	&	0.0182	\\
	&	-4.5453215	&		&	0.0021	&	0.0070	&		&		&		&	0.0025	&	0.0070	&		&		&		&	0.0039	&	0.0076	&		&		\\
VIDEO	&	35.6959814	&	39	&	18.9173	&	18.9030	&	0.0226	&	0.0148	&	40	&	18.3808	&	18.3622	&	0.0287	&	0.0180	&	48	&	18.1880	&	18.1696	&	0.0400	&	0.0228	\\
	&	-4.5264450	&		&	0.0036	&	0.0101	&		&		&		&	0.0043	&	0.0097	&		&		&		&	0.0049	&	0.0107	&		&		\\
VIDEO	&	35.7281547	&	39	&	18.8672	&	18.8681	&	0.0221	&	0.0121	&	40	&	18.4067	&	18.4025	&	0.0310	&	0.0172	&	48	&	18.1372	&	18.1443	&	0.0398	&	0.0244	\\
	&	-5.2037306	&		&	0.0029	&	0.0101	&		&		&		&	0.0041	&	0.0098	&		&		&		&	0.0053	&	0.0105	&		&		\\
VIDEO	&	35.8660837	&	39	&	18.6624	&	18.6644	&	0.0170	&	0.0100	&	40	&	18.1648	&	18.1651	&	0.0240	&	0.0175	&	48	&	17.9286	&	17.9385	&	0.0383	&	0.0229	\\
	&	-4.1075563	&		&	0.0024	&	0.0098	&		&		&		&	0.0042	&	0.0094	&		&		&		&	0.0050	&	0.0104	&		&		\\
VIDEO	&	35.9623324	&	39	&	18.3172	&	18.3222	&	0.0195	&	0.0089	&	40	&	17.8155	&	17.8129	&	0.0239	&	0.0142	&	48	&	17.5798	&	17.5857	&	0.0345	&	0.0208	\\
	&	-4.2237615	&		&	0.0021	&	0.0080	&		&		&		&	0.0034	&	0.0078	&		&		&		&	0.0045	&	0.0086	&		&		\\
VIDEO	&	35.9681759	&	39	&	18.3799	&	18.3917	&	0.0163	&	0.0116	&	40	&	17.9367	&	17.9951	&	0.0207	&	0.0116	&	48	&	17.7016	&	17.7283	&	0.0354	&	0.0193	\\
	&	-4.2573895	&		&	0.0028	&	0.0247	&		&		&		&	0.0028	&	0.0325	&		&		&		&	0.0042	&	0.0279	&		&		\\
VIDEO	&	35.9729949	&	39	&	18.2901	&	18.3199	&	0.0252	&	0.0092	&	40	&	17.7774	&	17.7690	&	0.0205	&	0.0120	&	48	&	17.5529	&	17.6046	&	0.0367	&	0.0182	\\
	&	-5.2047350	&		&	0.0022	&	0.0583	&		&		&		&	0.0028	&	0.0551	&		&		&		&	0.0039	&	0.0430	&		&		\\
VIDEO	&	36.0194131	&	39	&	18.4466	&	18.4416	&	0.0169	&	0.0098	&	40	&	17.9480	&	17.9333	&	0.0218	&	0.0089	&	48	&	17.6283	&	17.6006	&	0.0367	&	0.0203	\\
	&	-4.8728386	&		&	0.0024	&	0.0119	&		&		&		&	0.0021	&	0.0121	&		&		&		&	0.0044	&	0.0116	&		&		\\
VIDEO	&	36.7075066	&	38	&	18.4144	&	18.3932	&	0.0273	&	0.0091	&	40	&	17.9624	&	17.9428	&	0.0301	&	0.0142	&	50	&	17.8677	&	17.8408	&	0.0336	&	0.0189	\\
	&	-4.5302703	&		&	0.0022	&	0.0073	&		&		&		&	0.0034	&	0.0075	&		&		&		&	0.0040	&	0.0082	&		&		\\
UVISTA	&	149.7053794	&	131	&	17.8524	&	17.8119	&	0.0095	&	0.0066	&	232	&	17.3109	&	17.3465	&	0.0362	&	0.0105	&	199	&	17.0759	&	17.0883	&	0.0175	&	0.0093	\\
	&	2.4862294	&		&	0.0009	&	0.0012	&		&		&		&	0.0010	&	0.0012	&		&		&		&	0.0010	&	0.0012	&		&		\\
UVISTA	&	149.7087571	&	80	&	17.9656	&	17.9106	&	0.0086	&	0.0047	&	123	&	17.4313	&	17.4534	&	0.0155	&	0.0092	&	101	&	17.2116	&	17.2032	&	0.0182	&	0.0121	\\
	&	2.7888722	&		&	0.0008	&	0.0019	&		&		&		&	0.0012	&	0.0018	&		&		&		&	0.0018	&	0.0019	&		&		\\
UVISTA	&	149.7234241	&	80	&	18.6410	&	18.5871	&	0.0117	&	0.0071	&	123	&	18.1390	&	18.1468	&	0.0238	&	0.0139	&	101	&	17.8376	&	17.8220	&	0.0237	&	0.0143	\\
	&	2.7511805	&		&	0.0012	&	0.0027	&		&		&		&	0.0019	&	0.0023	&		&		&		&	0.0021	&	0.0024	&		&		\\
UVISTA	&	149.7264829	&	150	&	17.8180	&	17.7721	&	0.0094	&	0.0057	&	236	&	17.2913	&	17.3231	&	0.0750	&	0.0110	&	203	&	17.0327	&	17.0421	&	0.0169	&	0.0115	\\
	&	2.5902461	&		&	0.0007	&	0.0012	&		&		&		&	0.0011	&	0.0012	&		&		&		&	0.0012	&	0.0012	&		&		\\
UVISTA	&	149.7307359	&	80	&	18.1943	&	18.1414	&	0.0083	&	0.0045	&	123	&	17.6766	&	17.7059	&	0.0194	&	0.0100	&	101	&	17.4237	&	17.4190	&	0.0194	&	0.0129	\\
	&	2.7399894	&		&	0.0008	&	0.0021	&		&		&		&	0.0013	&	0.0020	&		&		&		&	0.0019	&	0.0020	&		&		\\
UVISTA	&	149.7313560	&	158	&	17.5590	&	17.5094	&	0.0075	&	0.0047	&	252	&	17.0082	&	17.0394	&	0.0394	&	0.0100	&	206	&	16.7933	&	16.8107	&	0.0159	&	0.0084	\\
	&	2.6928870	&		&	0.0006	&	0.0011	&		&		&		&	0.0009	&	0.0011	&		&		&		&	0.0009	&	0.0011	&		&		\\
UVISTA	&	149.7319650	&	80	&	19.2441	&	19.1790	&	0.0188	&	0.0118	&	123	&	18.7651	&	18.7666	&	0.0308	&	0.0186	&	101	&	18.4652	&	18.4630	&	0.0324	&	0.0225	\\
	&	2.7327131	&		&	0.0020	&	0.0031	&		&		&		&	0.0025	&	0.0027	&		&		&		&	0.0033	&	0.0031	&		&		\\
UVISTA	&	149.7332239	&	78	&	17.8730	&	17.8189	&	0.0084	&	0.0048	&	120	&	17.4382	&	17.4644	&	0.0167	&	0.0095	&	97	&	17.1655	&	17.1668	&	0.0183	&	0.0106	\\
	&	2.7947267	&		&	0.0008	&	0.0018	&		&		&		&	0.0013	&	0.0018	&		&		&		&	0.0016	&	0.0019	&		&		\\
UVISTA	&	149.7355822	&	150	&	18.3514	&	18.3066	&	0.0107	&	0.0055	&	236	&	17.7820	&	17.7993	&	0.0887	&	0.0123	&	203	&	17.5625	&	17.5589	&	0.0215	&	0.0130	\\
	&	2.3205009	&		&	0.0007	&	0.0015	&		&		&		&	0.0012	&	0.0015	&		&		&		&	0.0014	&	0.0015	&		&		\\
UVISTA	&	149.7360512	&	152	&	17.6720	&	17.6253	&	0.0087	&	0.0051	&	232	&	17.0809	&	17.1141	&	0.0622	&	0.0093	&	199	&	16.9594	&	16.9764	&	0.0186	&	0.0092	\\
	&	2.4872598	&		&	0.0006	&	0.0011	&		&		&		&	0.0009	&	0.0011	&		&		&		&	0.0010	&	0.0012	&		&		\\
UVISTA	&	149.7476065	&	80	&	18.6534	&	18.6047	&	0.0125	&	0.0087	&	123	&	18.1630	&	18.1692	&	0.0262	&	0.0129	&	101	&	17.9131	&	17.9004	&	0.0306	&	0.0182	\\
	&	2.7272746	&		&	0.0014	&	0.0023	&		&		&		&	0.0017	&	0.0021	&		&		&		&	0.0027	&	0.0022	&		&		\\
UVISTA	&	149.7564883	&	80	&	17.4583	&	17.4066	&	0.0078	&	0.0048	&	123	&	16.8964	&	16.9283	&	0.0164	&	0.0073	&	101	&	16.6443	&	16.6641	&	0.0150	&	0.0078	\\
	&	2.7279976	&		&	0.0008	&	0.0013	&		&		&		&	0.0010	&	0.0012	&		&		&		&	0.0012	&	0.0013	&		&		\\
UVISTA	&	149.7597522	&	80	&	17.8232	&	17.7737	&	0.0084	&	0.0053	&	123	&	17.2872	&	17.3208	&	0.0191	&	0.0106	&	101	&	16.9664	&	16.9830	&	0.0159	&	0.0100	\\
	&	2.7827541	&		&	0.0009	&	0.0018	&		&		&		&	0.0014	&	0.0017	&		&		&		&	0.0015	&	0.0018	&		&		\\
UVISTA	&	149.7742789	&	104	&	17.8107	&	17.7615	&	0.0100	&	0.0056	&	251	&	17.2140	&	17.2393	&	0.0799	&	0.0111	&	206	&	17.0529	&	17.0630	&	0.0172	&	0.0093	\\
	&	2.6687115	&		&	0.0008	&	0.0013	&		&		&		&	0.0010	&	0.0012	&		&		&		&	0.0010	&	0.0013	&		&		\\
UVISTA	&	149.7772394	&	80	&	18.3290	&	18.2735	&	0.0110	&	0.0069	&	123	&	17.7934	&	17.8162	&	0.0248	&	0.0105	&	101	&	17.5463	&	17.5351	&	0.0218	&	0.0121	\\
	&	2.7783479	&		&	0.0011	&	0.0023	&		&		&		&	0.0014	&	0.0021	&		&		&		&	0.0018	&	0.0021	&		&		\\
UVISTA	&	149.7923194	&	150	&	17.5830	&	17.5310	&	0.0076	&	0.0046	&	237	&	17.1304	&	17.1662	&	0.0826	&	0.0114	&	203	&	16.8712	&	16.8883	&	0.0161	&	0.0096	\\
	&	2.5832843	&		&	0.0006	&	0.0011	&		&		&		&	0.0011	&	0.0011	&		&		&		&	0.0010	&	0.0011	&		&		\\
UVISTA	&	149.8156044	&	80	&	17.7772	&	17.7230	&	0.0097	&	0.0057	&	123	&	17.2745	&	17.3082	&	0.0174	&	0.0074	&	101	&	17.0546	&	17.0646	&	0.0184	&	0.0090	\\
	&	2.7697500	&		&	0.0010	&	0.0018	&		&		&		&	0.0010	&	0.0017	&		&		&		&	0.0013	&	0.0018	&		&		\\
UVISTA	&	149.8172140	&	80	&	17.8016	&	17.7487	&	0.0074	&	0.0049	&	123	&	17.2609	&	17.2904	&	0.0202	&	0.0098	&	101	&	17.0372	&	17.0491	&	0.0165	&	0.0096	\\
	&	2.7717426	&		&	0.0008	&	0.0018	&		&		&		&	0.0013	&	0.0017	&		&		&		&	0.0014	&	0.0018	&		&		\\
UVISTA	&	149.8251519	&	80	&	17.8417	&	17.7810	&	0.0074	&	0.0041	&	123	&	17.3732	&	17.3980	&	0.0165	&	0.0089	&	101	&	17.1457	&	17.1466	&	0.0187	&	0.0097	\\
	&	2.7608222	&		&	0.0007	&	0.0018	&		&		&		&	0.0012	&	0.0017	&		&		&		&	0.0014	&	0.0018	&		&		\\
  		\hline
	\end{tabular}
	\\
Note: UltraVISTA source table aperture magnitudes are systematically brighter than the variability table mean magnitudes by  $\approx 0.05$ mag at $J$, 
and fainter  by  $\approx 0.02$ mag at $H$, 
due to differences in the aperture correction and zeropoints. All mean magnitudes given in this paper are self-consistent.
\end{tiny}
\end{table*}
\end{landscape}
\restoregeometry

 \begin{table*}
\begin{scriptsize}
	\caption{Standard Stars with VISTA-System $ZYJHKs$. The first and second columns give a unique running number for convenience (VFS~=~`Very Faint Standard') and a formal IAU designation for each star.\label{tab:final}}
			\setlength\tabcolsep{2pt}
	\begin{tabular}{ccrrrrrrrrrrr}
		\hline
VFS & IAU &  RA$^{\circ}$     & 
    n $Z$ & $Z$  & 
  n $Y$ & $Y$  & 
  n $J$ & $J$  & 
  n $H$ & $H$  & 
  n $Ks$ & $Ks$ \\
 & name &  Decl.$^{\circ}$	&
      & $\pm$ mag	& 
       & $\pm$ mag	& 
    & $\pm$  mag	&
          & $\pm$ mag	& 
    & $\pm$  mag \\
\hline	
1 & UUDS J021608.22-051022.4 & 	34.0342392	&		&		&		&		&	349	&	18.3094	&	225	&	17.8626	&	513	&	17.5573	\\
   &    & 	-5.1728814	&		&		&		&		&		&	0.0020	&		&	0.0018	&		&	0.0032	\\
2 & UUDS J021614.05-052110.9 & 	34.0585392	&		&		&		&		&	350	&	18.4943	&	220	&	18.0184	&	522	&	17.7527	\\
   &    & 	-5.3530358	&		&		&		&		&		&	0.0020	&		&	0.0020	&		&	0.0032	\\
3 & UUDS J021621.77-051113.9 & 	34.0906904	&		&		&		&		&	351	&	18.3070	&	225	&	17.8586	&	516	&	17.5862	\\
   &    & 	-5.1871909	&		&		&		&		&		&	0.0020	&		&	0.0020	&		&	0.0031	\\
4 & UUDS J021627.99-052559.1 & 	34.1166206	&		&		&		&		&	350	&	18.3581	&	221	&	17.8624	&	522	&	17.6619	\\
   &    & 	-5.4330768	&		&		&		&		&		&	0.0020	&		&	0.0018	&		&	0.0032	\\
5 & UUDS J021635.02-052509.3 & 	34.1459342	&		&		&		&		&	352	&	18.3878	&	221	&	17.8711	&	524	&	17.5971	\\
   &    & 	-5.4192513	&		&		&		&		&		&	0.0021	&		&	0.0018	&		&	0.0033	\\
6 & UUDS J021637.43-051111.4 & 	34.1559576	&		&		&		&		&	352	&	18.4886	&	224	&	17.9790	&	516	&	17.7385	\\
   &    & 	-5.1864870	&		&		&		&		&		&	0.0021	&		&	0.0020	&		&	0.0033	\\
7 & UUDS J021647.80-045651.5 & 	34.1991747	&		&		&		&		&	353	&	18.3392	&	221	&	17.8223	&	524	&	17.5819	\\
   &    & 	-4.9476391	&		&		&		&		&		&	0.0021	&		&	0.0019	&		&	0.0033	\\
8 & UUDS J021653.58-044456.9 & 	34.2232606	&		&		&		&		&	386	&	18.5122	&	257	&	17.9708	&	604	&	17.7333	\\
   &    & 	-4.7491396	&		&		&		&		&		&	0.0022	&		&	0.0021	&		&	0.0034	\\
9 & UUDS J021655.40-052328.0 & 	34.2308454	&		&		&		&		&	351	&	18.3580	&	222	&	17.9076	&	535	&	17.6218	\\
   &    & 	-5.3911233	&		&		&		&		&		&	0.0021	&		&	0.0020	&		&	0.0032	\\
10 & UUDS J021714.48-052502.2 & 	34.3103343	&		&		&		&		&	348	&	18.5593	&	227	&	18.0758	&	533	&	17.7539	\\
   &    & 	-5.4172864	&		&		&		&		&		&	0.0022	&		&	0.0018	&		&	0.0034	\\
11 & UUDS J021717.78-052600.6 & 	34.3240632	&		&		&		&		&	345	&	18.4768	&	227	&	17.9876	&	530	&	17.6696	\\
   &    & 	-5.4334966	&		&		&		&		&		&	0.0023	&		&	0.0020	&		&	0.0034	\\
12 & UUDS J021723.41-052904.2 & 	34.3475622	&		&		&		&		&	345	&	18.3072	&	227	&	17.8462	&	531	&	17.5767	\\
   &    & 	-5.4844981	&		&		&		&		&		&	0.0020	&		&	0.0018	&		&	0.0032	\\
13 & UUDS J021730.02-052905.5 & 	34.3750969	&		&		&		&		&	345	&	18.4748	&	227	&	17.9338	&	531	&	17.6445	\\
   &    & 	-5.4848715	&		&		&		&		&		&	0.0022	&		&	0.0019	&		&	0.0034	\\
14 & UUDS J021745.34-052912.8 & 	34.4389254	&		&		&		&		&	339	&	18.3653	&	222	&	17.8549	&	517	&	17.6085	\\
   &    & 	-5.4868910	&		&		&		&		&		&	0.0021	&		&	0.0020	&		&	0.0033	\\
15 & UUDS J021807.68-050605.4 & 	34.5319836	&		&		&		&		&	439	&	18.3762	&	270	&	17.8516	&	627	&	17.6315	\\
   &    & 	-5.1015107	&		&		&		&		&		&	0.0020	&		&	0.0018	&		&	0.0033	\\
16 & UUDS J021840.81-050443.6 & 	34.6700469	&		&		&		&		&	374	&	18.3469	&	247	&	17.8485	&	576	&	17.5109	\\
   &    & 	-5.0787756	&		&		&		&		&		&	0.0024	&		&	0.0020	&		&	0.0036	\\
17 & UUDS J021842.63-045801.7 & 	34.6776197	&		&		&		&		&	404	&	18.4426	&	244	&	17.9281	&	594	&	17.6586	\\
   &    & 	-4.9671258	&		&		&		&		&		&	0.0022	&		&	0.0019	&		&	0.0035	\\
18 & UUDS J021849.90-045414.2 & 	34.7079081	&		&		&		&		&	347	&	18.1506	&	229	&	17.6618	&	533	&	17.5555	\\
   &    & 	-4.9039390	&		&		&		&		&		&	0.0017	&		&	0.0017	&		&	0.0029	\\
19 & UUDS J021858.20-052841.2 & 	34.7425103	&		&		&		&		&	345	&	18.2738	&	229	&	17.7916	&	529	&	17.6957	\\
   &    & 	-5.4781043	&		&		&		&		&		&	0.0019	&		&	0.0018	&		&	0.0029	\\
20 & UUDS J021905.89-045320.3 & 	34.7745424	&		&		&		&		&	371	&	18.4493	&	260	&	17.9604	&	595	&	17.6400	\\
   &    & 	-4.8889668	&		&		&		&		&		&	0.0022	&		&	0.0020	&		&	0.0034	\\
21 & UUDS J021916.14-044739.4 & 	34.8172698	&		&		&		&		&	354	&	18.6426	&	220	&	18.1886	&	521	&	17.9029	\\
   &    & 	-4.7942724	&		&		&		&		&		&	0.0022	&		&	0.0019	&		&	0.0033	\\
22 & UUDS J021916.37-050622.6 & 	34.8182276	&		&		&		&		&	404	&	18.5033	&	249	&	17.9004	&	619	&	17.7308	\\
   &    & 	-5.1062739	&		&		&		&		&		&	0.0021	&		&	0.0019	&		&	0.0034	\\
23 & UUDS J021929.06-050655.7 & 	34.8710949	&		&		&		&		&	366	&	18.3481	&	228	&	17.9747	&	536	&	17.8797	\\
   &    & 	-5.1154861	&		&		&		&		&		&	0.0016	&		&	0.0019	&		&	0.0029	\\
24 & VIDEO J022013.01-042853.0 & 	35.0542019	&		&		&	78	&	19.0018	&	39	&	18.5465	&	40	&	18.0267	&	48	&	17.8310	\\
   &    & 	-4.4813805	&		&		&		&	0.0021	&		&	0.0028	&		&	0.0047	&		&	0.0043	\\
25 & VIDEO J022027.67-045401.2 & 	35.1152774	&		&		&	78	&	19.0704	&	39	&	18.5271	&	40	&	18.0857	&	48	&	17.8255	\\
   &    & 	-4.9003215	&		&		&		&	0.0020	&		&	0.0021	&		&	0.0032	&		&	0.0049	\\
26 & VIDEO J022131.08-043620.7 & 	35.3795003	&		&		&	78	&	19.1025	&	39	&	18.6101	&	40	&	18.1286	&	48	&	17.9108	\\
   &    & 	-4.6057616	&		&		&		&	0.0023	&		&	0.0023	&		&	0.0028	&		&	0.0049	\\
27 & VIDEO J022238.37-043806.5 & 	35.6598904	&		&		&	77	&	18.7513	&	39	&	18.3307	&	40	&	17.7733	&	48	&	17.6104	\\
   &    & 	-4.6351333	&		&		&		&	0.0016	&		&	0.0020	&		&	0.0030	&		&	0.0032	\\
28 & VIDEO J022238.65-043243.2 & 	35.6610332	&		&		&	78	&	19.0260	&	39	&	18.5364	&	40	&	18.0755	&	48	&	17.8501	\\
   &    & 	-4.5453215	&		&		&		&	0.0019	&		&	0.0021	&		&	0.0025	&		&	0.0039	\\
29 & VIDEO J022247.04-043135.2 & 	35.6959814	&		&		&	78	&	19.3279	&	39	&	18.9173	&	40	&	18.3808	&	48	&	18.1880	\\
   &    & 	-4.5264450	&		&		&		&	0.0026	&		&	0.0036	&		&	0.0043	&		&	0.0049	\\
30 & VIDEO J022254.76-051213.4 & 	35.7281547	&		&		&	78	&	19.3492	&	39	&	18.8672	&	40	&	18.4067	&	48	&	18.1372	\\
   &    & 	-5.2037306	&		&		&		&	0.0027	&		&	0.0029	&		&	0.0041	&		&	0.0053	\\
31 & VIDEO J022327.86-040627.2 & 	35.8660837	&		&		&	78	&	19.1724	&	39	&	18.6624	&	40	&	18.1648	&	48	&	17.9286	\\
   &    & 	-4.1075563	&		&		&		&	0.0024	&		&	0.0024	&		&	0.0042	&		&	0.0050	\\
32 & VIDEO J022350.96-041325.5 & 	35.9623324	&		&		&	78	&	18.8035	&	39	&	18.3172	&	40	&	17.8155	&	48	&	17.5798	\\
   &    & 	-4.2237615	&		&		&		&	0.0017	&		&	0.0021	&		&	0.0034	&		&	0.0045	\\
33 & VIDEO J022352.36-041526.6 & 	35.9681759	&		&		&	77	&	18.8680	&	39	&	18.3799	&	40	&	17.9367	&	48	&	17.7016	\\
   &    & 	-4.2573895	&		&		&		&	0.0021	&		&	0.0028	&		&	0.0028	&		&	0.0042	\\
34 & VIDEO J022353.52-051217.0 & 	35.9729949	&		&		&	78	&	18.7733	&	39	&	18.2901	&	40	&	17.7774	&	48	&	17.5529	\\
   &    & 	-5.2047350	&		&		&		&	0.0025	&		&	0.0022	&		&	0.0028	&		&	0.0039	\\
35 & VIDEO J022404.66-045222.2 & 	36.0194131	&		&		&	78	&	19.1804	&	39	&	18.4466	&	40	&	17.9480	&	48	&	17.6283	\\
   &    & 	-4.8728386	&		&		&		&	0.0022	&		&	0.0024	&		&	0.0021	&		&	0.0044	\\
36 & VIDEO J022649.80-043149.0 & 	36.7075066	&	74	&	18.9232	&	75	&	18.7517	&	38	&	18.4144	&	40	&	17.9624	&	50	&	17.8677	\\
   &    & 	-4.5302703  	&		&	0.0017	&		&	0.0016	&		&	0.0022	&		&	0.0034	&		&	0.0040	\\
37 & UCAL J055056.46+155253.9 & 	87.7352518	&	106	&	17.2730	&	121	&	17.0544	&	136	&	16.6104	&	132	&	16.2224	&	136	&	16.0999	\\
   &    & 	15.8816295	&		&	0.0070	&		&	0.0035	&		&	0.0031	&		&	0.0041	&		&	0.0060	\\
38 & UCAL J055210.55+161245.3 & 	88.0439399	&	86	&	17.4837	&	103	&	17.2840	&	130	&	16.7693	&	125	&	16.1954	&	133	&	16.0186	\\
   &    & 	16.2125856	&		&	0.0095	&		&	0.0044	&		&	0.0048	&		&	0.0036	&		&	0.0051	\\
39 & UltraVISTA J095849.29+022910.4 & 	149.7053794	&		&		&	90	&	18.3682	&	131	&	17.8524	&	232	&	17.3109	&	199	&	17.0759	\\
   &    & 	2.4862294	&		&		&		&	0.0013	&		&	0.0009	&		&	0.0010	&		&	0.0010	\\
40 & UltraVISTA J095850.10+024719.9 & 	149.7087571	&		&		&	75	&	18.4555	&	80	&	17.9656	&	123	&	17.4313	&	101	&	17.2116	\\
   &    & 	2.7888722	&		&		&		&	0.0011	&		&	0.0008	&		&	0.0012	&		&	0.0018	\\
 		\hline
	\end{tabular}
\end{scriptsize}
\end{table*}

\begin{table*}
\begin{scriptsize}
	\contcaption{Standard Stars with VISTA-System $ZYJHKs$}
			\setlength\tabcolsep{2pt}
	\begin{tabular}{ccrrrrrrrrrrr}
		\hline
VFS & IAU &  RA$^{\circ}$     & 
    n $Z$ & $Z$  & 
  n $Y$ & $Y$  & 
  n $J$ & $J$  & 
  n $H$ & $H$  & 
  n $Ks$ & $Ks$ \\
 & name &  Decl.$^{\circ}$	&
      & $\pm$ mag	& 
       & $\pm$ mag	& 
    & $\pm$  mag	&
          & $\pm$ mag	& 
    & $\pm$  mag \\
\hline
41 & UltraVISTA J095853.62+024504.2 & 	149.7234241	&		&		&	75	&	19.2576	&	80	&	18.6410	&	123	&	18.1390	&	101	&	17.8376	\\
   &    & 	2.7511805	&		&		&		&	0.0012	&		&	0.0012	&		&	0.0019	&		&	0.0021	\\
42 & UltraVISTA J095854.36+023524.9 & 	149.7264829	&		&		&	140	&	18.3791	&	150	&	17.8180	&	236	&	17.2913	&	203	&	17.0327	\\
   &    & 	2.5902461	&		&		&		&	0.0013	&		&	0.0007	&		&	0.0011	&		&	0.0012	\\
43 & UltraVISTA J095855.38+024424.0 & 	149.7307359	&		&		&	75	&	18.7503	&	80	&	18.1943	&	123	&	17.6766	&	101	&	17.4237	\\
   &    & 	2.7399894	&		&		&		&	0.0010	&		&	0.0008	&		&	0.0013	&		&	0.0019	\\
44 & UltraVISTA J095855.53+024134.4 & 	149.7313560	&		&		&	143	&	18.0241	&	158	&	17.5590	&	252	&	17.0082	&	206	&	16.7933	\\
   &    & 	2.6928870	&		&		&		&	0.0008	&		&	0.0006	&		&	0.0009	&		&	0.0009	\\
45 & UltraVISTA J095855.67+024357.8 & 	149.7319650	&		&		&	75	&	19.8434	&	80	&	19.2441	&	123	&	18.7651	&	101	&	18.4652	\\
   &    & 	2.7327131	&		&		&		&	0.0022	&		&	0.0020	&		&	0.0025	&		&	0.0033	\\
46 & UltraVISTA J095855.97+024741.0 & 	149.7332239	&		&		&	71	&	18.4389	&	78	&	17.8730	&	120	&	17.4382	&	97	&	17.1655	\\
   &    & 	2.7947267	&		&		&		&	0.0009	&		&	0.0008	&		&	0.0013	&		&	0.0016	\\
47 & UltraVISTA J095856.54+021913.8 & 	149.7355822	&		&		&	140	&	18.8246	&	150	&	18.3514	&	236	&	17.7820	&	203	&	17.5625	\\
   &    & 	2.3205009	&		&		&		&	0.0010	&		&	0.0007	&		&	0.0012	&		&	0.0014	\\
48 & UltraVISTA J095856.65+022914.1 & 	149.7360512	&		&		&	147	&	18.0802	&	152	&	17.6720	&	232	&	17.0809	&	199	&	16.9594	\\
   &    & 	2.4872598	&		&		&		&	0.0008	&		&	0.0006	&		&	0.0009	&		&	0.0010	\\
49 & UltraVISTA J095859.43+024338.2 & 	149.7476065	&		&		&	75	&	19.1804	&	80	&	18.6534	&	123	&	18.1630	&	101	&	17.9131	\\
   &    & 	2.7272746	&		&		&		&	0.0016	&		&	0.0014	&		&	0.0017	&		&	0.0027	\\
50 & UltraVISTA J095901.56+024340.8 & 	149.7564883	&		&		&	75	&	17.9949	&	80	&	17.4583	&	123	&	16.8964	&	101	&	16.6443	\\
   &    & 	2.7279976	&		&		&		&	0.0011	&		&	0.0008	&		&	0.0010	&		&	0.0012	\\
51 & UltraVISTA J095902.34+024657.9 & 	149.7597522	&		&		&	75	&	18.4764	&	80	&	17.8232	&	123	&	17.2872	&	101	&	16.9664	\\
   &    & 	2.7827541	&		&		&		&	0.0012	&		&	0.0009	&		&	0.0014	&		&	0.0015	\\
52 & UltraVISTA J095905.83+024007.4 & 	149.7742789	&		&		&	79	&	18.2324	&	104	&	17.8107	&	251	&	17.2140	&	206	&	17.0529	\\
   &    & 	2.6687115	&		&		&		&	0.0011	&		&	0.0008	&		&	0.0010	&		&	0.0010	\\
53 & UltraVISTA J095906.54+024642.1 & 	149.7772394	&		&		&	75	&	18.8640	&	80	&	18.3290	&	123	&	17.7934	&	101	&	17.5463	\\
   &    & 	2.7783479	&		&		&		&	0.0011	&		&	0.0011	&		&	0.0014	&		&	0.0018	\\
54 & UltraVISTA J095910.16+023459.8 & 	149.7923194	&		&		&	140	&	18.1225	&	150	&	17.5830	&	237	&	17.1304	&	203	&	16.8712	\\
   &    & 	2.5832843	&		&		&		&	0.0012	&		&	0.0006	&		&	0.0011	&		&	0.0010	\\
55 & UltraVISTA J095915.75+024611.1 & 	149.8156044	&		&		&	75	&	18.2690	&	80	&	17.7772	&	123	&	17.2745	&	101	&	17.0546	\\
   &    & 	2.7697500	&		&		&		&	0.0009	&		&	0.0010	&		&	0.0010	&		&	0.0013	\\
56 & UltraVISTA J095916.13+024618.3 & 	149.8172140	&		&		&	75	&	18.2823	&	80	&	17.8016	&	123	&	17.2609	&	101	&	17.0372	\\
   &    & 	2.7717426	&		&		&		&	0.0009	&		&	0.0008	&		&	0.0013	&		&	0.0014	\\
57 & UltraVISTA J095918.04+024539.0 & 	149.8251519	&		&		&	75	&	18.3345	&	80	&	17.8417	&	123	&	17.3732	&	101	&	17.1457	\\
   &    & 	2.7608222	&		&		&		&	0.0010	&		&	0.0007	&		&	0.0012	&		&	0.0014	\\
58 & UCAL J182653.41+040331.5 & 	276.7225557	&	133	&	17.4972	&	131	&	17.2159	&	156	&	16.6992	&	157	&	16.1551	&	167	&	15.9994	\\
   &    & 	4.0587376	&		&	0.0081	&		&	0.0042	&		&	0.0036	&		&	0.0034	&		&	0.0052	\\
59 & UCAL J182718.74+035838.7 & 	276.8280740	&	140	&	17.2663	&	138	&	17.0325	&	158	&	16.5879	&	157	&	16.1728	&	169	&	16.0038	\\
   &    & 	3.9774041	&		&	0.0074	&		&	0.0035	&		&	0.0036	&		&	0.0033	&		&	0.0051	\\
60 & UCAL J182857.59+040542.8 & 	277.2399424	&	135	&	17.5724	&	137	&	17.2598	&	156	&	16.7456	&	154	&	16.3002	&	168	&	16.0376	\\
   &    & 	4.0952285	&		&	0.0085	&		&	0.0041	&		&	0.0037	&		&	0.0037	&		&	0.0052	\\
61 & UCAL J182904.80+043119.7 & 	277.2699877	&	135	&	17.6120	&	133	&	17.3401	&	157	&	16.7872	&	156	&	16.3344	&	167	&	16.1180	\\
   &    & 	4.5221425	&		&	0.0081	&		&	0.0040	&		&	0.0037	&		&	0.0037	&		&	0.0056	\\
62 & UDXS J221217.08+004949.3 & 	333.0711763	&		&		&		&		&	26	&	17.3587	&		&	&	22	&	16.5811	\\
   &    & 	0.8303505	&		&		&		&		&		&	0.0027	&		&	&		&	0.0046	\\
63 & UDXS J221217.17+005458.4 & 	333.0715557	&		&		&		&		&	26	&	18.1551	&		&		&	22	&	17.3931	\\
   &    & 	0.9162355	&		&		&		&		&		&	0.0035	&		&		&		&	0.0054	\\
64 & UDXS J221217.24+011120.2 & 	333.0718253	&		&		&		&		&	26	&	18.2026	&		&		&	22	&	17.4465	\\
   &    & 	1.1889522	&		&		&		&		&		&	0.0042	&		&		&		&	0.0059	\\
65 & UDXS J221308.15-002438.1 & 	333.2839533	&		&		&		&		&	24	&	17.7879	&		&		&	27	&	17.0097	\\
   &    & 	-0.4105822	&		&		&		&		&		&	0.0030	&		&		&		&	0.0048	\\
66 & UDXS J221310.28+004403.5 & 	333.2928434	&		&		&		&		&	24	&	17.4070	&	29	&	16.9648	&	24	&	16.6916	\\
   &    & 	0.7343005	&		&		&		&		&		&	0.0028	&		&	0.0033	&		&	0.0048	\\
67 & UDXS J221312.69+004516.1 & 	333.3028790	&		&		&		&		&	24	&	17.5557	&	29	&	17.1132	&	24	&	16.8090	\\
   &    & 	0.7544734	&		&		&		&		&		&	0.0035	&		&	0.0037	&		&	0.0054	\\
68 & UDXS J221326.27+004504.4 & 	333.3594622	&		&		&		&		&	24	&	18.3002	&	29	&	17.8259	&	24	&	17.5602	\\
   &    & 	0.7512320	&		&		&		&		&		&	0.0034	&		&	0.0034	&		&	0.0057	\\
69 & UDXS J221331.78+004432.8 &  333.3824122   &		&		&		&		&   24  & 17.8077 & 28 & 17.3309 & 	24	& 16.9979 \\
   &    &  0.7424491     &		&		&		&		&       & 0.0028  &    & 0.0037 &  & 0.0048 \\
70 & UDXS J221348.36+004340.2 & 	333.4515020	&		&		&		&		&	24	&	17.6807	&	29	&	17.2226	&	24	&	16.9545	\\
   &    & 	0.7278254	&		&		&		&		&		&	0.0044	&		&	0.0032	&		&	0.0049	\\
71 & UDXS J221359.70+004405.8 & 	333.4987510	&		&		&		&		&	24	&	17.9313	&	29	&	17.4356	&	24	&	17.1440	\\
   &    & 	0.7349581	&		&		&		&		&		&	0.0034	&		&	0.0043	&		&	0.0052	\\
72 & UDXS J221450.27+003128.5 & 	333.7094552	&		&		&		&		&	25	&	17.2650	&	27	&	16.6699	&	22	&	16.5297	\\
   &    & 	0.5245924	&		&		&		&		&		&	0.0024	&		&	0.0032 &		&	0.0064	\\
73 & UDXS J221544.78-005649.9 & 	333.9366041	&		&		&		&		&	24	&	17.3086	&		&		&	24	&	16.4942	\\
   &    & 	-0.9471891	&		&		&		&		&		&	0.0024	&		&		&		&	0.0050	\\
74 & UDXS J221835.85+012255.2 & 	334.6493784	&		&		&		&		&	33	&	17.8448	&	20	&	17.2731	&	44	&	16.9818	\\
   &    & 	1.3819891	&		&		&		&		&		&	0.0039	&		&	0.0054	&		&	0.0051	\\
75 & UDXS J221916.37+002334.2 & 	334.8182190	&		&		&		&		&	32	&	17.6622	&		&		&	34	&	16.8747	\\
   &    & 	0.3928450	&		&		&		&		&		&	0.0034	&		&		&		&	0.0054	\\
76 & UDXS J221942.03+004338.8 & 	334.9251383	&		&		&		&		&	31	&	17.4172	&	21	&	16.8889	&	34	&	16.7754	\\
   &    & 	0.7274321	&		&		&		&		&		&	0.0030	&		&	0.0016	&		&	0.0047	\\
77 & UDXS J222007.27+002643.6 & 	335.0303119	&		&		&		&		&	29	&	17.4399	&	22	&	17.0128	&	30	&	16.9227	\\
   &    & 	0.4454373	&		&		&		&		&		&	0.0040	&		&	0.0035	&		&	0.0049	\\
78 & UDXS J222153.93+003033.1 & 	335.4746997	&		&		&		&		&	45	&	17.3604	&	33	&	16.7993	&	46	&	16.5495	\\
   &    & 	0.5091841	&		&		&		&		&		&	0.0031	&		&	0.0034	&		&	0.0049	\\
79 & UDXS J222205.63+001714.8 & 	335.5234567	&		&		&		&		&	31	&	17.1721	&	26	&	16.6742	&	32	&	16.5423	\\
   &    & 	0.2874459	&		&		&		&		&			&	0.0036	&		&	0.0036	&		&	0.0052	\\
80 & UDXS J222219.75-000839.9 & 	335.5823067	&		&		&		&		&		28	&	17.3728	&		&		&	33	&	16.6027	\\
   &    & 	-0.1444080	&		&		&		&		&			&	0.0031	&		&		&		&	0.0042	\\
81 & UDXS J222227.61+003032.8 & 	335.6150576	&		&		&		&		&		31	&	17.2914	&	26	&	16.8690	&	32	&	16.7655	\\
   &    & 	0.5091194	&		&		&		&		&			&	0.0027	&		&	0.0036	&		&	0.0054	\\
		\hline
	\end{tabular}
\end{scriptsize}
\end{table*}

\section*{Acknowledgements}

The authors gratefully acknowledge the work of Mike Irwin and the Cambridge Astronomical Survey Unit (CASU) along with Mike Read and the Edinburgh Wide Field Astronomy Unit (WFAU) for their tireless work in generating and serving the wide field infrared public surveys from UKIRT and VISTA. We also acknowledge the selfless work by the respective public survey leads: Alastair Edge (DXS), Omar Almaini (UDS), Marijn Franx and Jim Dunlop (UltraVISTA) and Matt Jarvis (VIDEO). 

The Pan-STARRS Survey and science archive have been made possible through contributions by the Institute for Astronomy, the University of Hawaii, the Pan-STARRS Project Office, the Max-Planck Society and its participating institutes, the Max Planck Institute for Astronomy, Heidelberg and the Max Planck Institute for Extraterrestrial Physics, Garching, The Johns Hopkins University, Durham University, the University of Edinburgh, the Queen's University Belfast, the Harvard-Smithsonian Center for Astrophysics, the Las Cumbres Observatory Global Telescope Network Incorporated, the National Central University of Taiwan, the Space Telescope Science Institute, the National Aeronautics and Space Administration under Grant No. NNX08AR22G issued through the Planetary Science Division of the NASA Science Mission Directorate, the National Science Foundation Grant No. AST-1238877, the University of Maryland, Eotvos Lorand University (ELTE), the Los Alamos National Laboratory, and the Gordon and Betty Moore Foundation.

This work presents results from the European Space Agency (ESA) space mission Gaia. Gaia data are being processed by the Gaia Data Processing and Analysis Consortium (DPAC). Funding for the DPAC is provided by national institutions, in particular the institutions participating in the Gaia MultiLateral Agreement (MLA). The Gaia mission website is \burl{https://www.cosmos.esa.int/gaia}. The Gaia archive website is \burl{https://archives.esac.esa.int/gaia}.





\bibliographystyle{mnras}
\bibliography{standards}



\appendix
\clearpage
\section{SQL Queries}

\indent SQL for the DXS\\
{\tt SELECT 
s.ra, s.dec,  s.jAperMag3,v.jMeanMag,v.jnGoodObs,  s.jAperMag3err, v.jMagRms, v.jMagMAD, s.hAperMag3,v.hMeanMag,
 v.hnGoodObs, s.hAperMag3err,v.hMagRms, v.hMagMAD, s.kAperMag3,  v.kMeanMag, v.knGoodObs,s.kAperMag3err, v.kMagRms, v.kMagMAD  
FROM 
 dxsSource AS s, dxsVariability AS v  
WHERE 
s.sourceID=v.sourceID AND 
 s.mergedClass=-1 AND v.variableClass=0 AND  s.dec > -30 AND  
 v.kMeanMag>16.5 AND 
v.jnGoodObs>20 AND  v.hnGoodObs>5 AND  v.knGoodObs>20 AND 
(v.jMagMAD/(SQRT(v.jnGoodObs -  1)))<=0.004 AND  
(v.hMagMAD/(SQRT(v.hnGoodObs - 1)))<=0.004 AND  
 (v.kMagMAD/(SQRT(v.knGoodObs - 1)))<=0.004 AND  
 (v.jMagRms/SQRT(v.jnGoodObs - 1))<=0.006 AND  
(v.hMagRms/SQRT(v.hnGoodObs  - 1))<=0.006 AND  
(v.kMagRms/SQRT(v.knGoodObs - 1))<=0.006 AND 
 ((s.jAperMag3 - v.jMeanMag) < 2.5 * (SQRT(s.jAperMag3err *  s.jAperMag3err + (v.jMagRms/SQRT(v.jnGoodObs - 1)) * (v.jMagRms/SQRT(v.jnGoodObs - 1))))) AND  
((s.hAperMag3 - v.hMeanMag) <  2.5 * (SQRT(s.hAperMag3err * s.hAperMag3err + (v.hMagRms/SQRT(v.hnGoodObs - 1)) * (v.hMagRms/SQRT(v.hnGoodObs - 1)))))  AND  
((s.kAperMag3 - v.kMeanMag) < 2.5 * (SQRT(s.kAperMag3err *  s.kAperMag3err + (v.kMagRms/SQRT(v.knGoodObs - 1)) *  (v.kMagRms/SQRT(v.knGoodObs - 1)))))}
\\
\\

\indent SQL for the UDS\\
{\it NOTE: there is an offset between {\tt AperMag3} and {\tt MeanMag} due to the former not being aperture-corrected.}\\
{\tt SELECT 
s.ra, s.dec,  s.jAperMag3,v.jMeanMag,v.jnGoodObs,  s.jAperMag3err, v.jMagRms, v.jMagMAD, s.hAperMag3,v.hMeanMag,
 v.hnGoodObs, s.hAperMag3err,v.hMagRms, v.hMagMAD, s.kAperMag3, v.kMeanMag, v.knGoodObs,s.kAperMag3err, v.kMagRms, v.kMagMAD  
FROM 
 udsSource AS s, udsVariability AS v  
WHERE 
 s.sourceID=v.sourceID AND  s.mergedClass=-1 AND v.variableClass=0 AND  v.kMeanMag>17.5 AND 
v.jnGoodObs>100 AND v.hnGoodObs>100 AND  v.knGoodObs>100 AND   
(v.jMagMAD/(SQRT(v.jnGoodObs - 1)))<=0.004 AND  
(v.hMagMAD/(SQRT(v.hnGoodObs - 1)))<=0.004 AND  
(v.kMagMAD/(SQRT(v.knGoodObs - 1)))<=0.004 AND  
(v.jMagRms/SQRT(v.jnGoodObs - 1))<=0.006 AND  
(v.hMagRms/SQRT(v.hnGoodObs  - 1))<=0.006 AND  
(v.kMagRms/SQRT(v.knGoodObs - 1))<=0.006 AND 
s.jAperMag3err<=0.003 AND  
s.hAperMag3err<=0.003 AND  
s.kAperMag3err<=0.003 AND  
(s.jAperMag3 - v.jMeanMag < 0.20) AND 
(s.jAperMag3 - v.jMeanMag > 0.18) AND 
(s.hAperMag3 - v.hMeanMag <  0.21) AND 
(s.hAperMag3 - v.hMeanMag > 0.19) AND 
(s.kAperMag3 -  v.kMeanMag < 0.19) AND 
 (s.kAperMag3 - v.kMeanMag > 0.17)
}
\\
\\

\indent SQL for the UltraVISTA\\
{\it NOTE: there is an offset between {\tt AperMag3} and {\tt MeanMag} due to differences in the aperture correction and zeropoints.}\\
{\tt SELECT 
v.sourceID,	 s.ra,	 s.dec,	 v.variableClass,	s.yclassStat,	 s.jclassStat,	 s.hclassStat,	 s.ksclassStat,	 
s.yAperMag3,	 v.ymeanMag,	 v.ynGoodObs,	 s.yAperMag3err,	 v.yMagRms,	 v.yMagMAD,	 s.jAperMag3,	 v.jmeanMag,	 
v.jnGoodObs,	 s.jAperMag3err,	 v.jMagRms,	 v.jMagMAD,	s.hAperMag3,	v.hmeanMag,	v.hnGoodObs,	s.hAperMag3err,	 v.hMagRms,	 v.hMagMAD,	 
s.ksAperMag3,	 v.ksmeanMag,	 v.ksnGoodObs,	 s.ksAperMag3err,	 v.ksMagRms,	 v.ksMagMAD 
FROM ultravistaSource AS s,	 ultravistaVariability AS v 				
WHERE s.sourceID=v.sourceID AND v.variableClass=0 AND v.ksMeanMag > 16.5 AND 
s.hclassStat>0.7 AND s.ksclassStat>0.7 AND v.ynGoodObs>=20 	AND  v.jnGoodObs>=20 AND v.hnGoodObs>=20 AND v.ksnGoodObs>=20 AND   					(v.yMagMAD/(SQRT(v.ynGoodObs- 1)))<=0.004 AND 					
(v.jMagMAD/(SQRT(v.jnGoodObs- 1)))<=0.004 AND 					
(v.hMagMAD/(SQRT(v.hnGoodObs -1)))<=0.004 AND 					
(v.ksMagMAD/(SQRT(v.ksnGoodObs- 1)))<=0.004 AND  					
(v.yMagRms/SQRT(v.ynGoodObs- 1))<=0.006 AND 					
(v.jMagRms/SQRT(v.jnGoodObs- 1))<=0.006 AND 					
(v.hMagRms/SQRT(v.hnGoodObs- 1))<=0.006 AND 					
(v.ksMagRms/SQRT(v.ksnGoodObs- 1))<=0.006 AND 
((s.yAperMag3 - v.yMeanMag) <= 0.0) AND 
((s.yAperMag3 - v.yMeanMag) >= -0.04) AND 					
((s.jAperMag3 - v.jMeanMag) <= -0.04) AND 
((s.jAperMag3 - v.jMeanMag)  >= -0.08) AND 
((s.hAperMag3 - v.hMeanMag) <= 0.04) AND 
((s.hAperMag3 - v.hMeanMag) >= 0.00) AND  
((s.ksAperMag3 - v.ksMeanMag) <= 0.02) 					
AND ((s.ksAperMag3 - v.ksMeanMag) >= -0.02)					
}
\\
\\

\indent SQL for VIDEO\\
{\tt SELECT 
v.sourceID, s.ra, s.dec, s.mergedClass,  s.zAperMag3, v.zMeanMag,v.znGoodObs, s.zAperMag3err, v.zMagRms, v.zMagMAD,  
 s.yAperMag3,v.ymeanMag,v.ynGoodObs,s.yAperMag3err, v.yMagRms,  v.yMagMAD, s.jAperMag3,v.jmeanMag,v.jnGoodObs, s.jAperMag3err, 
 v.jMagRms, v.jMagMAD,  s.hAperMag3,v.hmeanMag,v.hnGoodObs,  s.hAperMag3err, v.hMagRms, v.hMagMAD,  s.ksAperMag3, v.ksmeanMag, 
 v.ksnGoodObs,s.ksAperMag3err, v.ksMagRms, v.ksMagMAD  
FROM 
 videoSource AS s, videoVariability AS v 
WHERE 
 s.sourceID=v.sourceID 
 AND s.mergedClass in (-1, -2) AND v.variableClass=0 AND s.dec > -30 AND  
v.ksMeanMag > 17.5 AND  v.ynGoodObs>20 AND v.jnGoodObs> 20 AND v.hnGoodObs > 20 AND v.ksnGoodObs > 20 AND  
(v.yMagMAD/(SQRT(v.ynGoodObs - 1)))<=0.004 AND 
 (v.jMagMAD/(SQRT(v.jnGoodObs - 1)))<=0.004 AND  
 (v.hMagMAD/(SQRT(v.hnGoodObs - 1)))<=0.004 AND  
 (v.ksMagMAD/(SQRT(v.ksnGoodObs - 1)))<=0.004 AND  
 (v.yMagRms/SQRT(v.ynGoodObs - 1))<=0.006 AND 
(v.jMagRms/SQRT(v.jnGoodObs - 1))<=0.006 AND  
(v.hMagRms/SQRT(v.hnGoodObs - 1))<=0.006 AND  
 (v.ksMagRms/SQRT(v.ksnGoodObs - 1))<=0.006 AND 
((s.yAperMag3 -  v.yMeanMag) < 2.5 * (SQRT(s.yAperMag3err * s.yAperMag3err + (v.yMagRms/SQRT(v.ynGoodObs - 1)) * (v.yMagRms/SQRT(v.ynGoodObs - 1)))))  AND 
((s.jAperMag3 - v.jMeanMag) < 2.5 * (SQRT(s.jAperMag3err *  s.jAperMag3err + (v.jMagRms/SQRT(jnGoodObs - 1)) * (v.jMagRms/SQRT(v.jnGoodObs - 1))))) AND  
((s.hAperMag3 - v.hMeanMag) <  2.5 * (SQRT(s.hAperMag3err * s.hAperMag3err +  (v.hMagRms/SQRT(v.hnGoodObs - 1)) * (v.hMagRms/SQRT(v.hnGoodObs - 1)))))  AND 
((s.ksAperMag3 - v.ksMeanMag) < 2.5 * (SQRT(s.ksAperMag3err *  s.ksAperMag3err + (v.ksMagRms/SQRT(ksnGoodObs - 1)) *  (v.ksMagRms/SQRT(ksnGoodObs - 1)))))}
\\
\\

\indent SQL for WFCAMCAL\\
{\tt SELECT
s.ra, s.dec,  s.zAperMag3,v.zMeanMag,v.znGoodObs,  s.zAperMag3err, v.zMagRms, v.zMagMAD,  s.yAperMag3,v.yMeanMag,
 v.ynGoodObs, s.yAperMag3err, v.yMagRms, v.yMagMAD,  s.jAperMag3,
 v.jMeanMag,v.jnGoodObs, s.jAperMag3err, v.jMagRms, v.jMagMAD,  
 s.hAperMag3,v.hMeanMag,v.hnGoodObs, s.hAperMag3err,v.hMagRms, 
 v.hMagMAD,  s.kAperMag3, v.kMeanMag, v.knGoodObs,s.kAperMag3err, 
 v.kMagRms, v.kMagMAD  
FROM 
calSource AS s, calVariability AS v  
WHERE 
 s.sourceID=v.sourceID AND s.mergedClass=-1 AND v.variableClass=0 AND   v.kMeanMag>16 AND 
v.ynGoodObs>100 AND  v.jnGoodObs>100 AND v.hnGoodObs>100 AND v.knGoodObs>100 AND   
 (v.yMagMAD/(SQRT(v.ynGoodObs - 1)))<=0.004 AND 
 (v.jMagMAD/(SQRT(v.jnGoodObs - 1)))<=0.004 AND  
 (v.hMagMAD/(SQRT(v.hnGoodObs - 1)))<=0.004 AND  
 (v.kMagMAD/(SQRT(v.knGoodObs - 1)))<=0.004 AND  
(v.yMagRms/SQRT(v.ynGoodObs -  1))<=0.006 AND 
(v.jMagRms/SQRT(v.jnGoodObs - 1))<=0.006 AND  
(v.hMagRms/SQRT(v.hnGoodObs - 1))<=0.006 AND  
(v.kMagRms/SQRT(v.knGoodObs  - 1))<=0.006 AND 
((s.yAperMag3 -  v.yMeanMag) < 2.5 * (SQRT(s.yAperMag3err * s.yAperMag3err +  (v.yMagRms/SQRT(v.ynGoodObs - 1)) * (v.yMagRms/SQRT(v.ynGoodObs - 1))))) AND 
((s.jAperMag3 - v.jMeanMag) < 2.5 * (SQRT(s.jAperMag3err *  s.jAperMag3err + (v.jMagRms/SQRT(v.jnGoodObs - 1)) *  (v.jMagRms/SQRT(v.jnGoodObs - 1))))) AND  
((s.hAperMag3 - v.hMeanMag) <  2.5 * (SQRT(s.hAperMag3err * s.hAperMag3err +  (v.hMagRms/SQRT(v.hnGoodObs - 1)) * (v.hMagRms/SQRT(v.hnGoodObs - 1)))))  AND 
((s.kAperMag3 - v.kMeanMag) < 2.5 * (SQRT(s.kAperMag3err *  s.kAperMag3err + (v.kMagRms/SQRT(v.knGoodObs - 1)) *  (v.kMagRms/SQRT(v.knGoodObs - 1)))))
}
\\
\\

\section{Possible Galaxies}

\newgeometry{width=20cm,height=13cm}
\begin{landscape}
\begin{table*}
\begin{tiny}
	\caption{Data for Sources Identified as Possible Galaxies}
			\setlength\tabcolsep{1pt}
	\begin{tabular}{rrcrrrrrrrrrrrrrrrr}
		\hline
 \multicolumn{2}{c}{IR Survey} & Flag &    \multicolumn{6}{c}{Gaia} &    \multicolumn{5}{c}{Pan-STARRS meanPSF AB mags} & 
   \multicolumn{5}{c}{VISTA Vega mags}\\
   Name & RA$^{\circ}$     & 		&	RA$^{\circ}$ 	&	Decl.$^{\circ}$	& Parallax  & $\mu$ RA		      &	$\mu$ Decl.	 	& G			&	$g$ &	$r$ 	&	$i$	& $z$ 	& $y$ 	& $Z$		& $Y$	&	$J$	&	$H$	& $Ks$			\\
     & Decl.$^{\circ}$	&		&	$\pm$ mas		&	$\pm$ mas   	& $\pm$  mas & $\pm$ mas yr$^{-1}$ &	$\pm$ mas yr$^{-1}$ 	& 		& $\pm$ 	& $\pm$ 	& $\pm$ 	& $\pm$  	& $\pm$ 	&  $\pm$	&	 $\pm$		&	 $\pm$				&	 $\pm$		&	 $\pm$			\\
\hline
UDS	&	34.0387095		&	1	&	34.038750344750	&	-5.393224535432	&	2.0828	&	4.0926				&	-8.9745		& 20.7447	&	21.9886	&	20.8532	&	20.0351	&	19.6112	&	19.3296	&			&			&	18.2804	&		17.7371	&	17.5162			\\
	&	-5.3931529		&       &	1.43711         &    1.25846		&	1.7445	&	2.8440				&	2.8502		&           &	0.0532	& 0.0207	&	0.0182	& 0.0205	&  0.0493	&			&			&0.0020	&0.0018	&	0.0032	\\
UDS	&	34.0632333		&	2	&					&					&			&						&				&			&			&			&	20.9925	&	20.2328	&	19.7778	&			&			&	18.5751	&		18.0561	&		17.7469		\\	
	&	-5.4123534		&       &					&					&			&						&				&			&			&			&	0.0323	&	0.0234	&	0.0550	&			&			&0.0022	&0.0022	&0.0034	\\
UDS	&	34.0725841		&	1	&	34.072628552341	&	-5.367278457759	&	0.8299	&	-0.2061				&		-4.3827	& 19.6567	&	19.9357	&	19.6258	&	19.5138	&	19.4769	&	19.4584	&			&			&	18.6107	&		18.3250	&	18.2648		\\	
	&	-5.3672215		&       &	0.62595			&	0.50292			&	0.8537	&	1.0136				&	0.8118		&	        &	0.0124	& 0.0226	&	0.0115	&	0.0154	& 0.0358	&			&			&0.0017	&0.0021	&	0.0027	\\
UDS	&	34.2131217		&	1	&	34.213113369956	&	-4.741901511918	&	0.2797	&		-0.3908			&		-3.9373	&	19.0469	&	19.3203	&	19.0049	&	18.8999	&	18.9062	&	18.9296	&			&			&	18.0473	&		17.7772	&		17.6997		\\	
	&	-4.7418757		&       &	0.37411	  		&	0.26570			& 0.4289	&	0.5696				&	0.4495		&			&	0.0056	&	0.0104	&	0.0060	& 0.0124	& 0.0333	&			&			&0.0015	&0.0018	&0.0026	\\
UDS	&	34.2836251		&	1	&					&					&			&						&				&			&			&	21.3821	&	20.3480	&	19.9655	&	19.5100	&			&			&	18.4806	&		17.9502	&	17.6859		\\	
	&	-5.5122054		&       &					&					&			&						&				&			&			& 0.0442	&	0.0143	& 0.0380	& 0.1018	&			&			&0.0021	&0.0019	&	0.0034	\\
UDS	&	34.3328267		&	1	&					&					&			&						&				&			&			&	21.9449	&	20.7901	&	20.1041	&	19.6980	&			&			&	18.5164	&		17.9305	&		17.6470		\\	
	&	-5.2005163		&       &					&					&			&						&				&			&			&	0.1826	& 0.0256	& 0.0299	& 0.0670	&			&			&0.0024	&0.0021	&0.0036	\\
UDS	&	34.3905494		&	1	&					&					&			&						&				&			&			& 21.5139	&	20.3717	&	19.8806	&	19.4405	&			&			&	18.3636	&		17.8970	&		17.6138		\\
	&	-5.1081487		&       &					&					&			&						&				&			&			& 0.0772	&	0.0156	&	0.0365	& 0.0262	&			&			&0.0022	&0.0023	&	0.0033	\\
UDS	&	34.4226317		&	2	&	34.422696476887	& -5.319394990941	&			&						&				&	20.9700	&	21.9304	&	21.1561	&	20.1452	&	19.6900	& 19.4092	&			&			&	18.3216	&		17.8349	&		17.5908		\\	
	&	-5.3193711		&       &	2.88829			&	7.68729			&			&						&				&			&	 0.1454	&	0.0364	& 0.0123	&	0.0222	&	0.0706	&			&			&0.0022	&0.0019	&0.0033	\\
UDS	&	34.4476113		&	1	&	34.447624608681	&	-5.115979177256	&	-0.4065	&		-0.5852			&	-1.9265		&	19.8064	&	20.2081	&	19.7500	&	19.5395	&	19.4742	&	19.3670	&			&			&	18.4760	&		18.0645	&		17.9778		\\	
	&	-5.1159795		&       &  0.65716			&	0.53976			& 1.0029	&	0.8452				&	0.8947		&			&	0.0176	& 0.0209	& 0.0161	&	0.0194	& 0.0826	&			&			&0.0021	&0.0024	&0.0032	\\
UDS	&	34.4614311		&	1	&					&					&			&						&				&			&			& 21.7428	&	20.6132	&	20.0122	&	19.5653	&			&			&	18.4719	&		17.9032	&	17.6350			\\
	&	-5.1427924		&       &					&					&			&						&				&			&			& 0.0493	& 0.0315	& 0.0264	& 0.0470	&			&			&	0.0023	&0.0022	&	0.0035	\\
UDS	&	34.5115289		&	2	&					&					&			&						&				&			&			& 21.8823	&	20.7066	&	20.0423	&	19.6703	&			&			&	18.4488	&		18.0040	&		17.6905			\\
	&	-5.2450022		&       &					&					&			&						&				&			& 			& 0.1836	& 0.0242	& 0.0221	& 0.0723	&			&			&	0.0023	&0.0022	&0.0033	\\
UDS	&	34.5259644		&	1,2	&					&					&			&						&				&			&	21.8031	&	21.7587	&	20.5598	&	19.8926	&	19.4451	&			&			&	18.3814	&		17.9104	&		17.6043		\\	
	&	-5.2628373		&       &					&					&			&						&				&			& 0.1946	& 0.0513	& 0.0334	& 0.0416	&	0.0253	&			&			&0.0022	&0.0021	&0.0033	\\	
UDS	&	34.5434314		&	1	&	34.543463866893	& -5.171139515300	&	0.6485	&		-0.1422			&	-2.7368		&	19.5637	&	20.1365	&	19.4735	&	19.2855	&	19.1939	&	19.1399	&			&			&	18.1998	&		17.8036	&17.7009		\\	
	&	-5.1711112		&       &	0.50344			&	0.35905			& 0.5815	& 0.7438				&	0.6398		&  			& 0.0043	& 0.0370	& 0.0119	& 0.0140	& 0.0194	&			&			&0.0016	&	0.0017	&	0.0028	\\
UDS	&	34.5471946		&	1	&	34.547230921295	&	-5.176806400064	&	0.3174	&		6.3978			&	-9.4345		&	19.6890	&	20.1476	&	19.6217	&	19.3895	&	19.2849	&	19.1582	&			&			&	18.2589	&		17.8400	&		17.7102			\\
	&	-5.1767668		&       &	0.55339			& 0.39306			& 0.6321	& 0.8295				&	0.7060		&			& 0.0146	& 0.0144	& 0.0098	& 0.0106	& 0.0339	&			&			&0.0018	&0.0017	&0.0029	\\
UDS	&	34.5487686		&	1	&					&					&			&						&				&			&			& 21.3331	&	20.3608	& 19.9194	&	19.4469	&			&			&	18.4316	&		17.9402	&		17.7101		\\	
	&	-5.3929635		&       &					&					&			&						&				&			&		&	0.0609		& 0.0229	&	0.0202	& 0.0669	&			&			&0.0021	&0.0019	&0.0032	\\
UDS	&	34.5590572		&	1	&	34.559105809238	& -4.927423298600	&	-0.1536	&		3.8946			&		-3.5982	&	19.2729	&	19.5643	&	19.2749	&	19.1408	&	19.1434	&	19.1397	&			&			&	18.2892	&		18.0363	&	17.9828		\\	
	&	-4.9273779		&       &  0.60308			&	0.45118			& 0.7568	& 0.9018				&	0.7822		&			& 0.0208	& 0.0020	&	0.0108	&	0.0169	&	0.0479	&			&			&0.0014	&	0.0020	&0.0027	\\
UDS	&	34.6711628		&	1	&					&					&			&						&				&			&			& 22.0633	&	20.8571	&	20.2795	&	19.8136	&			&			&	18.7770	&		18.3086	&		18.0321	\\	
	&	-5.1675618		&       &					&					&			&						&				&			&			& 0.1102	& 0.0530	& 0.0278	& 0.0633	&			&			&0.0023	&0.0023	&0.0035	\\	
UDS	&	34.6858172		&	1	&	34.685894904232	& -4.703061471509	&	0.3458	&		11.8361			&		-6.0594	&	19.4035	&	19.7527	&	19.3522	&	19.1732	&	19.1381	&	19.0562	&			&			&	18.1939	&	17.8580	&		17.7599		\\	
	&	-4.7030450		&       &	0.63685			&		0.48470		& 0.9393	& 0.8860				&	0.7466		&			&	0.0115	& 0.0128	& 0.0096	& 0.0110	& 0.0295	&			&			&0.0016	&	0.0017	&0.0027	\\
UDS	&	34.7683910		&	2	&					&					&			&						&				&			&			& 21.8062	&	20.8612	&	20.2771	&	19.8818	&			&			&	18.7346	&		18.2226	&		17.9366	\\	
	&	-5.0398575		&       &					&					&			&						&				&			& 			& 0.0160	& 0.0548	& 0.0322	& 0.0947	&			&			&0.0022	&0.0021	&0.0034	\\	
UDS	&	34.7885620		&	1,2	&					&					&			&						&				&			&	22.1346	&	21.3754	&	20.3912	&	19.8497	&	19.4869	&			&			&	18.4368	&		17.9835	&		17.7201			\\
	&	-4.8819019		&       &					&					&			&						&				&			& 0.1739	&	0.0588	& 0.0258	& 0.0361	& 0.0416	&			&			&0.0020	&0.0018	&0.0033	\\
UDS	&	34.8300954		&	1	&					&					&			&						&				&			&	21.9653	&	21.7962	&	20.5947	&	20.0403	&	19.6512	&			&			&	18.5714	&		18.1156	&		17.8530			\\
	&	-5.0119407		&       &					&					&			&						&				&			& 0.1578	& 0.0556	& 0.0144	&	0.0215	& 0.0364	&			&			&0.0021	&0.0020	&0.0032	\\
UDS	&	34.8397670		&	2	&					&					&			&						&				&			&	22.6037	&	21.9676	&	20.7494	&	20.0257	&	19.6058	&			&			&	18.3350	&	17.8180	&		17.5136		\\	
	&	-4.7351863		&       &					&					&			&						&				&			&	0.2478	& 0.0095	& 0.0279	& 0.0278	& 0.0594	&			&			&0.0022	&	0.0018	&0.0034	\\
UDS	&	34.8474938		&	1	&	34.847507314304	&	-5.335409824465	&			&						&				&	20.6009	&	21.8297	&	20.5566	&	19.9651	&	19.6698	&	19.3245	&			&			&	18.3256	&		17.7770	&		17.5709		\\	
	&	-5.3353341		&       &	2.19216			&	2.17627			&			&						&				&			& 0.0467	& 0.0499	& 0.0117	& 0.0189	& 0.0746	&			&			&0.0021	&0.0019	&0.0033	\\
UDS	&	34.8707629		&	1	&					&					&			&						&				&			&			&			&	21.2460	&	20.3159	&	19.6511	&			&			&	18.4190	&		17.9452	&		17.5810	\\	
	&	-5.1977157		&       &					&					&			&						&				&			&			&			& 0.0373	& 0.0206	& 0.0313	&			&			&0.0022	&0.0020	&0.0035	\\
UDS	&	34.8890054		&	2	&	34.889028664017	&	-4.822057442111	&			&						&				&	21.0061	&	22.1342	&	21.2954	&	20.2237	&	19.7842	&	19.5169	&			&			&	18.3800	&	17.8426	&		17.6186		\\
	&	-4.8219876		&       &			3.44884	&	10.04841		&			&						&				&			& 0.1309	& 0.0398	& 0.0166	& 0.0195	& 0.0354	&			&			&	0.0023	&0.0022	&0.0035	\\
VIDEO	&	35.0364084	&	2	&					&					&			&						&				&			&			& 			& 21.0837	&	20.2244	&	19.7591	&			&	18.9637	&		18.3366	&		17.8180	&	17.5239	\\
	&	-4.9878868		&       &					&					&			&						&				&			&			& 			& 0.0253	& 0.0486	&	0.0730	&			&0.0022	&0.0017	&	0.0028	&	0.0039	\\
VIDEO	&	35.0634277	&	1	&					&					&		    &						&				&			&			&			&	20.8514	&	20.2694	&	19.7768	&			&	19.3151	&		18.7851	&		18.1603	&	17.9089		\\
	&	-4.8899093		&       &					&					&	  	    &						&				&			&			& 			& 0.0269	& 0.0319	&	0.0136	&			&0.0023	&0.0025	&	0.0032	&0.0040	\\
VIDEO	&	35.1404271	&	1,2	&					&					&			&						&				&			&	21.9319	&	21.0028	&	20.1999	&	19.8095	&	19.3154	&			&	18.8489	&		18.3912	&		17.9152	&		17.7179	\\
	&	-4.5489657		&       &					&					&			&						&				&			& 0.0104	&	0.0309	&0.0243		& 0.0215	& 0.0698	&			&0.0028	&0.0024	&0.0025	&	0.0050	\\
VIDEO	&	35.1942372	&	1	&	35.194280993865	&	-4.866190049713	&			&	15.1823				& -5.4891		& 20.0758	&	20.6159	&	19.9907	&	19.7274	&	19.6449	&	19.4795	&			&	18.9034	&		18.5732	&		18.1256	&		18.0160		\\
	&	-4.8661564		&       &	0.68076			&	0.48087			&			&	1.0299				& 0.8569		&			& 0.0223	& 0.0123	& 0.0141	& 0.0150	& 0.0444	&			&0.0015	&0.0030	&0.0035	&0.0045	\\
VIDEO	&	35.2729996	&	1	&					&					&			&		 				&				&			&	22.3620	&	21.9127	&	20.8624	&	20.4448	&	20.0550	&			&	19.4404	&		18.9522	&		18.5055	&		18.2381	\\	
	&	-5.0046632		&       &					&					&			&		 				&				&			& 0.2410	& 0.1926	& 0.0450	& 0.0543	& 0.0634	&			&	0.0021	&0.0030	&0.0054	&0.0047	\\
VIDEO	&	35.3783901	&	1	&					&					&			&						&				&			&			&	21.7415	&	20.8513	&	20.2565	&	19.7882	&			&	19.2379	&		18.7073	&	18.2273	&		17.9349		\\
	&	-5.2565457		&       &					&					&			&						&				&			&			&	 0.0291	&	0.0190	& 0.0312	& 0.1079	&			&	0.0022	&0.0026	&	0.0051	&0.0045	\\
VIDEO	&	35.3829208	&	1	&					&					&			&						&				&		    &	21.8337	&	20.9823	&	20.4489	&	20.1954	&	19.8148	&			&	19.3882	&		18.9576	&	18.3868	&		18.1676	\\
	&	-4.9597282		&       &					&					&			&					    &			 	&		    & 0.0144	& 0.0513	& 0.0243	& 0.0340	& 0.1005	&			&0.0023	&0.0030	&	0.0048	&	0.0056	\\
VIDEO	&	35.3846956	&	1	&					&					&			&						&				&			&			&	21.7485	&	20.7684	&	20.2319	&	19.8866	&			&	19.2649	&		18.7794	&	18.3458	&		18.0929		\\
	&	-4.4494825		&       &					&					&			&						&				&			&			&	0.1597	&	0.0302	& 0.0201	&	0.0325	&			&0.0019	&0.0022	&	0.0044	&0.0024	\\
VIDEO	&	35.4517777	&	1	&	35.451819126350	&	-4.484091070150	&			&		1.9364			&	-9.6414		&	20.6038	&	21.3819	&	20.5229	&	20.1307	&	19.8683	&	19.7105	&			&	19.1223	&		18.7165	&	18.1694	&		18.0295	\\
	&	-4.4840691		&       &	0.94865			&	1.11158			&			& 	2.0494				& 2.9685		&			& 0.0351	& 0.0226	& 0.0149	& 0.0245	& 0.0322	&			&0.0025	&	0.0027	&0.0029	&	0.0044	\\
VIDEO	&	35.5506321	&	1	&					&					&			&						&				&			&			&	21.7361	&	20.9951	&	20.5266	&	20.0104	&			&	19.4807	&		18.9698	&		18.4913	&		18.2201	\\
	&	-4.6345957		&       &					&					&			&						&				&			&			& 0.1500	& 0.0219	& 0.0151	& 0.0008	&			&0.0022	&0.0024	& 0.0045&	0.0047	\\
VIDEO	&	35.5527346	&	1	&					&					&			&						&				&			&			&			& 20.9396 	&	20.2919	&	19.8182	&			&	19.2588	&		18.7094	&		18.2230	&		17.9311	\\	
	&	-4.9423113		&       &					&					&			&						&				&			&			&			& 0.0302  	& 0.0286	& 0.0443	&			&0.0031	&0.0024	&0.0049	&0.0053	\\
VIDEO	&	35.6795871	&	1,2 &					&					&			&						&				&			&			& 21.4877	&	20.8680	&	20.5088	&	20.1017	&			&	19.6523	&		19.2109	&		18.7295	&		18.4770		\\
	&	-5.2156567		&       &					&					&			&						&				&			&			&	0.0726	&	0.0352	& 0.0326	& 0.0976	&			&0.0028	&0.0038	&0.0039	&0.0047	\\
VIDEO	&	35.7087718	&	1	&	35.708793461198	&	-4.837947978546	&	-1.8486	&		7.7207			&	-3.5925		&	19.6779	&	20.0911	&	19.6083	&	19.3864	&	19.2842	&	19.2764	&			&   18.6218	&		18.3506	&		17.9893	&		17.9061		\\
	&	-4.8379052		&       &	0.67135			&	0.49583			& 0.6994	&	1.3335				&	1.0385		&			& 0.0190	&	0.0145	& 0.0158	& 0.0125	& 0.0424	&			&0.0016	&0.0026	&0.0053	&0.0055	\\
VIDEO	&	35.7342835	&	2	&					&					&			&						&				&			&	21.4823	&	21.5245	&	20.2976	&	19.6781	&	19.5094	&			&	18.7360	&		18.2636	&		17.7840	&		17.5673	\\	
	&	-5.3693787		&       &					&					&			&						&				&			& 0.1503	& 0.0620	& 0.0183	& 0.0622	& 0.0402	&			&0.0013	&0.0021	&0.0026	&0.0038	\\

		\hline
	\end{tabular}
	
	Flag as possible galaxy: (1) $HKyJ$ and/or $HKYJ$ colour (2) one or more Pan-STARRS filters have Kron magnitudes brighter than the PSF magnitude by 0.05 mag. \\
\end{tiny}
\end{table*}
\end{landscape}
\restoregeometry

\newgeometry{width=20cm,height=13cm}
\begin{landscape}
\begin{table*}
\begin{tiny}
	\contcaption{Data for Sources Identified as Possible Galaxies}
			\setlength\tabcolsep{1pt}
	\begin{tabular}{rrcrrrrrrrrrrrrrrrr}
		\hline
 \multicolumn{2}{c}{IR Survey} & Flag &    \multicolumn{6}{c}{Gaia} &    \multicolumn{5}{c}{Pan-STARRS meanPSF AB mags} & 
   \multicolumn{5}{c}{VISTA Vega mags}\\
   Name & RA$^{\circ}$     & 		&	RA$^{\circ}$ 	&	Decl.$^{\circ}$	& Parallax  & $\mu$ RA		      &	$\mu$ Decl.	 	& G			&	$g$ &	$r$ 	&	$i$	& $z$ 	& $y$ 	& $Z$		& $Y$	&	$J$	&	$H$	& $Ks$			\\
     & Decl.$^{\circ}$	&		&	$\pm$ mas		&	$\pm$ mas   	& $\pm$  mas & $\pm$ mas yr$^{-1}$ &	$\pm$ mas yr$^{-1}$ 	& 		& $\pm$ 	& $\pm$ 	& $\pm$ 	& $\pm$  	& $\pm$ 	&  $\pm$	&	 $\pm$		&	 $\pm$	&	 $\pm$		&	 $\pm$			\\
\hline
VIDEO	&	35.7408174	&	1	&					&					&			&						&				&			&			&	21.5238	&	20.6134	&	20.1719	&	19.8212	&			&	19.2800	&		18.8185	&		18.3103	&		18.0684	\\	
	&	-5.2873244		&       &					&					&			&						&				&			&			&	 0.0496	& 0.0220	& 0.0486	& 0.0911	&			&0.0020	&0.0034	&0.0036	&	0.0044	\\
VIDEO	&	35.7414119	&	1	&					&					&			&						&				&			&			&	21.6836	&	20.7530	&	20.1849	&	19.7888	&			&	19.2312	&		18.7566	&	18.2657	&	18.0543		\\
	&	-4.8313848		&       &					&					&			&						&				&			&			&	0.0472	&	0.0258	&	0.0118	& 0.0540	&			&0.0028	&	0.0027	&	0.0054	&0.0052	\\
VIDEO	&	35.7655027	&	2	&					&					&			&						&				&		    &	21.2297	&	21.2518	&	20.2632	&	19.7629	&	19.3965	&			&	18.7797	&		18.3147	&	17.8308	&		17.6399	\\
	&	-5.0096975		&       &					&					&			&						&				&		    &	0.0982	&	0.0172	& 0.0159	&	0.0173	& 0.0239	&			&0.0023	&	0.0027	&0.0022	&	0.0044	\\
VIDEO	&	35.8265344	&	1	&					&					&			&						&				&			&	22.2128	&	21.0445	&	20.4301	&	20.0052	&	19.7425	&			&	19.2129	&		18.7693	&		18.2653	&		18.0430	\\
	&	-4.9284675		&       &					&					&			&						&				&		    &	0.0194	& 0.0200	& 0.0164	& 0.0264	&	0.0822	&			&0.0024	&0.0024	&0.0046	&	0.0052	\\
VIDEO	&	35.9375707	&	1	&					&					&			&						&				&			&			&			& 21.1609	&	20.5428	&	19.9431	&			&	19.4630	&	18.9046	&		18.4344	&	18.1389	\\
	&	-4.4698025		&       &					&					&			&						&				&			&			&			&	0.0336	&   0.0366	&  	0.1410	&			&0.0025	&	0.0032	&	0.0047	&	0.0055	\\
VIDEO	&	35.9460049	&	1	&					&					&			&						&				&		 	&			& 21.6890	&	20.8613	&	20.2493	&	19.5264	&			&	19.2133	&		18.6971	&		18.2172	&		17.9746	\\
	&	-4.5784901		&       &					&					&	 		&						&				&		 	&			& 0.1515	&	0.0313	& 0.0432	&	0.0814	&			& 0.0024	&0.0026	&0.0041	&	0.0054	\\
VIDEO	&	36.0710161	&	1	&	36.071091123688	& -4.413983619399	&			&	9.7112		  		& 5.4933		&	20.5758	&	21.6183	&	20.5739	&	19.9287	&	19.5959	&	19.3285	&	19.0171	&	18.7109	&	18.2809	&	17.7485	&	17.5633	\\
	&	-4.4139791		&       &	1.28068			&	0.78531			&			&	 2.0081				& 1.8851		&			& 0.0369	& 0.0174	& 0.0093	&	0.0242	&	0.0322	&	0.0017	&	0.0015	& 0.0031	&	0.0024	&	0.0042	\\
VIDEO	&	36.3757309	&	1	&	36.375784685588	&	-4.090708837809	&	0.7755	&		2.5459			&	-12.1766	&	19.6292	&	20.0069	& 19.5620	&	19.3690	&	19.3078	& 19.2324	&	18.7250	&	18.5751	&	18.2832	&	17.9134	&17.8111		\\
	&	-4.0906468		&       &	0.55848			&	0.44575			& 0.6330	& 1.0837				& 0.7780		&			&	0.0108	& 0.0097	&	0.0117	&	0.0124	&	0.0491	&0.0018	&	0.0012	&	0.0021	&	0.0028	&	0.0033	\\
VIDEO	&	36.6440048	&	1,2	&	36.644038728037	&	-4.812678556252	&			&						&				&	20.7420	&	21.7597	&	20.8992	&	20.1040	&	19.6922	&	19.4314	&	19.1277	&		18.8228	&		18.3785	&		17.8625	&	17.6593		\\
	&	-4.8126421		&       &	1.46427			&	2.36397			&			&						&				&		 	& 0.0488	& 0.0123	&	0.0094  & 0.0117	& 0.0428	&0.0020	&	0.0013	&		0.0019	&		0.0033	&		0.0035	\\
VIDEO	&	36.6880308	&	1	&					&					&			&						&				&			&			&	21.8353	&	21.0651	&	20.4204	&	19.8794	&	19.9126	&		19.4749	&		18.9476	&	18.4462	&	18.1874		\\
	&	-4.8435704		&       &					&					&			&						&				&			&			&	0.2784	&	0.0366	& 0.0608	& 0.1596	&0.0031	&		0.0031	&		0.0022	&		0.0039	&	0.0041	\\
VIDEO	&	36.7097084	&	1	&	36.709737717167	&	-4.524403538704	&	-1.0689	&		1.9018			&		-6.7427	&	20.1873	&	20.5807	& 20.0892	&	19.8732	&	19.7632	&	19.6094	&	19.2129	&		19.0658	&		18.7609	&		18.3770		&	18.2744		\\
	&	-4.5243931		&       &	0.67107			&	0.72915			& 0.7793	& 1.5715				& 2.0535		&			&	0.0185	&	0.0116	& 0.0072	& 0.0237	& 0.0367	&0.0016	&		0.0019	&		0.0026	&		0.0043	&		0.0049	\\
VIDEO	&	52.0097291	&	1	&					&					&			&						&				&			&			&			&	20.6892	&	20.0223	&	19.5554	&	19.3653	&		18.9275 &	18.4124	&		18.0183	&		17.7420\\
	&	-27.6189720		&       &					&					&			&						&				&			&			&			& 0.0364	& 0.0276	& 0.0276	&0.0035	&		0.0024	&	0.0028	&			0.0042	&		0.0049	\\
VIDEO	&	52.4462564	&	1	&					&					&			&						&				&			&			&			&	20.6798	&	19.9899	&	19.4606	&	19.3498	&		18.8782	&		18.3040	&		17.8475	&		17.5704		\\
	&	-27.8237538		&       &					&					&			&						&				&			&			&			& 0.0262	& 0.0360	&	0.0296	&0.0034	&		0.0030	&		0.0019	&		0.0033	&		0.0049	\\
VIDEO	&	52.7557893	&	1	&	52.755781300291	& -27.448804052908	&	-0.0858	&		0.9688			&	-6.4672		&	20.1116	&	20.8253	&	20.0564	&	19.6631	&	19.5744	&	19.3967	&	18.9411	&		18.7801	&	18.4311	&		17.9267	&		17.8077	\\
	&	-27.4488061		&       &	0.27729			&	0.43118			& 0.5462	&	0.6486				&	0.8882		&			& 0.0423	& 0.0191	& 0.0086	&	0.0158	& 0.0304	&0.0040	&		0.0024	&	0.0033	&		0.0032	&		0.0056	\\
VIDEO	&	52.9274555	&	2	&					&					&			&						&				&			&			& 21.6304	&	20.7160	&	20.3278	&	19.8924	&	19.6453	&		19.2771	&		18.8137		&	18.3209		&	18.0889		\\
	&	-27.4700603		&       &					&					&			&						&				&			&			&	0.0417	& 0.0457	& 0.0611	& 0.1073	&0.0030	&		0.0049	&	0.0041	&		0.0038		&	0.004\\
VIDEO	&	53.2112772	&	1	&	53.211345751370	& -27.207439654263	&	0.7372	&		18.1557			&		-9.0152	&	20.8317	&			& 20.9198	& 20.1251	&	19.7043	& 19.5700	&	19.0776	&			18.8092	&	18.3537	&		17.9459		&	17.6940		\\
	&	-27.2074247		&       &	0.81033			&		1.24769		& 1.6953	& 1.7435				&	2.4402		&           &           & 	0.0388	&	0.0212	&	0.0165	&	0.0340	&0.0022	&		0.0030	&		0.0025	&		0.0048		&	0.0046	\\
VIDEO	&	53.3409644	&	1,2	&	53.340983559831	&-27.845025165778	&	-0.7650	&		12.5581			&	-17.2877	&	20.8951	&	21.8998	&	20.8341	&	20.2670	&	19.9444	&	19.6204	&	19.3216	&		19.0852	&		18.6317		&	18.1686	&		17.9531		\\
	&	-27.8449838		&       &	0.51344			&	0.97876			& 1.1896	& 1.3332				& 2.4270		&			&	0.1610	& 0.0330	& 0.0475	& 0.0201	&	0.0523	&0.0042	&		0.0043	&		0.0018		&	0.0047		&	0.0040	\\
CAL	&	87.6810876		&	1	&	87.681107197032	&	16.372256410953	&	0.5648	&	0.7724				&		-2.9455	&	18.3802	&	19.0977	&	18.3361	&17.9998	&	17.8359	&	17.6862	&	17.2938	&		17.1385		&	16.6670		&	16.2310	&		16.0786		\\
	&	16.3722372		&       &	0.22657			&	0.19663			& 0.2468	& 0.4921				&	0.4134		&			& 0.0037	&	0.0058	&	0.0031	& 0.0078	& 0.0102	&0.0078	&		0.0038	&		0.0037		&	0.0043	&		0.0053	\\
CAL	&	87.7003018		&	1	&	87.700299075751	& 15.824548535954	&	0.2557	&		2.3478			&	-1.9968		&	17.9648	&	18.5295	&	17.9302	&	17.6653	&	17.5282	&	17.4332	&	17.0166	&		16.8713	&		16.4748	&		16.1133	&		15.9915		\\
	&	15.8245989		&       &	0.14842			&	0.13937			&	0.1558	& 0.3193				&	0.2695		&			& 0.0045	&	0.0029	& 0.0027	& 0.0066	& 0.0091	&0.0070	&		0.0034		&	0.0034	&		0.0031	&	0.0043	\\
UVISTA	&	149.3964243	&	1	&	149.396426037047	&	1.627980712026	&		&		&		&	18.0515	&	18.3400	&	18.0239	&	17.9150	&	17.8861	&	17.8455	&		&	17.2622	&	17.0280	&	16.7440	&	16.6915	\\
	&	1.6279765	&		&	0.57794	&	0.98644	&		&		&		&		&	0.0118	&	0.0064	&	0.0055	&	0.0105	&	0.0158	&		&	0.0018	&	0.0016	&	0.0024	&	0.0041	\\
UVISTA	&	149.7445300	&	1	&		&		&		&		&		&		&		&	22.7023	&	21.1733	&	20.5041	&	19.9158	&		&	19.3617	&	18.8135	&	18.2684	&	18.0102	\\
	&	2.7571103	&		&		&		&		&		&		&		&		&	0.0249	&	0.0378	&	0.0375	&	0.0424	&		&	0.0012	&	0.0015	&	0.0018	&	0.0025	\\
UVISTA	&	149.7772837	&	1	&		&		&		&		&		&		&		&	22.3306	&	20.9640	&	20.4401	&	19.9358	&		&	19.3635	&	18.8459	&	18.3268	&	18.0769	\\
	&	2.7455382	&		&		&		&		&		&		&		&		&	0.0212	&	0.0270	&	0.0452	&	0.0252	&		&	0.0012	&	0.0008	&	0.0017	&	0.0028	\\
UVISTA	&	149.7833198	&	1,2	&		&		&		&		&		&		&		&	22.6901	&	21.1603	&	20.3877	&	19.8489	&		&	19.2958	&	18.7427	&	18.2693	&	17.9968	\\
	&	2.7839207	&		&		&		&		&		&		&		&		&	0.1463	&	0.0437	&	0.0358	&	0.0869	&		&	0.0018	&	0.0014	&	0.0020	&	0.0023	\\
UVISTA	&	149.7947682	&	2	&	149.794759440482	&	2.765319258775	&		&		&		&	20.7674	&	21.7339	&	20.6624	&	20.1608	&	19.8799	&	19.7512	&		&	19.0779	&	18.6551	&	18.1258	&	17.9319	\\
	&	2.7653208	&		&	3.37503	&	5.25135	&		&		&		&		&	0.0749	&	0.0210	&	0.0212	&	0.0273	&	0.0567	&		&	0.0012	&	0.0013	&	0.0018	&	0.0027	\\
UVISTA	&	149.8276357	&	2	&		&		&		&		&		&		&		&	21.4163	&	20.0852	&	19.3989	&	19.1021	&		&	18.3504	&	17.8068	&	17.3179	&	17.0589	\\
	&	2.7542743	&		&		&		&		&		&		&		&		&	0.1219	&	0.0230	&	0.0182	&	0.0311	&		&	0.0009	&	0.0007	&	0.0012	&	0.0014	\\
CAL	&	312.6852216		&	1	& 312.685178912608	&	6.185061145961	&	-0.5758	&	-10.5551			&		-9.0282	&	18.1012	&	18.5962	&	18.0192	&	17.7478	&	17.6417	&	17.5611	&	17.1128	&	16.9912	&		16.6060	&	16.1981	&		16.0897		\\
	&	6.1850892		&       &	0.17048			&	0.12374			&	0.2360	& 0.3446				& 0.2020		&			&	0.0043	& 0.0043	& 0.0047	& 0.0144	& 0.0052	&	0.0069	&	0.0037	&		0.0041		&	0.0035	&	0.0049	\\
DXS	&	333.0678519		&	1	& 312.685178912608	&	6.185061145961	&	-0.5758	&		-10.5551		&	-9.0282		&	18.1012	&	18.7337	&	18.3275	&	18.1735	&	18.1308	&	18.0981	&			&		&	17.2188	&		16.9352		&	16.8494	\\
	&	0.2463832		&       &	0.17048			&	0.12374			&	0.2360	&	0.3446				& 0.2020		&			& 0.0094	& 0.0072	& 0.0032	& 0.0050	& 0.0163	&			&		&0.0035		&	0.0031		&	0.0050	\\
DXS	&	333.1814666		&	1	& 333.071813320305	&	1.188906051043	&	1.2559	&	-0.023				&		-4.0006	&	20.2384	&	19.2513	&	18.8487	&	18.6976	&	18.6483	&	18.5967	&			&		&	17.7038	&		17.3603		&	17.2893		\\
	&	0.5377071		&       &	1.68843			&	1.04016			& 1.9038	& 1.6055				& 1.9193		&			&	0.0059	& 0.0068	& 0.0041	& 0.0108	& 0.0176	&			&		&0.0048	&		0.0059		&	0.0057	\\
DXS	&	333.1895155		&	1	& 333.181502744943	&	0.537661770120	&	0.0649	&	3.5358				&		-4.0912	&	18.8806	&	19.5321	& 	19.0986	&	18.9363	&	18.8497	&	18.8137	&			&		&	17.9090	&		17.5613	&		17.4914		\\
	&	0.7402520		&       &	0.39660			&	0.45081			&	0.4259	& 0.6231	& 0.6999	&							& 0.0093	&	0.0038	& 0.0066	& 0.0198	& 0.0471	&			&		&0.0035		&	0.0059		&	0.0059	\\
DXS	&	333.2850037		&	1	& 333.283923388425	&	-0.410674164956	&			&						&				&	20.7836	&	22.1518	&	21.4454	&	20.0104	&	19.3589	&	18.8466	&			&		&	17.7170	&		17.1759		&	16.9041	\\
	&	-0.9465384		&       &	1.74142			&	1.61698			&			&						&				&			& 0.1829	& 0.0420	& 0.0108	& 0.0166	&	0.0378	&			&		&0.0036	&		0.0053	&	0.0055	\\
DXS	&	333.4320262		&	2	&					&					&			&						&				&			&			& 21.7068	&	20.6360	&	19.8744	&	19.3903	&			&		&	18.0916		&	17.5841		&	17.2524	\\
	&	0.7382165		&       &					&					&			&						&				&			&			& 0.0222	& 0.0136	& 0.0269	&	0.0019	&			&		&0.0051	&	0.0052		&	0.0059	\\
DXS	&	333.6445754		&	1	& 333.644571087727	&	0.534256767502	&	0.2749	&		-3.1602			&		-3.8142	&	18.4809	&	18.7319	&	18.4562	&	18.3525	&	18.3092	&	18.2465	&			&		&	17.4536	&		17.1791		&	17.1329		\\
	&	0.5342898		&       &	0.25131	 		&	0.26301			& 0.2872	&	0.4120				& 0.4355		&			& 0.0114	& 0.0071	& 0.0029	& 0.0143	& 0.0251	&			&		&0.0028		&	0.0038		&	0.0054	\\
DXS	&	334.8159877		&	1	& 334.815996502933	&	1.097322542853	&	0.0244	&	1.5929				&	-10.8697	&	18.1228	&	18.4275	&	18.1034	&	17.9505	&	17.8859	&	17.8456	&			&		&	16.9836	&		16.6704	&	16.5741		\\
	&	1.0973756		&       &	0.20974			&	0.22680			&	0.2655	& 0.5427				& 0.4169		&			& 0.0037	& 0.0061	& 0.0040	&	0.0050	& 0.0162	&			&		&0.0020	&		0.0029	&		0.0047	\\
DXS	&	334.9984584		&	1,2	& 334.998484432193	&	0.731938922959	&			&						&				&	21.0597	&	21.9666	&	21.3736	&	20.1511	&	19.4318	&	18.9772	&			&		&	17.8653	&		17.3730		&	17.0625		\\
	&	0.7319926		&       &	9.42378			&1	1.34213			&			&						&				&			&	0.1602	& 0.0499	&	0.0107	& 0.0106	& 0.0313	&			&		&	0.0032	&		0.0049	&		0.0055	\\
DXS	&	335.0317979		&	2	& 335.031828446221	&	-0.109437792442	&	-3.5425	&		2.4789			&	7.4609		&	20.5124	&	22.1153	&	21.1476	&	19.7651	&	19.0880	&	18.7638	&			&		&	17.4985	&		17.0236	&		16.7454		\\
	&	-0.1094196		&       &	2.07486			&	2.32689			& 2.5997	&	3.6136				& 3.9026		&			& 0.0245	& 0.0381	& 0.0109	&	0.0128	& 0.0480	&			&		&0.0032	&		0.0038		&	0.0058	\\
DXS	&	335.1704830		&	1	& 335.170502659949	&	0.510427567859	&	0.5604	&		0.1249			&		-8.3115	&	18.3110	&	18.7558	&	18.2612	&	18.0592	&	17.9900	&	17.9099	&			&		&	16.9948		&	16.6466	&		16.5527		\\
	&	0.5104702		&       &	0.29755			&	0.28425			&0.3486		& 0.3769				& 0.4869		&			& 0.0089	& 0.0090	& 0.0032	& 0.0066	& 0.0137	&			&		&0.0027	&		0.0028	&	0.0054	\\
DXS	&	335.3820794		&	2	& 335.382075808196	&	1.162262771954	&			&						&				&	20.2360	&	21.5394	&	20.2713	&	19.5524	&	19.1903	&	18.9633	&			&		&	17.7985	&		17.2037	&	16.9959		\\
	&	1.1622935		&       &	2.11766			&	2.10523			&			&						&				&			& 0.0340	& 0.0228	& 0.0081	& 0.0130	& 0.0554	&			&		&0.0040	&		0.0038		&	0.0046	\\
DXS	&	335.6191436		&	1	& 335.619172934901	&	1.161866766328	&	0.1049	&		9.2640			&	-10.4955	&	18.6493	&	19.1149	& 18.6094	&	18.3917	&	18.2874	&	18.2377	&			&		&	17.2880	&		16.9008	&		16.8051	\\
	&	1.1619136		&       &	0.32319			&	0.41983			& 0.3448	& 0.5976				& 0.6150		&			& 0.0163	& 0.0081	& 0.0038	& 0.0119	&	0.0160	&			&		&0.0029	&		0.0034	&0.0047	\\
		\hline
	\end{tabular}
	Flag as possible galaxy: (1) $HKyJ$ and/or $HKYJ$ colour (2) one or more Pan-STARRS filters have Kron magnitudes brighter than the PSF magnitude by 0.05 mag. \\
\end{tiny}
\end{table*}
\end{landscape}
\restoregeometry

\section{Stars Excluded from the Final Sample}

\begin{table*}
\begin{scriptsize}
	\caption{Data for Sources Identified as Likely Stars with Atypical Colours}
	\begin{tabular}{rrrrrrrrrrrr}
		\hline
 \multicolumn{2}{c}{IR Survey} &      \multicolumn{5}{c}{Pan-STARRS meanPSF AB mags} & 
   \multicolumn{5}{c}{VISTA Vega mags}\\
   Name & RA$^{\circ}$     & 		$g$ &	$r$ 	&	$i$	& $z$ 	& $y$ 	& $Z$		& $Y$	&	$J$	&	$H$	& $Ks$			\\
     & Decl.$^{\circ}$ & $\pm$  mag	& $\pm$  mag	& $\pm$  mag	& $\pm$   mag	& $\pm$  mag	&  $\pm$ mag	&	 $\pm$ mag		&	 $\pm$		 mag		&	 $\pm$	 mag	&	 $\pm$	 mag		\\
\hline
UDS	&	34.2123189	&			21.7954	&	21.6093	&	20.3610	&	19.8250	&		19.4105	&		&		&	18.3512	&	17.8466	&	17.6092	\\
	&	-4.8947515	&		0.1793	&	0.0835	&	0.0147	&	0.0315	&		0.1145	&		&		&	0.0021	&	0.0019	&	0.0033	\\
UDS	&	34.4109253	&				&	21.7650	&	20.7175	&	19.9836	&		19.7015	&		&		&	18.5133	&	18.0228	&	17.7496	\\
	&	-5.2700687	&			&	0.1675	&	0.0290	&	0.0255	&		0.0731	&		&		&	0.0022	&	0.0019	&	0.0033	\\
UDS	&	34.5890235	&			&	21.8035	&	20.8665	&	20.1967	&		19.7849	&		&		&	18.6272	&	18.1619	&	17.9030	\\
	&	-4.9424929	&				&	0.0724	&	0.0345	&	0.0278	&		0.0966	&		&		&	0.0021	&	0.0020	&	0.0033	\\
VIDEO	&	35.1128282	&		21.6591	&	21.9368	&	20.8698	&	20.1845	&		19.8103	&		&	19.1399	&	18.6294	&	18.1766	&	17.9161	\\
	&	-4.9552726	&			0.1655	&	0.1798	&	0.0538	&	0.0102	&		0.0657	&		&	0.0020	&	0.0025	&	0.0028	&	0.0024	\\
VIDEO	&	52.0756099	&			&		&		&	20.8638	&		19.9288	&	20.2966	&	19.4553	&	18.6419	&	18.1248	&	17.7045	\\
	&	-27.8683759	&			&		&		&	0.0682	&		0.0791	&	0.0063	&	0.0043	&	0.0026	&	0.0037	&	0.0050	\\
DXS	&	333.3096800	&			&	21.6666	&	20.8225	&	20.1457	&		19.7224	&		&		&	18.3903	&	17.8713	&	17.5302	\\
	&	0.7372219	&			&	0.1489	&	0.0248	&	0.0391	&		0.0516	&		&		&	0.0040	&	0.0053	&	0.0051	\\
DXS	&	333.8609666	&			&	21.4346	&	20.2861	&	19.5578	&		19.1589	&		&		&	18.0057	&	17.4381	&	17.1374	\\
	&	0.7496622	&			&	0.1123	&	0.0126	&	0.0266	&	0.2435	&		&		&	0.0055	&	0.0036	&	0.0047	\\
		\hline
			\end{tabular}
\\
\end{scriptsize}
\end{table*}

\begin{table*}
\begin{scriptsize}
	\caption{Likely Stars with Deviant Mean and Aperture Magnitudes}
	\begin{tabular}{rrrrrrrrrrrr}
		\hline
 		Survey & RA$^{\circ}$     & 
 		$Z$ mean & $Z$ Aper & $Y$ mean & $Y$ Aper &   $J$ mean & $J$ Aper &
		$H$ mean & $H$ Aper & $K(s)$ mean & $K(s)$ Aper \\ 
		    & Decl.$^{\circ}$ & $\pm$ mag	& $\pm$  mag	&  $\pm$ mag &  $\pm$ mag & $\pm$ mag	& $\pm$  mag	&     $\pm$ mag &  $\pm$ mag
		 &   $\pm$ mag &  $\pm$ mag \\
\hline
VIDEO	&	35.1191358	&		&		&	19.2971	&	19.2615	&	18.8033	&	18.8031	&	18.3092	&	18.3053	&	18.0712	&	18.0743	\\
	&	-4.4549895	&		&		&	0.0025	&	0.0093	&	0.0037	&	0.0086	&	0.0035	&	0.0082	&	0.0045	&	0.0090	\\
CAL	&	87.6786696	&	17.4104	&	17.5649	&	17.2602	&	17.2153	&	16.7709	&	16.7628	&	16.1723	&	16.1782	&	16.0033	&	16.0350	\\
	&	15.8554744	&	0.0039	&	0.0183	&	0.0031	&	0.0153	&	0.0037	&	0.0172	&	0.0034	&	0.0138	&	0.0036	&	0.0211	\\
CAL	&	276.8783681	&	17.3634	&	17.3458	&	17.0501	&	17.0870	&	16.5950	&	16.6020	&	16.0855	&	16.0882	&	16.0014	&	15.9022	\\
	&	4.5214510	&	0.0031	&	0.0159	&	0.0028	&	0.0144	&	0.0030	&	0.0167	&	0.0027	&	0.0148	&	0.0040	&	0.0218	\\
CAL	&	277.1729349	&	17.5923	&	17.6030	&	17.3352	&	17.3172	&	16.8237	&	16.8150	&	16.3529	&	16.2602	&	16.1989	&	16.0637	\\
	&	4.5807735	&	0.0035	&	0.0184	&	0.0028	&	0.0159	&	0.0035	&	0.0180	&	0.0031	&	0.0161	&	0.0050	&	0.0242	\\
		\hline
			\end{tabular}
\\
\end{scriptsize}
\end{table*}

\clearpage
\section{Finding Charts}

\begin{figure*}
	\includegraphics[scale=0.9]{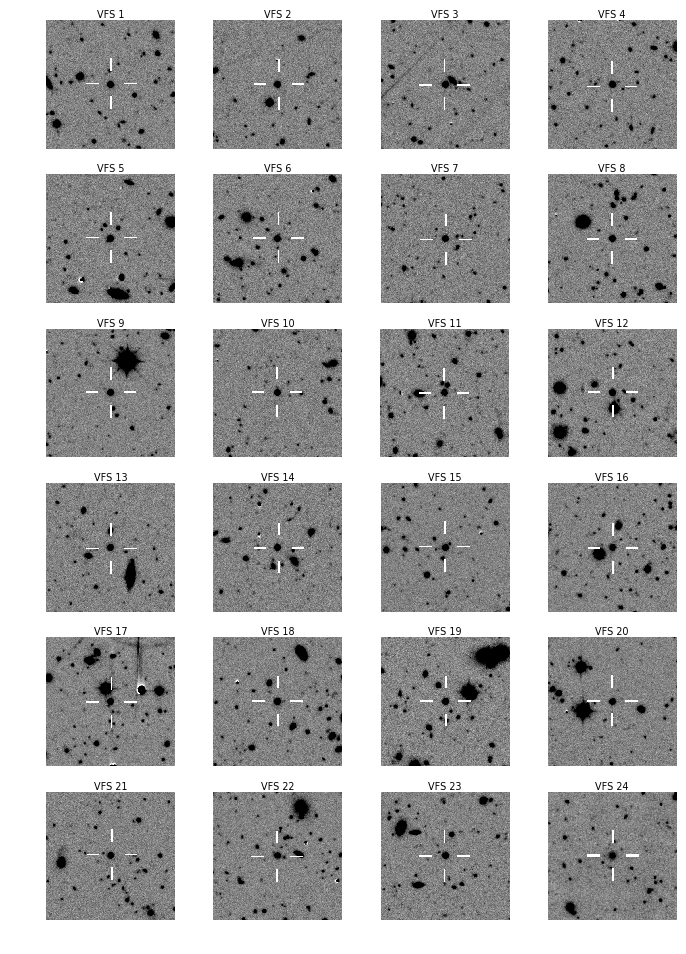}
    \caption{J--band finding charts, 1 arcminute field--of--view, for the standards presented above in Table~\ref{tab:final}. North is up and east to the left.}
    \label{finders1}
\end{figure*}

\setcounter{figure}{0}
\begin{figure*}
	\includegraphics[scale=0.9]{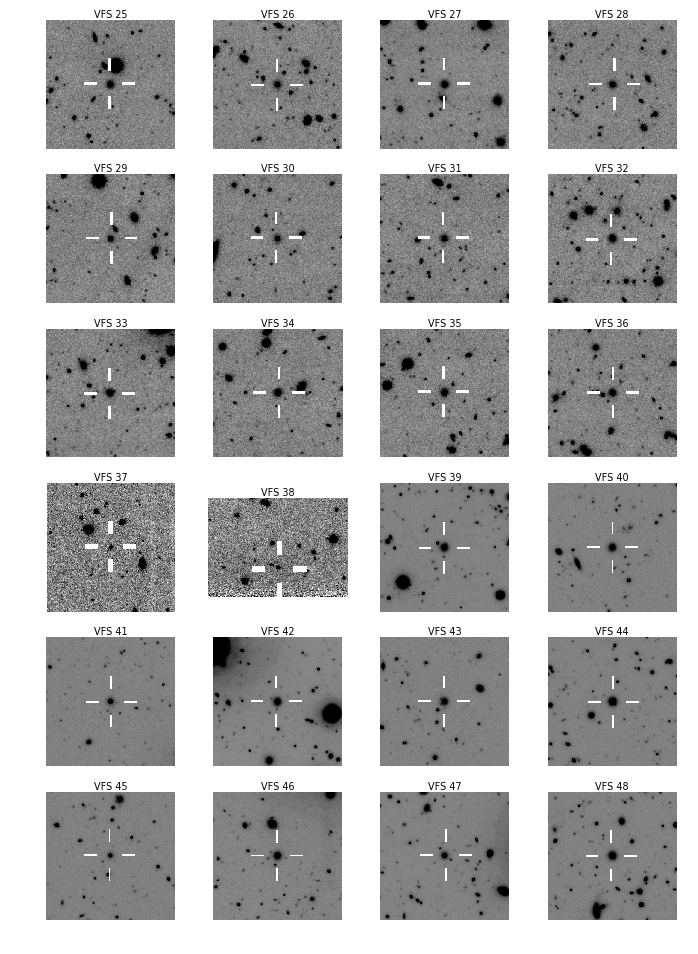}
    \caption{(cont.) J--band finding charts, 1 arcminute field--of--view, for the standards presented above in Table~\ref{tab:final}. North is up and east to the left.}
    \label{finders2}
\end{figure*}

\setcounter{figure}{0}
\begin{figure*}
	\includegraphics[scale=0.9]{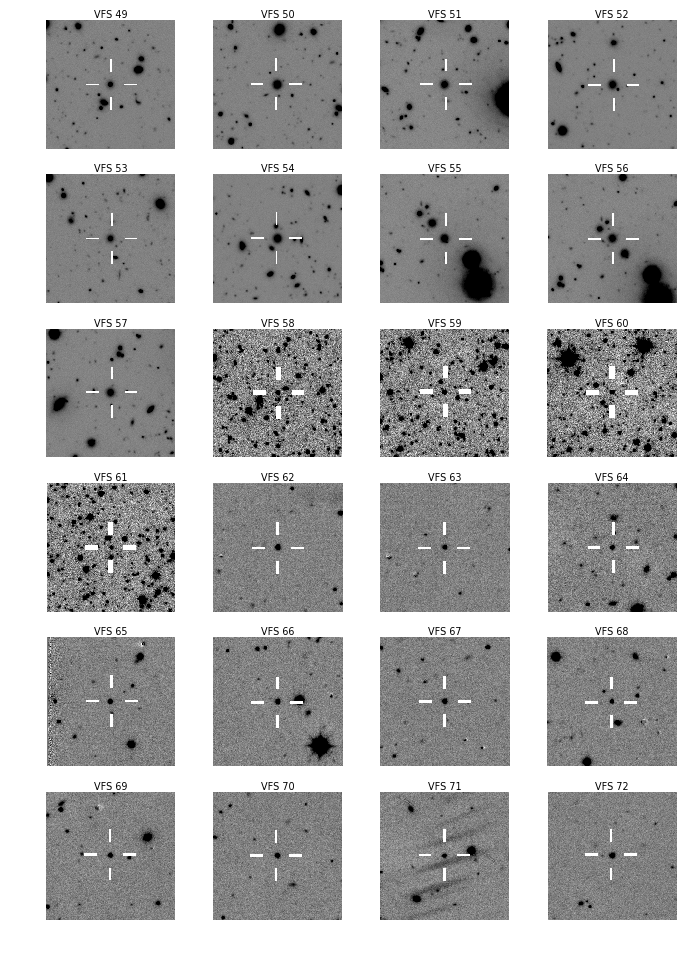}
    \caption{(cont.) J--band finding charts, 1 arcminute field--of--view, for the standards presented above in Table~\ref{tab:final}. North is up and east to the left.}
    \label{finders3}
\end{figure*}

\setcounter{figure}{0}
\begin{figure*}
	\includegraphics[scale=0.9]{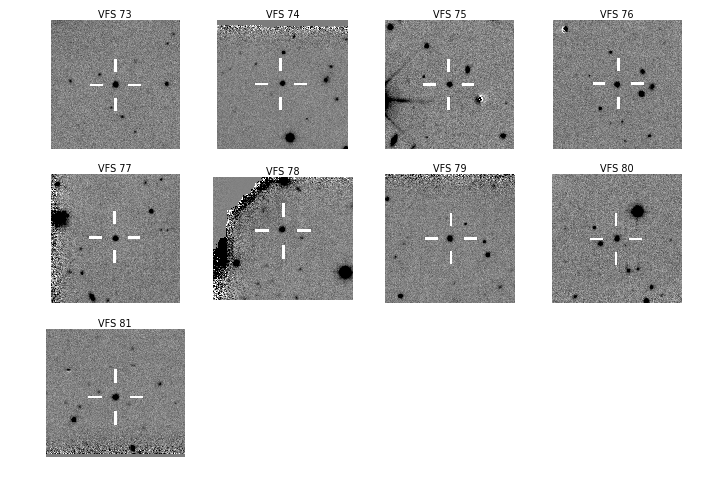}
   \caption{(cont.) J--band finding charts, 1 arcminute field--of--view, for the standards presented above in Table~\ref{tab:final}. North is up and east to the left.}
    \label{finders4}
\end{figure*}


\bsp	
\label{lastpage}
\end{document}